\documentclass[
    journal,
    twocolumn,
]{IEEEtran}

\usepackage{graphicx}
\usepackage{amsmath,amssymb,amsfonts}
\usepackage{epsfig}
\usepackage{epstopdf}
\usepackage{amsthm}
\usepackage{afterpage}
\usepackage{verbatim}
\usepackage{psfrag}
\usepackage{color}
\usepackage{cite}
\usepackage{setspace}
\usepackage{adjustbox}
\usepackage{footmisc}
\usepackage{fmtcount}
\usepackage{stackrel}
\usepackage{float}
\usepackage{breqn}
\usepackage{hyperref}
\usepackage{enumerate}
\usepackage{subcaption}
\usepackage{algorithm}
\usepackage{algpseudocode}
\usepackage{flushend}
\usepackage{soul}
\usepackage{tikz}
\usetikzlibrary{arrows.meta,positioning,calc,shapes.geometric,decorations.pathmorphing}
\usepackage{xcolor}
\usepackage{multirow}
\usepackage[T1]{fontenc}
\usepackage{tabularx,booktabs}

\begin{document}

\title{AI-Empowered UAV-Assisted Backscatter Localization and ISAC for Zero-Energy IoT: A Comprehensive Survey}

\author{Ruhul Amin Khalil, \IEEEmembership{Member, IEEE}
\thanks{R. A. Khalil is with the Engineering Requirements Unit (ERU), College of Engineering, United Arab Emirates University, Al Ain 15551, UAE. e-mail: ruhulamin@uaeu.ac.ae}
}

\maketitle

\begin{abstract}
Zero-energy Internet of Things (IoT) is emerging as a sustainable paradigm for massive sensing networks, in which passive or near-passive devices operate on harvested energy rather than conventional batteries. Backscatter communication (BackCom) enables such operation by allowing passive tags to transmit information through reflection and modulation of incident RF signals. However, BackCom suffers from weak reflected signals, double path loss, limited coverage, direct-link interference, and dependence on external RF sources. Unmanned aerial vehicles (UAVs) can alleviate these limitations by serving as mobile carrier emitters, data collectors, relays, aerial receivers, mobile anchors, sensing platforms, and edge-intelligence nodes. Meanwhile, integrated sensing and communication (ISAC) allows the same wireless resources to support data transmission, localization, target sensing, and environmental awareness. This article presents a comprehensive survey with tutorial elements on RF-based AI-empowered UAV-assisted backscatter localization and ISAC for zero-energy IoT networks, covering foundational and recent developments. We review the enabling technologies, describe a structured, PRISMA-informed review methodology, and propose a unified taxonomy that covers network architectures, UAV roles, backscatter modes, RF sources, localization and sensing functions, AI techniques, and performance metrics. We further discuss UAV-assisted BackCom, passive localization, ISAC-enabled UAV-backscatter systems, and AI-driven optimization, supported by comparative tables, quantitative trend analysis, coverage evaluation, and tutorial-style numerical illustrations. Finally, we identify open challenges and future directions for realistic channel modeling, energy-neutral operation, benchmarking, reproducibility, scalable and trustworthy AI, security and privacy, hardware validation, and integration with RIS, MEC, digital twins, and 6G technologies.
\end{abstract}


\section{Introduction}
\label{sec:introduction}

\IEEEPARstart{T}{he} rapid growth of the Internet of Things (IoT) is creating strong demand for massive, low-cost, and long-lifetime sensing devices. Early IoT visions emphasized the integration of identification, sensing, communication, and intelligence into everyday objects \cite{Atzori2010IoTSurvey,Gubbi2013IoTVision}, while later surveys highlighted the role of RFID, smart sensors, communication protocols, and cloud/edge support in large-scale IoT services \cite{AlFuqaha2015IoTSurvey}. However, conventional battery-powered IoT nodes face limitations in maintenance cost, device size, deployment flexibility, and environmental sustainability. Future wireless systems are therefore expected to support sustainable and ubiquitous connectivity, where sensing, communication, energy efficiency, and intelligence are jointly designed \cite{ITUR2023IMT2030,Saad2020Vision6G,Tataria2021SixGWireless,Letaief2019Roadmap6G}. In this context, zero-energy IoT enables passive or near-passive devices to operate on harvested energy rather than dedicated batteries. RF energy harvesting and wireless-powered communication networks provide foundations for remote energy replenishment and joint information-energy transfer \cite{Lu2015RFEH,Zhou2013SWIPT,Bi2016WPCN,Kamalinejad2015WirelessEnergyIoT}. Recent 3GPP discussions on Ambient IoT further highlight the importance of low-complexity and battery-free devices for future cellular ecosystems \cite{Qu2024AmbientIoT3GPP,ThreeGPP2025AmbientIoT,Lin2024Evolution5G6G}. Backscatter communication (BackCom) is a key enabler because passive tags can transmit information by reflecting and modulating incident RF signals. Ambient and WiFi backscatter demonstrated the feasibility of battery-free communication using existing RF signals \cite{Liu2013AmbientBackscatter,Kellogg2014WiFiBackscatter}, while later surveys summarized BackCom principles, architectures, and challenges \cite{Boyer2014BackscatterRFID,VanHuynh2018AmbientSurvey,Zhang2019NextGenBackscatter}. Since backscatter devices avoid power-hungry active RF transmission, they are highly suitable for large-scale zero-energy IoT deployments \cite{Xu2023AIEmpoweredBackscatter}.

Despite these advantages, BackCom suffers from limited range, weak reflected signals, double path loss, direct-link interference, and dependence on external RF sources \cite{VanHuynh2018AmbientSurvey,Zhang2019NextGenBackscatter,Zhong2024InterferenceUAVBackscatter}. UAVs can mitigate these limitations by acting as mobile carriers, data collectors, relays, aerial access points, mobile anchors, or sensing platforms for passive IoT devices. Compared with fixed infrastructure, UAVs can dynamically adjust altitude, trajectory, and service position to improve link quality, line-of-sight probability, and coverage \cite{Zeng2016UAVComm,Mozaffari2019UAVTutorial,shafiq2026deterministic,Li2019UAV5G,Fotouhi2019UAVCellularSurvey,Shakhatreh2019UAVSurvey,Motlagh2016UAVIoT}. This flexibility is valuable for smart agriculture, disaster monitoring, industrial inspection, environmental sensing, logistics, and remote-area IoT. In UAV-assisted BackCom, the UAV can illuminate passive tags, collect backscattered data, shorten propagation distances, and improve localization and sensing geometry. Existing studies show that trajectory planning, scheduling, power control, and reflection-coefficient design strongly affect throughput, fairness, and energy efficiency \cite{Yang2021UAVBackscatter,Wang2022UAVBackscatterRA}. However, UAV-assisted passive networks introduce additional challenges, including limited UAV battery capacity, propulsion energy, time-varying air-to-ground channels, Doppler effects, and direct-link interference.

Localization is essential in zero-energy IoT because the positions of passive devices are often unknown, time-varying, or difficult to obtain using conventional methods. Accurate localization supports UAV trajectory design, device association, beam control, sensing, and data collection. Wireless localization has been widely studied using RSS, ToA, TDoA, AoA, CSI, and fingerprinting methods \cite{Patwari2005LocatingNodes,Mao2007Localization,Liu2007IndoorPositioning,Zafari2019IndoorLocalization}, while cooperative localization exploits network geometry and shared measurements \cite{Wymeersch2009CooperativeLocalization}. In passive backscatter systems, localization is harder than in active wireless systems because the received signal is weak, indirect, and affected by tag orientation, cascaded propagation, multipath, and the relative geometry among the RF source, tag, UAV, and receiver. Recent localization surveys also emphasize robustness, privacy, and security as key requirements for practical location-aware IoT services \cite{Pettorru2024TrustworthyLocalization}. UAV mobility can improve localization by collecting measurements from diverse spatial viewpoints, but it also couples localization accuracy with trajectory, energy, and communication constraints.

\begin{figure*}[!t]
    \centering
    \includegraphics[width=0.785\textwidth]{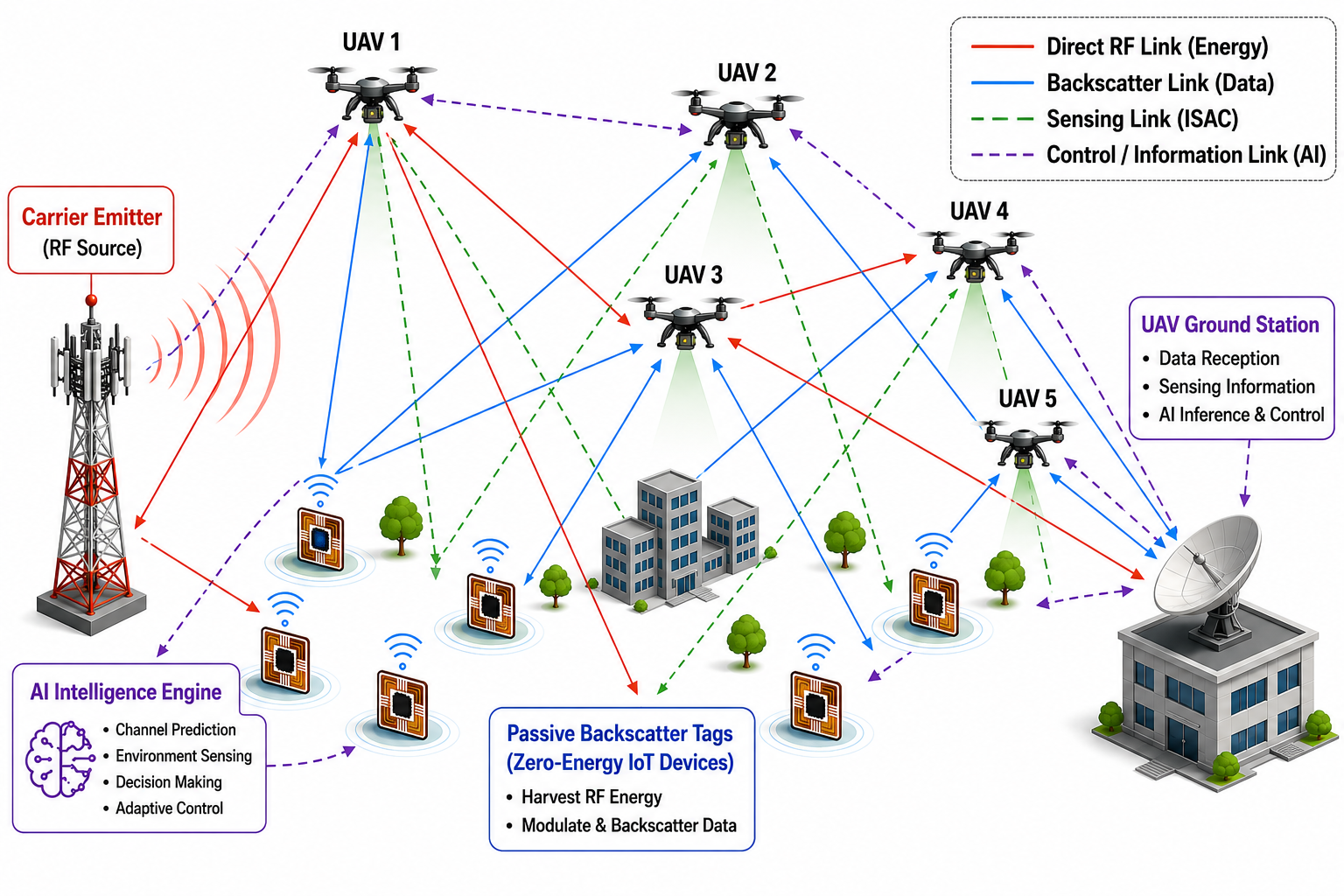}
    \caption{Representative AI-empowered UAV-assisted backscatter localization and ISAC scenario for zero-energy IoT.}
    \label{fig:intro_ai_uav_backscatter_scenario}
\end{figure*}

Integrated sensing and communication (ISAC) further extends wireless systems by enabling simultaneous data transmission and environmental sensing. Instead of designing communication and sensing separately, ISAC enables shared use of spectrum, hardware, waveforms, beamforming, and signal processing \cite{Liu2020JointRadarComm,Zhang2021JCRSignalProcessing,Liu2022ISACSurvey,Liu2022FundamentalLimitsISAC,althobaiti2023robust,Wei2023ISACSignals,Wen2024ISCCSurvey,khalil2024robust}. This is highly relevant to backscatter-based IoT, where the same reflected signal can carry device data and provide sensing or localization information. Recent works on sensing and BackCom integration show that BackCom can support low-power sensing, tag detection, parameter estimation, and environment-aware communication \cite{Zargari2023BackscatterSensingIntegration,Zhao2024BISAC}. When UAVs are introduced, the system gains additional flexibility because the aerial platform can adapt its sensing viewpoint, illumination direction, and communication geometry. UAV-assisted ISAC has been studied in IoT scenarios involving three-dimensional trajectory optimization, sensing task scheduling, and resource allocation \cite{Liu2024UAVISACIoT,Ahmed2025UAVISACSurvey,althobaiti2023robust}. Nevertheless, combining UAV mobility, BackCom, localization, ISAC, and zero-energy operation creates complex tradeoffs among sensing accuracy, communication rate, localization error, harvested energy, interference suppression, and UAV flight energy. Fig.~\ref{fig:intro_ai_uav_backscatter_scenario} illustrates a representative AI-empowered UAV-assisted backscatter localization and ISAC scenario.

AI is becoming an important tool for addressing these coupled, dynamic design problems. At the physical layer, deep learning has been used for transceiver design, channel estimation, signal detection, and end-to-end wireless optimization \cite{OShea2017DeepLearningPhysical,Zhang2019DeepLearningWireless,Chen2020TowardsAI6G}. In AI-empowered BackCom, learning techniques can support channel estimation, signal detection, interference suppression, tag classification, reflection control, and resource allocation \cite{Xu2023AIEmpoweredBackscatter}. For UAV-assisted systems, machine learning and reinforcement learning can optimize UAV trajectory, scheduling, transmit power, data collection, and sensing decisions under uncertain channel and mobility conditions \cite{Mozaffari2019UAVTutorial,Sutton2018RL}. For localization and ISAC, deep learning can exploit signal fingerprints, spatial correlations, CSI patterns, and temporal features to improve estimation accuracy in complex environments \cite{Zhang2025IntelligentISAC}. However, AI-based solutions also face challenges related to training data scarcity, limited generalization, computational complexity, explainability, and robustness. Edge computing, federated learning, RISs, and digital twins can further support scalable and adaptive zero-energy IoT systems, but they also introduce new coordination and design issues \cite{Mao2017MEC,Kairouz2021FederatedLearning,Basar2019RIS,Khan2022DigitalTwinWireless}. Motivated by these opportunities and challenges, this paper presents a systematic survey of AI-empowered UAV-assisted backscatter localization and ISAC for zero-energy IoT.

The scope of this survey is deliberately cross-layer. At the device level, it considers passive tags, reflection coefficients, and harvested energy. At the aerial-network level, it considers UAV trajectory, placement, hovering, and multi-UAV cooperation. At the signal-processing level, it considers localization features, sensing gains, beamforming, interference cancellation, and cascaded-channel estimation. At the intelligence level, it considers supervised learning, deep learning, reinforcement learning, edge intelligence, federated learning, and trustworthy AI. This layered perspective is useful because practical zero-energy IoT systems cannot be optimized from only one viewpoint. A design that maximizes communication rate may be energy-infeasible, while a design that maximizes sensing gain may not provide reliable tag data collection. The survey is organized to make these dependencies explicit.

\subsection{Related Surveys}
\label{subsec:related_surveys}
This comparison clarifies why the topic under consideration should not be treated as a simple extension of any single existing survey stream. UAV-assisted passive IoT has a cascaded propagation structure, passive reflection control, and strict energy-causality constraints that are not present in most active-radio UAV studies. Likewise, backscatter-ISAC must extract communication, sensing, and localization information from weak reflected signals while suppressing direct-link interference. These characteristics justify a dedicated taxonomy and a separate comparative analysis.

The role of AI also differs from many conventional wireless surveys. In the considered setting, AI is not only a resource-allocation tool; it can influence trajectory control, tag scheduling, reflection design, sensing-task selection, localization inference, channel estimation, and interference mitigation. Therefore, this survey treats AI as a cross-layer intelligence mechanism that interacts with physical-layer signals, UAV mobility, passive energy constraints, and network-level optimization.

Several surveys have examined individual components of the topic under consideration, including IoT, RF energy harvesting, BackCom, UAV-assisted networks, localization, ISAC, and AI-enabled wireless systems. Early IoT surveys established the foundations of pervasive sensing, identification, communication protocols, and large-scale architectures \cite{Atzori2010IoTSurvey,Gubbi2013IoTVision,AlFuqaha2015IoTSurvey}. Similarly, surveys on RF energy harvesting and wireless-powered communication explained how low-power devices harvest RF energy and jointly optimize energy and information transfer \cite{Lu2015RFEH,Bi2016WPCN}. These works are important for zero-energy IoT, but they do not focus on passive backscatter localization, UAV assistance, or ISAC-enabled sensing and communication.

Backscatter communication has been surveyed from different viewpoints. Ambient BackCom surveys discuss passive architectures, ambient RF sources, interference, and IoT applications \cite{VanHuynh2018AmbientSurvey,Zhang2019NextGenBackscatter}. Recent surveys focus on AI-empowered BackCom, where machine learning supports signal detection, channel estimation, interference mitigation, resource allocation, and intelligent passive communication \cite{Xu2023AIEmpoweredBackscatter}. Millimeter-wave BackCom surveys highlight directional beams, large bandwidth, and compact arrays for high-rate and sensing-aware passive systems \cite{Chen2024MMWaveBackscatter,Han2024THzISAC,Elbir2024THzISAC}. However, these surveys primarily focus on BackCom itself and do not provide a unified treatment of UAV-assisted localization and ISAC for zero-energy IoT.

\begin{table*}[!t]
\centering
\caption{Comparison of Related Surveys and the Present Survey.}
\label{tab:related_surveys}
\footnotesize
\renewcommand{\arraystretch}{1.25}
\begin{tabularx}{\textwidth}{|p{0.15\textwidth} |p{0.23\textwidth}| X |X|}
\hline
\hline
\textbf{Survey / Reference} & \textbf{Main Focus} & \textbf{What is Missing with Respect to This Topic} & \textbf{Coverage in This Survey} \\
\hline
\hline
IoT surveys \cite{Atzori2010IoTSurvey,Gubbi2013IoTVision,AlFuqaha2015IoTSurvey} & IoT architectures, enabling technologies, protocols, applications, and cloud/edge-supported services & Do not focus on zero-energy passive devices, UAV-assisted BackCom, localization, or ISAC & Positions zero-energy IoT as the target domain and links it with BackCom, UAVs, AI, localization, and ISAC \\
\hline
RF energy harvesting and WPCN surveys \cite{Lu2015RFEH,Bi2016WPCN} & RF energy harvesting, wireless-powered communication, and energy-information transfer & Limited discussion of backscatter modulation, UAV mobility, localization, and sensing-communication integration & Explains energy-neutral operation alongside passive BackCom, UAV-assisted carrier delivery, and ISAC energy tradeoffs \\
\hline
Ambient/backscatter communication surveys \cite{VanHuynh2018AmbientSurvey,Zhang2019NextGenBackscatter} & BackCom principles, ambient RF sources, architectures, modulation, interference, and IoT applications & UAV-assisted deployment, AI-based optimization, localization, and ISAC are not jointly covered & Extends BackCom analysis toward UAV-assisted zero-energy IoT, localization, ISAC, and AI-driven design \\
\hline
AI-empowered BackCom survey \cite{Xu2023AIEmpoweredBackscatter} & AI for BackCom, including channel estimation, detection, interference mitigation, and resource allocation & Does not systematically focus on UAV-assisted localization or UAV-backscatter-ISAC integration & Uses AI as a cross-layer tool for UAV trajectory control, passive localization, ISAC optimization, and resource management \\
\hline
mmWave BackCom survey \cite{Chen2024MMWaveBackscatter} & mmWave backscatter architectures, high-rate passive communication, directional links, and applications & Limited treatment of UAV-assisted zero-energy IoT, AI-enabled control, and integrated localization-ISAC design & Includes RF-source and spectrum perspectives, including mmWave opportunities for high-resolution sensing and localization \\
\hline
UAV communication surveys \cite{Zeng2016UAVComm,Mozaffari2019UAVTutorial,Li2019UAV5G,Gu2023UAVSurvey} & UAV base stations, relays, aerial data collection, trajectory optimization, channel modeling, and 5G/6G applications & Mostly focus on active-radio networks and do not jointly address passive BackCom, zero-energy IoT, localization, and ISAC & Classifies UAV roles as carrier emitter, collector, receiver, relay, anchor, sensing platform, and edge-intelligence node \\
\hline
Localization surveys \cite{Patwari2005LocatingNodes,Mao2007Localization,Wymeersch2009CooperativeLocalization,Zafari2019IndoorLocalization,Pettorru2024TrustworthyLocalization} & Wireless localization, cooperative localization, indoor positioning, security, privacy, and trustworthiness & Do not focus on passive backscatter localization under UAV mobility and cascaded channels & Reviews UAV-assisted backscatter localization using RSS, CSI, AoA, ToA, TDoA, fingerprints, cooperation, and AI methods \\
\hline
ISAC surveys \cite{Liu2020JointRadarComm,Zhang2021JCRSignalProcessing,Liu2022ISACSurvey} & Joint radar-communication design, ISAC waveforms, beamforming, signal processing, and tradeoffs & Backscatter-based passive IoT and UAV-assisted zero-energy operation are not the main focus & Connects ISAC with BackCom, UAV trajectory design, localization, sensing, reflection control, and energy constraints \\
\hline
UAV-ISAC survey \cite{Ahmed2025UAVISACSurvey} & UAV-enabled ISAC, including channel estimation, beam tracking, throughput, AoI, energy efficiency, and security & Limited emphasis on passive BackCom, zero-energy IoT, and backscatter localization & Extends UAV-ISAC toward passive tags, backscatter reflection, AI-based localization, and energy-neutral operation \\
\hline
Backscatter-sensing and B-ISAC works \cite{Zargari2023BackscatterSensingIntegration,Zhao2024BISAC} & Integration of BackCom with sensing and early B-ISAC frameworks & Do not provide broad systematic coverage of UAV-assisted architectures, AI techniques, localization, and zero-energy tradeoffs & Develops a unified taxonomy and comparative framework for AI-empowered UAV-assisted backscatter localization and ISAC \\
\hline
Ambient IoT and 3GPP-oriented surveys \cite{Qu2024AmbientIoT3GPP,ThreeGPP2025AmbientIoT} & Ambient IoT use cases, requirements, device types, connectivity topology, and standardization & Limited technical treatment of UAV-assisted BackCom, localization, ISAC, and AI-based optimization & Relates Ambient IoT trends to UAV-supported passive sensing, localization, and communication \\
\hline
\textbf{This survey} & AI-empowered UAV-assisted backscatter localization and ISAC for zero-energy IoT & -- & Provides a unified taxonomy, methodology, comparative analysis, AI-enabled design perspective, and future directions across BackCom, UAVs, localization, ISAC, and zero-energy IoT \\
\hline
\hline
\end{tabularx}
\end{table*}

UAV-assisted wireless communications have also been widely reviewed. Existing UAV surveys summarize aerial base stations, relays, data collection, trajectory optimization, air-to-ground channel modeling, energy constraints, and 5G/6G integration \cite{Zeng2016UAVComm,Mozaffari2019UAVTutorial,Li2019UAV5G,Gu2023UAVSurvey}. These studies provide a foundation for UAV mobility and network design, but mostly focus on active radio systems rather than passive backscatter IoT. Similarly, localization surveys cover RSS-, CSI-, AoA-, ToA-, TDoA-, fingerprinting-, and cooperative-localization methods, including security and trustworthiness issues \cite{Patwari2005LocatingNodes,Mao2007Localization,Wymeersch2009CooperativeLocalization,Zafari2019IndoorLocalization,Pettorru2024TrustworthyLocalization}. Nevertheless, localization in passive backscatter networks is more challenging due to cascaded channels, weak reflections, tag orientation, and the coupling between UAV trajectory and signal geometry.

ISAC surveys explain how wireless systems can jointly support data transmission and sensing using shared spectrum, hardware, waveforms, and signal processing \cite{Liu2020JointRadarComm,Zhang2021JCRSignalProcessing,Liu2022ISACSurvey}. Recent works also discuss intelligent ISAC, UAV-enabled ISAC, and backscatter-ISAC as emerging directions \cite{Zhang2025IntelligentISAC,Ahmed2025UAVISACSurvey,Zargari2023BackscatterSensingIntegration,Zhao2024BISAC,Fang2023MIMOISAC}. However, these works either focus on ISAC in general, UAV-ISAC without passive BackCom, or BackCom-sensing integration without a systematic UAV-assisted zero-energy IoT perspective. In contrast, this survey jointly considers AI, UAV mobility, BackCom, localization, ISAC, and zero-energy IoT under a unified taxonomy and comparative framework. Table~\ref{tab:related_surveys} summarizes the differences between existing surveys and this work.

\subsection{Main Contributions}
This survey is written with both review and tutorial objectives. Where appropriate, compact equations are included to show how harvested energy, cascaded backscatter channels, UAV--tag distance, localization measurements, sensing gain, learning policies, and multi-objective rewards are commonly represented. These equations are not intended to define a single universal system model; instead, they provide a common notation that helps connect otherwise separate studies. This is useful for readers who wish to move from a survey-level understanding to formulating new optimization or learning problems.

Beyond summarizing previous studies, this survey emphasizes the relationships among the main system functions. In particular, it highlights how UAV altitude, trajectory, RF-source placement, reflection coefficient, array directivity, localization feature selection, and AI complexity jointly affect communication, sensing, localization, and energy performance. This helps move the discussion beyond a paper-by-paper review toward a design-oriented understanding of why particular assumptions, metrics, and optimization variables matter.

This paper presents a systematic survey of AI-empowered UAV-assisted backscatter localization and ISAC for zero-energy IoT networks. The main contributions are summarized as follows:

\begin{itemize}
    \item Firstly, we review the key enabling technologies, including zero-energy IoT, RF energy harvesting, BackCom, UAV-assisted wireless networks, localization, ISAC, and AI-driven wireless optimization.
    \item Secondly, we develop a unified taxonomy that classifies existing studies according to network architecture, UAV role, backscatter mode, RF source, localization method, sensing function, AI technique, and performance metric.
    \item Then, we review representative works on UAV-assisted BackCom, backscatter localization, ISAC-enabled UAV-backscatter systems, and AI-based design optimization, highlighting their assumptions, objectives, techniques, and limitations.
    \item Finally, we provide comparative tables, quantitative trends, coverage analysis, tutorial-style numerical illustrations, and open research directions toward scalable, sustainable, and sensing-aware zero-energy IoT systems.
\end{itemize}

\begin{figure*}[!t]
    \centering
    \includegraphics[width=0.95\textwidth]{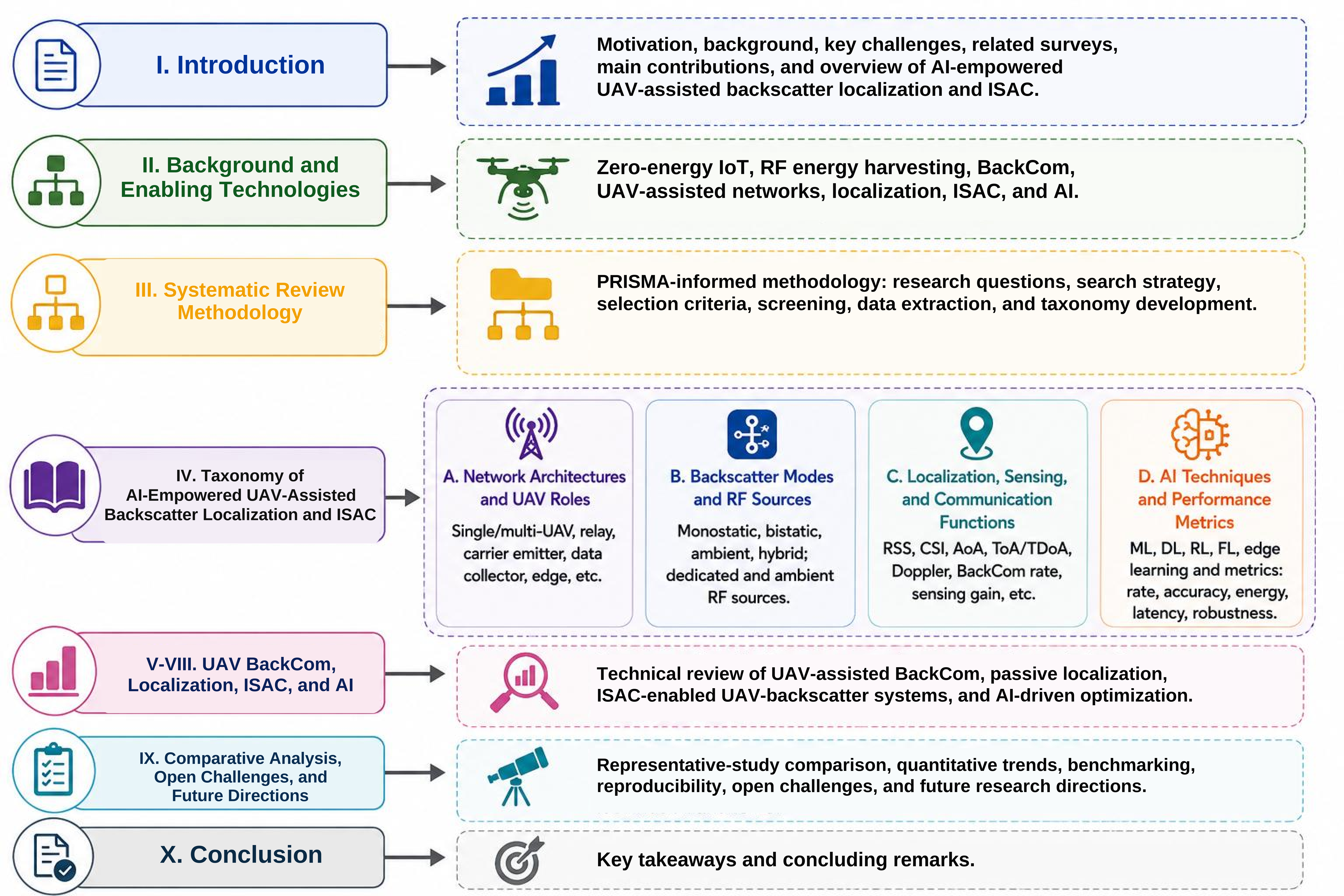}
    \caption{Organization of this survey paper.}
    \label{fig:survey_organization}
\end{figure*}

\subsection{Organization of the Paper}
The remainder of this paper is organized as follows. Section~II presents the background and enabling technologies, including zero-energy IoT, BackCom, UAV-assisted networks, localization, ISAC, and AI-based wireless optimization. Section~III describes the systematic review methodology. Section~IV introduces the proposed taxonomy. Section~V reviews UAV-assisted BackCom for zero-energy IoT. Section~VI discusses backscatter localization in UAV-assisted IoT networks. Section~VII examines ISAC-enabled UAV-backscatter systems. Section~VIII discusses AI-empowered design and optimization. Section~IX provides comparative analysis, open challenges, and future research directions. Finally, Section~X concludes the paper. Fig.~\ref{fig:survey_organization} illustrates the organization of the survey.

\section{Background and Enabling Technologies}
The integration of these technologies is attractive because each one addresses a different limitation. BackCom reduces active transmission power, UAVs improve geometry and coverage, ISAC increases spectrum and hardware reuse, localization enables position-aware data collection, and AI supports adaptation under uncertainty \cite{khalil2024robust}. At the same time, their integration creates new coupling effects. For instance, a UAV trajectory that improves harvested energy may not be optimal for sensing, and a reflection coefficient that improves communication may reduce the energy available for future tag operation.

The research area addressed in this survey combines zero-energy IoT, BackCom, UAV-assisted networking, localization, ISAC, and AI. These technologies are individually important for future wireless networks, but their integration is especially promising for sustainable and intelligent IoT systems. Large-scale IoT systems rely on pervasive sensing, identification, interoperability, and intelligent services \cite{Atzori2010IoTSurvey,Gubbi2013IoTVision,AlFuqaha2015IoTSurvey}, while the IMT-2030 framework identifies ubiquitous connectivity, ISAC, AI-native operation, and sustainability as major future directions \cite{ITUR2023IMT2030}. In this context, UAV-assisted backscatter localization and ISAC can provide flexible connectivity, low-power operation, and environment-aware sensing for passive IoT devices. Table~\ref{tab:background_enabling_technologies} summarizes the main enabling technologies.

\begin{table*}[!t]
\centering
\caption{Summary of Background and Enabling Technologies.}
\label{tab:background_enabling_technologies}
\footnotesize
\renewcommand{\arraystretch}{1.25}
\setlength{\tabcolsep}{3.5pt}
\begin{tabularx}{\textwidth}{|p{0.13\textwidth}| X| X| X| p{0.14\textwidth}|}
\hline\hline
\textbf{Enabling Technology} & \textbf{Core Concept} & \textbf{Key Functions in the Considered System} & \textbf{Main Challenges and Relevance to This Survey} & \textbf{Key References} \\
\hline\hline
Zero-energy IoT & Passive or near-passive IoT devices operate using harvested energy instead of conventional batteries & Battery-free sensing, massive low-cost deployment, long-term monitoring, and sustainable operation & Limited harvested energy, intermittent operation, hardware constraints, and reliability motivate energy-aware UAV-assisted BackCom and ISAC & \cite{Lu2015RFEH,Bi2016WPCN,Zhou2013SWIPT} \\
\hline
Backscatter communication & Tags transmit by modulating and reflecting an incident RF signal & Ultra-low-power communication, passive data transfer, ambient RF reuse, and energy-efficient IoT connectivity & Weak reflections, double path loss, short range, and direct-link interference motivate UAV mobility and AI & \cite{Liu2013AmbientBackscatter,Kellogg2014WiFiBackscatter,VanHuynh2018AmbientSurvey,Xu2023AIEmpoweredBackscatter,Zhong2024InterferenceUAVBackscatter} \\
\hline
UAV-assisted networks & UAVs act as aerial base stations, relays, collectors, emitters, anchors, or sensing platforms & Flexible coverage, LoS links, adaptive data collection, RF illumination, and improved localization geometry & Flight energy, trajectory constraints, air-to-ground channels, and regulation require joint optimization & \cite{Zeng2016UAVComm,Mozaffari2019UAVTutorial,Gu2023UAVSurvey,Yang2021UAVBackscatter,Wang2022UAVBackscatterRA} \\
\hline
Localization & Positions are estimated using RSS, CSI, AoA, ToA, TDoA, Doppler, or fingerprints & Supports UAV trajectory planning, tag association, beam control, sensing, and data collection & Passive-tag localization is difficult due to weak cascaded channels, tag orientation, multipath, and UAV motion & \cite{Patwari2005LocatingNodes,Mao2007Localization,Wymeersch2009CooperativeLocalization,Zafari2019IndoorLocalization,Pettorru2024TrustworthyLocalization} \\
\hline
ISAC & Communication and sensing share spectrum, hardware, waveforms, and processing & Simultaneous data transmission, target detection, parameter estimation, localization, and environment awareness & Rate, sensing accuracy, localization precision, harvested energy, and UAV energy are coupled & \cite{Liu2020JointRadarComm,Zhang2021JCRSignalProcessing,Liu2022ISACSurvey,Zargari2023BackscatterSensingIntegration,Zhao2024BISAC,Liu2024UAVISACIoT} \\
\hline
AI & Learning supports adaptive decision-making in dynamic wireless environments & Channel estimation, detection, interference suppression, trajectory control, resource allocation, localization, and ISAC optimization & Data scarcity, generalization, complexity, explainability, and robustness remain major issues & \cite{OShea2017DeepLearningPhysical,Zhang2019DeepLearningWireless,Xu2023AIEmpoweredBackscatter,Zhang2025IntelligentISAC,Kairouz2021FederatedLearning} \\
\hline\hline
\end{tabularx}
\end{table*}

\subsection{Zero-Energy IoT and Backscatter Communication}
The practical circuit model of a zero-energy tag is also important. Even when RF energy is available, the tag may require a minimum activation threshold before it can sense, switch impedance, store data, or backscatter information. This means that the useful service region is determined not only by the communication SINR but also by the probability that harvested energy exceeds the circuit's requirements. In UAV-assisted systems, this threshold behavior affects whether the UAV should transmit a stronger carrier signal, hover longer, or return later once the tag has accumulated enough energy.

From a design perspective, the reflection coefficient is not only a communication variable but also an energy-management and sensing variable. A larger reflection coefficient can improve the received backscatter signal and make the tag more visible for localization or sensing, but it can reduce the fraction of incident RF power available for harvesting. Conversely, allocating more power to harvesting can improve long-term tag availability but reduce instantaneous communication rate. This reflection--harvesting coupling is one of the main differences between passive backscatter IoT and conventional active low-power IoT.

The source of the incident signal also changes the system behavior. Dedicated RF sources provide controllable power, waveform, and scheduling, which simplify optimization but increase infrastructure or UAV energy costs \cite{khalil2023convex}. Ambient RF sources reduce dedicated power consumption but introduce uncertainty in signal availability, waveform structure, and interference. Hybrid source architectures are therefore attractive because they can combine the robustness of dedicated illumination with the sustainability of ambient RF reuse.

Zero-energy IoT refers to passive or near-passive devices that operate with little or no battery support by harvesting energy from RF signals, solar energy, vibration, thermal gradients, or other ambient sources. Such devices are attractive for applications where battery replacement is costly or impractical, including smart agriculture, logistics, structural monitoring, industrial automation, and environmental sensing. RF energy harvesting and wireless-powered communication networks provide foundations for powering low-energy devices over the air \cite{Lu2015RFEH,Bi2016WPCN,Clerckx2019WIPTFundamentals}. Simultaneous wireless information and power transfer shows how information and energy delivery can be jointly designed, although practical receivers introduce rate-energy tradeoffs \cite{Zhou2013SWIPT}. Recent 3GPP discussions on Ambient IoT further highlight the need for low-complexity and low-power devices beyond conventional battery-powered NB-IoT and eMTC systems \cite{Qu2024AmbientIoT3GPP,ThreeGPP2025AmbientIoT}. The harvested RF energy at tag $k$ can be approximated as
\begin{equation}
    E_{k}^{\mathrm{har}}=\eta_k P_{k}^{\mathrm{in}} T,
\label{eq:harvested_energy_background}
\end{equation}
where $\eta_k$ is the RF-to-DC conversion efficiency, $P_{k}^{\mathrm{in}}$ is the incident RF power, and $T$ is the harvesting duration \cite{Lu2015RFEH,Bi2016WPCN}. 

BackCom is highly suitable for zero-energy IoT because it avoids active RF transmission. A backscatter device modulates information by varying its antenna impedance and reflecting the incident carrier. A simplified received signal is
\begin{equation}
    y(t)=h_{sr}x(t)+h_{st}\Gamma h_{tr}x(t)+n(t),
\label{eq:basic_backscatter_model}
\end{equation}
where $h_{sr}$ is the direct source-to-receiver channel, $h_{st}$ and $h_{tr}$ are the source-to-tag and tag-to-receiver channels, $\Gamma$ is the tag reflection coefficient, $x(t)$ is the incident signal, and $n(t)$ is noise \cite{Boyer2014BackscatterRFID,VanHuynh2018AmbientSurvey}. Classical RFID, ambient backscatter, WiFi backscatter, BackFi, passive WiFi, inter-technology backscatter, and LoRa backscatter demonstrated practical passive communication using dedicated or ambient RF signals \cite{Boyer2014BackscatterRFID,Liu2013AmbientBackscatter,Kellogg2014WiFiBackscatter,Bharadia2015BackFi,Kellogg2016PassiveWiFi,Iyer2016InterTechnologyBackscatter,Talla2017LoRaBackscatter}. BackCom systems are commonly categorized as monostatic, bistatic, or ambient depending on the relative placement of the emitter, tag, and receiver \cite{VanHuynh2018AmbientSurvey,Zhang2019NextGenBackscatter,Katanbaf2021FullDuplexLoRaBackscatter}. However, weak reflections, double path loss, short range, and direct-link interference motivate the integration of UAV mobility, ISAC, and AI-based optimization \cite{Zhong2024InterferenceUAVBackscatter}.

\subsection{UAV-Assisted Wireless Networks}
The UAV operating mode changes the mathematical structure of the design problem. If the UAV acts as a carrier emitter, its transmit power and position determine tag activation probability and harvested energy. If it acts as a receiver or collector, its trajectory determines backscatter link quality and data freshness. If it acts as a mobile anchor, its path determines localization geometry and measurement diversity. If it acts as an ISAC platform, all these functions must be balanced under flight-energy and mission-time constraints. Consequently, UAV-assisted passive IoT is naturally a cross-layer problem rather than a conventional link-budget problem.

UAV-assisted wireless networks use UAVs as aerial base stations, relays, data collectors, energy transmitters, sensing platforms, or mobile edge nodes. Compared with fixed infrastructure, UAVs can adjust altitude, trajectory, and service location to improve coverage and line-of-sight connectivity \cite{Zeng2016UAVComm,Mozaffari2019UAVTutorial}. UAV communications have been widely studied for 5G, beyond-5G, and non-terrestrial networks due to flexible deployment and controllable mobility \cite{Li2019UAV5G,Gu2023UAVSurvey,Rinaldi2020NTNBeyond,Azari2022EvolutionNTN,Giordani2021NTN6G}. In zero-energy IoT, UAVs can support passive devices by transmitting carrier signals, collecting backscattered data, reducing propagation distance, and improving localization/sensing geometry.

The integration of UAVs with BackCom creates a flexible but challenging framework. A UAV may serve as a carrier-emitter, backscatter receiver, relay, mobile anchor, or ISAC platform. Existing UAV-assisted BackCom studies show that trajectory planning, power control, scheduling, and reflection-coefficient design must be jointly optimized to improve throughput and energy efficiency \cite{Yang2021UAVBackscatter}. Resource-allocation studies further show that UAV transmit power, tag scheduling, reflection coefficients, and passive-device access strategies affect fairness and reliability \cite{Wang2022UAVBackscatterRA}. However, UAV-assisted systems remain constrained by flight time, propulsion energy, payload capacity, air-to-ground channel variation, and regulations \cite{Mozaffari2019UAVTutorial}. Thus, UAV-assisted BackCom requires cross-layer design across mobility, communication, energy harvesting, localization, and interference management.

\subsection{Localization and ISAC Fundamentals}
The coupling among localization, sensing, communication, and energy is especially strong in backscatter-enabled ISAC because the sensing object, communication terminal, and energy-harvesting device may be the same passive tag. A UAV movement that improves the communication channel may also improve localization geometry, but it may reduce coverage for other tags or increase propulsion energy. A waveform that improves sensing resolution may not be optimal for passive modulation. These interactions motivate joint utility functions and multi-objective optimization formulations.

Localization estimates the position of a device, object, or target using wireless measurements or environmental information. In IoT systems, localization supports asset tracking, device association, context-aware sensing, mobility management, and efficient data collection. Common features include RSS, CSI, AoA, ToA, TDoA, Doppler information, and RF fingerprints \cite{Patwari2005LocatingNodes,Mao2007Localization,Liu2007IndoorPositioning,Zafari2019IndoorLocalization}. Cooperative localization improves accuracy by allowing multiple nodes to share measurements and exploit spatial relationships \cite{Wymeersch2009CooperativeLocalization}. For passive backscatter devices, localization is more difficult because the received signal is weak, indirect, and dependent on tag orientation, multipath, and the emitter--tag--UAV--receiver geometry. Recent IoT localization surveys also emphasize robustness, privacy, and trustworthiness \cite{Pettorru2024TrustworthyLocalization}.  

ISAC integrates wireless sensing and data communication by sharing spectrum, hardware, waveform, beamforming, and signal-processing resources \cite{Liu2020JointRadarComm,Zhang2021JCRSignalProcessing,Liu2022ISACSurvey}. In backscatter-enabled IoT, the same reflected signal can carry tag information and provide sensing or localization cues. A simplified normalized ISAC utility can be represented as
\begin{equation}
    \mathcal{U}=\omega_c \bar{R}+\omega_s \bar{S},
\label{eq:isac_utility_background}
\end{equation}
where $\bar{R}$ and $\bar{S}$ denote normalized communication and sensing metrics, respectively, and $\omega_c$ and $\omega_s$ are non-negative weighting coefficients \cite{Liu2022ISACSurvey}. Normalization is important because communication rate and sensing quality may have different physical units. Recent backscatter-ISAC studies show that BackCom and ISAC can support low-power information transfer, tag detection, parameter estimation, and environment-aware operation \cite{Zargari2023BackscatterSensingIntegration,Zhao2024BISAC}. UAV-assisted ISAC for IoT has also been studied in the context of task scheduling, three-dimensional trajectory optimization, and resource allocation \cite{Liu2024UAVISACIoT}. The main difficulty is that communication rate, sensing accuracy, localization precision, harvested energy, and UAV flight energy are tightly coupled.

\subsection{AI for Intelligent Wireless Systems}
For zero-energy IoT, the role of AI must be matched to the device's capabilities. Passive tags generally cannot execute computationally intensive learning algorithms, exchange gradients frequently, or provide dense feedback. Therefore, most intelligence is expected to reside at UAVs, edge servers, access points, or cloud-assisted controllers. This makes edge inference, split learning, model compression, and federated learning particularly relevant. At the same time, AI-driven policies must respect safety, UAV energy, and tag energy-causality constraints.

AI has become important for wireless systems operating in dynamic, heterogeneous, and uncertain environments. Deep learning has been explored for physical-layer tasks such as end-to-end transceiver design, signal detection, and channel estimation \cite{OShea2017DeepLearningPhysical}, while broader machine-learning methods support resource management, mobility control, and network analytics \cite{Zhang2019DeepLearningWireless}. In BackCom, AI can assist channel estimation, signal detection, tag classification, interference cancellation, resource allocation, and reflection control when analytical models are incomplete or complex \cite{Xu2023AIEmpoweredBackscatter}. For UAV networks, learning can optimize trajectory planning, user association, transmit power, data collection, and energy management under time-varying air-to-ground channels \cite{Mozaffari2019UAVTutorial,Gu2023UAVSurvey}. Reinforcement learning is particularly useful when the UAV must sequentially decide where to move, which tags to serve, and how to balance sensing and communication objectives \cite{Sutton2018RL}. 

AI is also relevant to localization and ISAC. For localization, supervised learning and fingerprinting can map wireless features to positions, while deep learning can exploit spatial, temporal, and channel-domain correlations. For ISAC, AI can support waveform design, beamforming, sensing-task scheduling, interference suppression, and adaptive resource management \cite{Zhang2025IntelligentISAC}. In zero-energy IoT, lightweight learning, edge intelligence, and federated learning are important because passive devices have limited computation, memory, and energy \cite{Mao2017MEC,Kairouz2021FederatedLearning,Lim2020FederatedLearningMEC,Wang2020FederatedLearningWireless}. However, AI-based methods face training-data scarcity, generalization limits, computational overhead, explainability issues, and vulnerability to adversarial or environmental changes.

\section{Systematic Review Methodology}
Because this topic spans several research communities, the methodology separates foundational references from system-specific studies. Foundational references are retained when they introduce widely used models, metrics, or concepts, such as RF energy harvesting, cooperative localization, UAV communications, ISAC signal processing, or federated learning. System-specific studies are selected when they explicitly address UAV-assisted BackCom, passive localization, B-ISAC, AI-enabled BackCom, or reproducible wireless datasets. This distinction helps maintain broad coverage while keeping the review focused on the integrated UAV--backscatter--localization--ISAC theme.

This survey follows a structured, PRISMA-informed review procedure to improve transparency, reproducibility, and balanced coverage of the literature. The methodology is guided by PRISMA 2020 \cite{Page2021PRISMA}, PRISMA-S search-reporting recommendations \cite{Rethlefsen2021PRISMAS}, and systematic mapping principles for broad interdisciplinary fields \cite{Petersen2015Mapping}. The process includes defining research questions, constructing search strings, selecting relevant studies, extracting technical information, and developing a taxonomy for comparative analysis. This follows established systematic review practice, where research questions guide search, screening, extraction, and synthesis \cite{Kitchenham2007SLR}. Since the paper combines foundational studies, surveys, standards-oriented reports, datasets, and representative technical works, the quantitative synthesis should be interpreted as a curated mapping of representative literature rather than as a database-wide bibliometric census.

\subsection{Research Questions}
The research questions are designed to separate architectural, algorithmic, and evaluation-related issues. RQ1 focuses on topology and operating assumptions; RQ2 and RQ3 focus on localization and ISAC functionality; RQ4 captures the role of AI and learning-based control; and RQ5 addresses practical gaps such as benchmarking, reproducibility, security, and hardware validation. This structure avoids a purely chronological review and instead organizes the literature according to design dimensions that directly affect system performance.

The review is organized around research questions that capture both technical foundations and open gaps in AI-empowered UAV-assisted backscatter localization and ISAC. The questions identify how existing works model the system, what optimization techniques they use, and how they evaluate communication, localization, sensing, and energy performance.

\begin{itemize}
    \item \textbf{RQ1:} What are the main architectures, operating modes, and assumptions used in UAV-assisted backscatter communication for zero-energy IoT?
    \item \textbf{RQ2:} Which localization techniques are used for passive backscatter devices, and how do UAV mobility, channel conditions, and signal features affect localization accuracy?
    \item \textbf{RQ3:} How is ISAC integrated with UAV-assisted backscatter systems, and what tradeoffs exist among communication rate, sensing accuracy, localization error, and energy consumption?
    \item \textbf{RQ4:} What AI techniques are applied for channel estimation, signal detection, UAV trajectory control, resource allocation, localization, and sensing optimization?
    \item \textbf{RQ5:} What are the main limitations, benchmarking gaps, and future research directions for practical AI-empowered UAV-backscatter-ISAC systems?
\end{itemize}

\subsection{Search Strategy and Selection Criteria}
The search process also considers variations in terminology across communities. For example, zero-energy IoT may appear as passive IoT, battery-free IoT, ambient IoT, or wireless-powered IoT, while backscatter communication may be described as RFID, ambient backscatter, bistatic backscatter, or passive reflection modulation. Similarly, ISAC may appear as joint radar-communication, dual-functional radar-communication, sensing-assisted communication, or communication-assisted sensing. Using these related terms reduces the risk of missing relevant studies.

The literature search targets peer-reviewed journal articles, conference papers, early-access papers, technical standards, and highly relevant preprints. Recent works are prioritized to capture emerging developments in UAV-assisted BackCom, ISAC, Ambient IoT, and AI-enabled wireless optimization, while older works are included when they provide fundamental concepts, widely used models, or established methodology. The main databases include IEEE Xplore, ACM Digital Library, ScienceDirect, SpringerLink, Wiley Online Library, MDPI, Scopus, Web of Science, arXiv, and Google Scholar. Following PRISMA-S \cite{Rethlefsen2021PRISMAS}, the search terms combine keywords related to BackCom, UAV networks, zero-energy IoT, localization, ISAC, and AI. The resulting corpus is used for qualitative synthesis, taxonomy development, and manually coded trend analysis.

A representative Boolean search string is:
\begin{quote}
\small
(\textit{``backscatter communication'' OR ``ambient backscatter'' OR ``passive IoT'' OR ``zero-energy IoT'' OR ``ambient IoT''}) AND (\textit{``UAV'' OR ``drone'' OR ``aerial platform'' OR ``aerial base station''}) AND (\textit{``localization'' OR ``positioning'' OR ``sensing'' OR ``ISAC'' OR ``integrated sensing and communication''}) AND (\textit{``artificial intelligence'' OR ``machine learning'' OR ``deep learning'' OR ``reinforcement learning'' OR ``federated learning''}).
\end{quote}

Backward and forward snowballing are also applied to the most relevant articles \cite{Wohlin2014Snowballing}. Studies are included if they address UAV-assisted BackCom, passive-device localization, ISAC-enabled IoT, AI-based wireless optimization, or zero-energy IoT operation. Studies are excluded if they are unrelated to wireless IoT, do not consider backscatter or passive communication, focus only on conventional active-radio UAV networks, or lack sufficient technical detail. Non-English papers, duplicate records, inaccessible full texts, non-technical editorials, and papers with unclear evaluation methodology are also excluded.

\subsection{Screening, Data Extraction, and Taxonomy Development}
During extraction, special attention is paid to whether a study jointly considers multiple functions or treats them independently. For example, a UAV-assisted BackCom paper may optimize throughput and harvested energy but ignore localization and sensing, whereas a UAV-ISAC paper may optimize sensing and communication while assuming active devices. AI-enabled BackCom papers may report detection or estimation accuracy without accounting for UAV trajectories or passive energy constraints. Capturing these distinctions is essential for the dimension-level coverage matrix and for identifying why a fully integrated framework remains open.

The screening process has three stages: duplicate removal, title-and-abstract screening, and full-text assessment. First, duplicate records from multiple databases are removed. Second, titles and abstracts are examined to identify studies related to UAV-assisted BackCom, localization, ISAC, zero-energy IoT, or AI-based optimization. Third, full texts are reviewed to verify relevance and extract detailed information, following transparent selection logic recommended in systematic reviews \cite{Page2021PRISMA}. 

For each selected study, the extracted information includes publication year, application scenario, network architecture, UAV role, backscatter type, RF source, energy model, localization method, sensing function, AI technique, optimization objective, performance metrics, evaluation setup, and main limitations. The extracted data are organized using taxonomy-based synthesis \cite{Petersen2015Mapping}. The taxonomy classifies the literature into six dimensions: network architecture, backscatter mode, UAV functionality, localization and sensing method, AI paradigm, and performance objective. These dimensions distinguish monostatic, bistatic, ambient, cooperative, and multi-UAV settings; UAVs as emitters, collectors, relays, anchors, sensing platforms, or edge nodes; RSS-, CSI-, AoA-, ToA-, TDoA-, Doppler-, fingerprinting-, and learning-based localization \cite{Patwari2005LocatingNodes,Mao2007Localization,Wymeersch2009CooperativeLocalization}; and supervised learning, deep learning, reinforcement learning, federated learning, edge intelligence, and explainable AI \cite{OShea2017DeepLearningPhysical,Zhang2019DeepLearningWireless,Kairouz2021FederatedLearning}. The performance dimension includes communication rate, localization accuracy, sensing accuracy, energy efficiency, coverage, latency, reliability, scalability, and complexity.

The quantitative synthesis in this subsection is based on the coded corpus provided in Supplementary Tables~S1--S4, which makes the study-selection and coding process behind Figs.~\ref{fig:publication_trend_exact_corpus}--\ref{fig:dimension_coverage_matrix_heatmap} explicit and reproducible. The quantitative synthesis is intended to strengthen the paper's survey-style contribution by complementing the qualitative review with a structured mapping of the selected literature.

\section{Taxonomy of AI-Empowered UAV-Assisted Backscatter Localization and ISAC}
\label{sec:taxonomy}

The compact network representation in this section is useful for comparing apparently different studies. For example, a UAV-assisted wireless-powered BackCom study may emphasize the links among UAVs, tags, and RF sources, whereas a localization study may emphasize the geometric relationships among UAVs, passive tags, and receivers. A B-ISAC study may add sensing targets and shared waveform or beamforming variables. By expressing these cases within the same system view, the taxonomy can expose which entities and links are included or ignored in each research stream. This makes the later dimension-level coverage matrix easier to interpret.

The taxonomy is intended to support both literature classification and system design. A study may be strong in one dimension, such as UAV trajectory optimization, but weak in another, such as passive localization or reproducible evaluation. By separating the dimensions, the taxonomy helps identify which combinations are well studied and which remain underexplored. This is especially important because the system under consideration is multifunctional, and a gain in one function may entail a cost in another.

\begin{figure*}[!t]
    \centering
    \includegraphics[width=\textwidth]{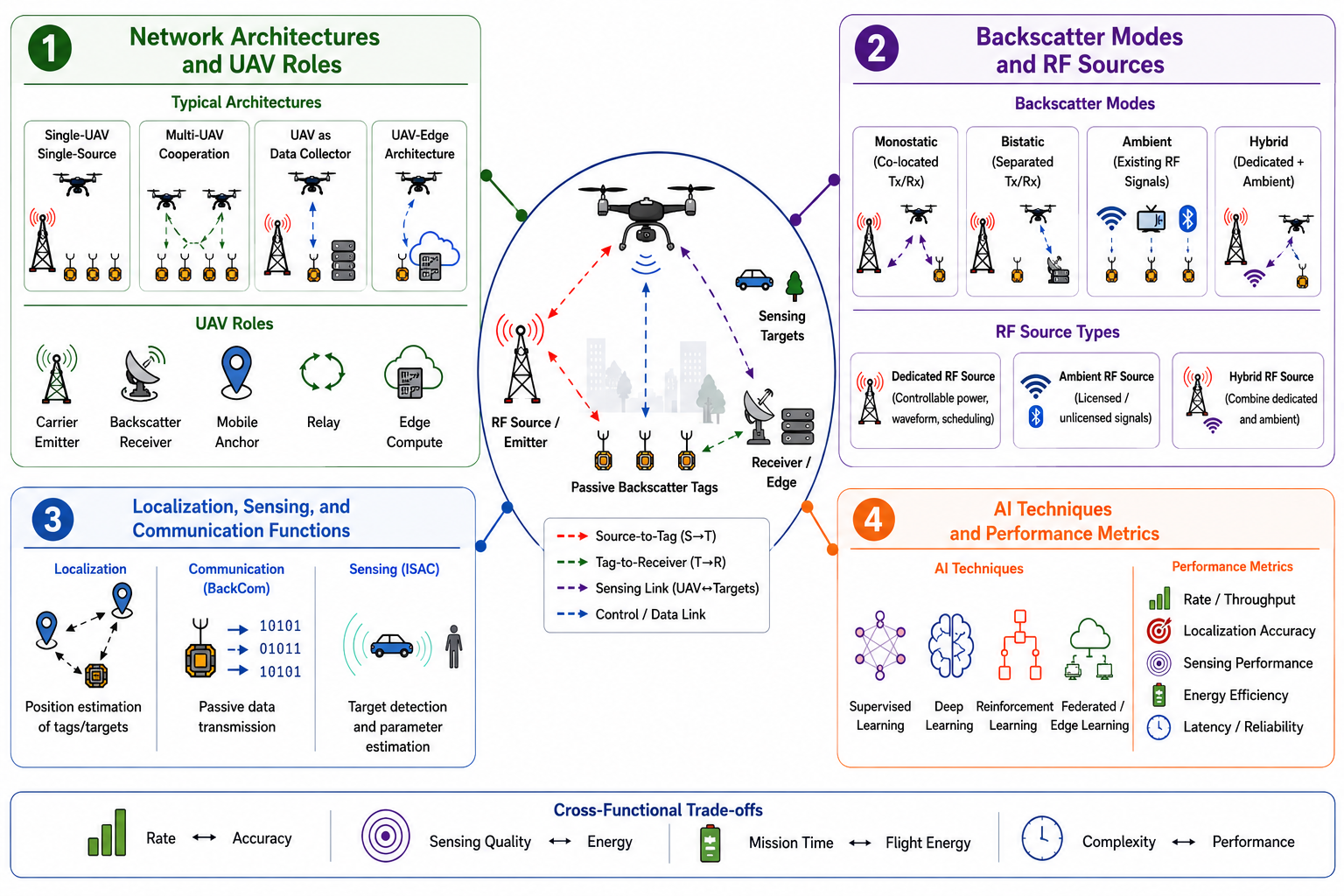}
    \caption{Taxonomy of AI-empowered UAV-assisted backscatter localization and ISAC.}
    \label{fig:taxonomy_visual_refined}
\end{figure*}

A clear taxonomy is needed to organize the emerging literature on AI-empowered UAV-assisted backscatter localization and ISAC. The considered systems combine passive communication, aerial mobility, localization, sensing, and learning-based optimization. Unlike conventional IoT networks, where communication, localization, and sensing are often treated separately, UAV-assisted backscatter-ISAC requires joint modeling of UAVs, passive tags, RF sources, receivers, sensing targets, propagation channels, and AI controllers. This survey classifies existing studies according to four dimensions: network architectures and UAV roles, backscatter modes and RF sources, localization--sensing--communication functions, and AI techniques with associated performance metrics. Fig.~\ref{fig:taxonomy_visual_refined} presents the proposed taxonomy.

A general UAV-assisted backscatter-ISAC network can be represented as
\begin{equation}
    \mathcal{N}=\{\mathcal{U},\mathcal{T},\mathcal{S},\mathcal{R},\mathcal{O},\mathcal{E}\},
\label{eq:taxonomy_network}
\end{equation}
where $\mathcal{U}$ denotes UAVs, $\mathcal{T}$ passive backscatter tags, $\mathcal{S}$ RF sources, $\mathcal{R}$ receivers or edge nodes, $\mathcal{O}$ sensing targets, and $\mathcal{E}$ the communication, sensing, energy, and control links. This representation highlights that the same RF signal may support tag communication, energy harvesting, localization, and environmental sensing.

\subsection{Network Architectures and UAV Roles}
\label{subsec:network_architectures_uav_roles}
The number of UAVs also changes the design space. A single-UAV system is simpler to coordinate but may suffer from limited coverage, poor localization geometry, and restricted mission time. Multi-UAV systems can provide spatial diversity, cooperative sensing, distributed data collection, and improved localization accuracy. However, they introduce issues such as inter-UAV interference, collision avoidance, synchronization, task allocation, and backhaul coordination. Therefore, UAV-assisted backscatter networks should be classified not only by UAV role but also by cooperation level and control architecture.

The network architecture defines how UAVs, passive tags, carrier emitters, receivers, edge servers, and sensing targets interact \cite{Zeng2016UAVComm}. A basic architecture uses a single UAV to transmit a carrier and receive the corresponding backscattered information. More advanced architectures include multi-UAV networks, UAV-relay-assisted systems, cooperative terrestrial--aerial systems, and UAV-edge architectures \cite{Mozaffari2019UAVTutorial,Gu2023UAVSurvey}. These settings are useful when passive devices are widely distributed or deployed in infrastructure-limited environments.

For a UAV located at $\mathbf{q}_{u}[n]=[x_u[n],y_u[n],H_u[n]]^T$, the distance between UAV $u$ and tag $k$ is
\begin{equation}
    d_{u,k}[n]=\left\|\mathbf{q}_{u}[n]-\mathbf{p}_{k}\right\|_2,
\label{eq:uav_tag_distance_taxonomy}
\end{equation}
where $\mathbf{p}_{k}$ is the tag position. A commonly used large-scale air-to-ground channel gain is
\begin{equation}
    g_{u,k}[n]=\beta_0 d_{u,k}^{-\alpha}[n],
\label{eq:uav_tag_channel_gain_taxonomy}
\end{equation}
where $\beta_0$ is the reference power gain and $\alpha$ is the path-loss exponent. When a complex baseband channel coefficient is needed, it can be written as
\begin{equation}
    h_{u,k}[n]=\sqrt{g_{u,k}[n]}\,\tilde{h}_{u,k}[n],
\label{eq:uav_tag_complex_channel_taxonomy}
\end{equation}
where $\tilde{h}_{u,k}[n]$ captures small-scale fading, phase, antenna response, or other channel variations. UAV placement and trajectory are central variables because UAV motion changes channel strength, harvested energy, localization geometry, and sensing visibility \cite{Yang2021UAVBackscatter}. The UAV may operate as a carrier emitter, receiver, mobile anchor, relay, sensing platform, or edge-computing node. As an ISAC platform, it must balance communication, sensing, localization, flight energy, and mission time \cite{Liu2024UAVISACIoT}.

\subsection{Backscatter Modes and RF Sources}
\label{subsec:backscatter_modes_rf_sources}
The RF-source dimension is closely related to deployment and standardization. Ambient IoT discussions in cellular networks suggest that future passive devices may use network-controlled carriers, sidelink signals, or existing downlink transmissions for connectivity and energy. In contrast, ad hoc UAV-assisted deployments may rely on temporary carrier emitters or UAV-mounted RF sources. These possibilities lead to different synchronization requirements, receiver designs, interference environments, and energy-harvesting profiles.

Backscatter modes are classified according to the relationship among the carrier emitter, passive tag, and receiver. In monostatic BackCom, the emitter and receiver are co-located; in bistatic BackCom, they are separated; and in ambient BackCom, passive devices reflect existing signals such as cellular, WiFi, TV, Bluetooth, or LoRa \cite{Liu2013AmbientBackscatter,Kellogg2014WiFiBackscatter,VanHuynh2018AmbientSurvey}. For UAV-assisted systems, these modes extend to aerial monostatic, aerial bistatic, hybrid terrestrial--aerial, and ambient-UAV architectures.

For tag $k$, the cascaded backscatter channel can be written as
\begin{equation}
    h_{k}^{\mathrm{bs}}[n]=h_{k,r}[n]\,\Gamma_k[n]\,h_{s,k}[n],
\label{eq:cascaded_backscatter_channel_taxonomy}
\end{equation}
where $h_{s,k}[n]$ is the RF-source-to-tag channel, $h_{k,r}[n]$ is the tag-to-receiver channel, and $\Gamma_k[n]$ is the tag reflection coefficient. The corresponding cascaded power gain is
\begin{equation}
    g_{k}^{\mathrm{bs}}[n]=|\Gamma_k[n]|^2 |h_{s,k}[n]|^2 |h_{k,r}[n]|^2.
\label{eq:cascaded_backscatter_power_gain_taxonomy}
\end{equation}
The reflection coefficient is modeled as
\begin{equation}
    \Gamma_k[n]=\sqrt{\rho_k[n]}e^{j\phi_k[n]}, \quad 0\leq \rho_k[n]\leq 1,
\label{eq:reflection_coefficient_taxonomy}
\end{equation}
where $\rho_k[n]$ controls the reflected-power ratio and $\phi_k[n]$ controls the reflection phase. Dedicated RF sources provide controllable carriers, while ambient sources reduce infrastructure and energy cost but are harder to control \cite{VanHuynh2018AmbientSurvey}. Millimeter-wave BackCom offers large bandwidth and directional beams for high-resolution sensing and higher data rates, but it is sensitive to blockage, beam misalignment, and UAV mobility \cite{Chen2024MMWaveBackscatter}.

\subsection{Localization, Sensing, and Communication Functions}
\label{subsec:localization_sensing_communication_functions}
This functional distinction is important when comparing performance metrics. Communication-only BackCom studies usually report rate, outage, fairness, or energy efficiency. Localization-assisted BackCom studies emphasize position error, CRLB, GDOP, or feature quality. Sensing-assisted BackCom may focus on tag detection, target detection, or parameter estimation. Fully integrated UAV-backscatter-ISAC systems should report several metrics together, including communication rate, localization error, sensing accuracy, harvested energy, latency, and UAV energy.

\begin{figure}[!t]
    \centering
    \includegraphics[width=1.0\columnwidth]{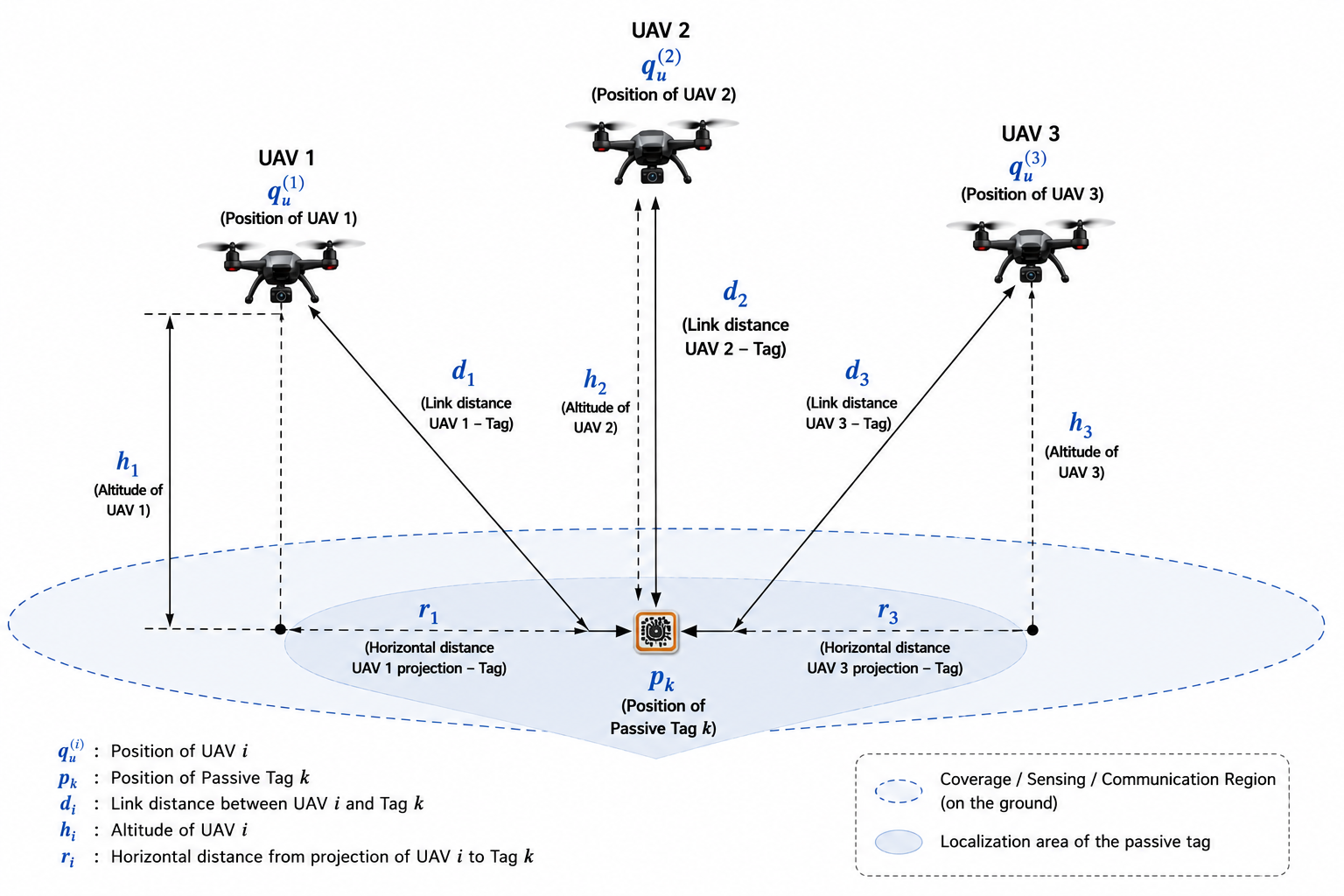}
    \caption{Multi-UAV localization geometry for passive backscatter tag positioning.}
    \label{fig:multi_uav_localization_geometry}
\end{figure}

In UAV-assisted backscatter IoT, localization, sensing, and communication are tightly coupled. Passive-device localization can use RSS, CSI, AoA, ToA, TDoA, Doppler, phase, or fingerprints \cite{Mao2007Localization,Liu2007IndoorPositioning,Zafari2019IndoorLocalization,Burghal2020MLLocalizationSurvey}. However, backscatter localization is challenging because the signal follows a cascaded emitter--tag--receiver path and is affected by weak reflections, tag orientation, multipath, RF-source uncertainty, and UAV mobility. Fig.~\ref{fig:multi_uav_localization_geometry} illustrates representative multi-UAV localization geometry.

For the geometry in Fig.~\ref{fig:multi_uav_localization_geometry}, the distance between UAV $i$ and tag $k$ is
\begin{equation}
    d_{i,k}[n]=\left\|\mathbf{q}_{u_i}[n]-\mathbf{p}_{k}\right\|_2,
\label{eq:uav_tag_distance_localization}
\end{equation}
and if $H_i[n]$ is the UAV altitude and $r_{i,k}[n]$ is the horizontal distance, then
\begin{equation}
    d_{i,k}[n]=\sqrt{r_{i,k}^{2}[n]+H_i^{2}[n]}.
\label{eq:distance_altitude_horizontal}
\end{equation}
A generic localization measurement collected by UAV $i$ from tag $k$ is
\begin{equation}
    \mathbf{z}_{k,i}[n]=\mathbf{f}\left(\mathbf{p}_{k},\mathbf{q}_{u_i}[n],\mathbf{s},\Gamma_k[n]\right)+\mathbf{v}_{k,i}[n],
\label{eq:localization_measurement_taxonomy}
\end{equation}
where $\mathbf{s}$ is the RF-source position and $\mathbf{v}_{k,i}[n]$ is noise. The function $\mathbf{f}(\cdot)$ may represent RSS, CSI, AoA, delay, Doppler, phase, or fingerprints. In backscatter-ISAC, the same reflected signal can support tag data transfer, passive localization, and target/environment sensing. For tag $k$, the backscatter rate is
\begin{equation}
    R_k[n] = B\log_2\left(1+\gamma_k[n]\right),
\label{eq:backscatter_rate_taxonomy}
\end{equation}
and a simplified sensing gain is
\begin{equation}
    G_{\mathrm{s}}(\theta,n)=\mathbf{a}^{H}(\theta)\mathbf{R}_{x}[n]\mathbf{a}(\theta),
\label{eq:sensing_gain_taxonomy}
\end{equation}
where $\mathbf{a}(\theta)$ is the steering vector and $\mathbf{R}_{x}[n]$ is the transmit covariance matrix \cite{Zhang2021JCRSignalProcessing,Liu2022ISACSurvey}. Recent B-ISAC studies show that BackCom and ISAC can be combined for tag detection, parameter estimation, communication enhancement, and spectrum sharing \cite{Zhao2024BISAC,Zargari2023BackscatterSensingIntegration}.

\subsection{AI Techniques and Performance Metrics}
\label{subsec:ai_techniques_metrics_taxonomy}
A complete performance evaluation should also consider the cost of intelligence. A learning-based method that improves throughput or localization accuracy may still be impractical if it requires large training datasets, frequent model updates, high inference latency, or excessive communication between UAVs and edge servers. Thus, AI-based schemes should be compared against strong model-based baselines using not only task performance but also training overhead, inference time, memory footprint, and robustness to distribution shift.

AI methods should also be interpreted in terms of where they are executed. A deep neural network used on a ground edge server faces different feasibility constraints than a lightweight model deployed onboard a UAV. Similarly, federated learning among multiple UAVs incurs different communication overhead than centralized training on a cloud server. For this reason, the taxonomy distinguishes learning paradigms not only by algorithm type but also by deployment layer, feedback requirement, and computational burden.

AI techniques address the nonlinear, dynamic, and high-dimensional nature of UAV-assisted backscatter localization and ISAC. Supervised learning supports signal classification, channel-feature mapping, fingerprint localization, and tag-state detection. Deep learning can exploit spatial, temporal, and channel-domain correlations for channel estimation, localization, beam selection, and sensing inference \cite{OShea2017DeepLearningPhysical,Zhang2019DeepLearningWireless}. Reinforcement learning is well-suited to sequential control problems such as UAV trajectory planning, tag scheduling, power adaptation, reflection control, and sensing task selection \cite{Sutton2018RL}. Federated and edge learning enable privacy-preserving and low-latency intelligence when raw data cannot be continuously sent to the cloud \cite{Mao2017MEC,Kairouz2021FederatedLearning}.

A generic learning-based decision model is
\begin{equation}
    \mathbf{a}[n]=\pi_{\boldsymbol{\theta}}\left(\mathbf{x}[n]\right),
\label{eq:ai_policy_taxonomy}
\end{equation}
where $\mathbf{x}[n]$ includes channel estimates, UAV position, tag energy states, localization uncertainty, and sensing requirements, while $\mathbf{a}[n]$ includes UAV movement, transmit power, scheduling, reflection coefficient, or sensing mode. A compact normalized multi-objective utility is
\begin{equation}
    \mathcal{U}[n]=
    \omega_c \bar{R}[n]
    +\omega_s \bar{S}[n]
    -\omega_l \bar{e}_{\mathrm{loc}}[n]
    -\omega_e \bar{E}[n]
    -\omega_{\tau}\bar{\tau}[n].
\label{eq:multiobjective_utility_taxonomy}
\end{equation}
Here, $\bar{R}[n]$, $\bar{S}[n]$, $\bar{e}_{\mathrm{loc}}[n]$, $\bar{E}[n]$, and $\bar{\tau}[n]$ denote normalized communication, sensing, localization-error, energy-consumption, and latency metrics, respectively. The weights should be chosen according to the application requirements or evaluated using Pareto-front analysis.

\begin{figure*}[!t]
    \centering
    \includegraphics[width=0.8\textwidth]{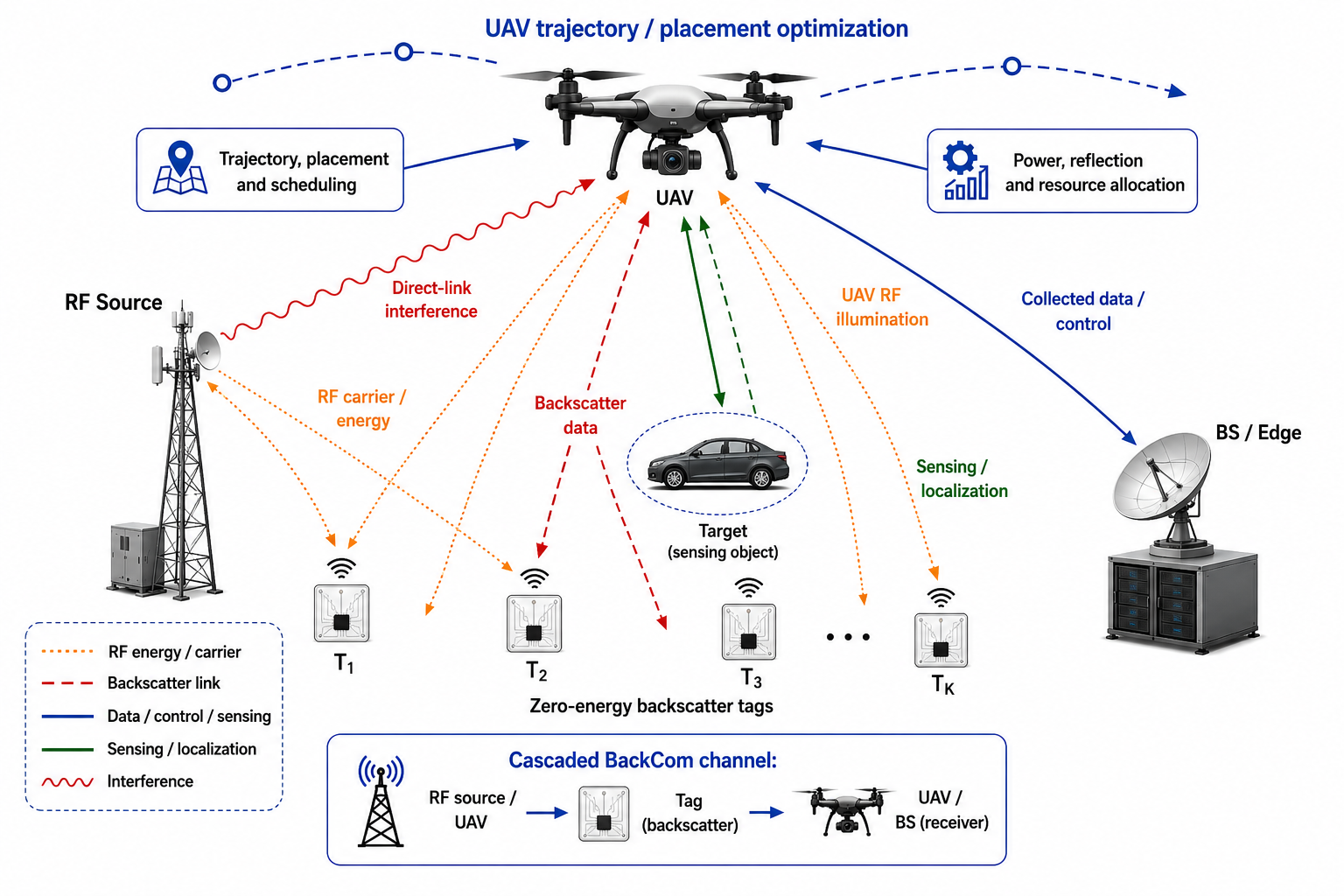}
    \caption{UAV-assisted backscatter communication for zero-energy IoT.}
    \label{fig:uav_backscatter_zero_energy_iot}
\end{figure*}

Communication metrics include throughput, spectral efficiency, BER, outage, latency, reliability, and fairness. Localization metrics include positioning error, CRLB, coverage probability, convergence time, and robustness \cite{Patwari2005LocatingNodes,Pettorru2024TrustworthyLocalization}. Sensing metrics include detection probability, false-alarm probability, estimation error, resolution, tracking accuracy, and radar estimation rate \cite{Liu2022ISACSurvey,Zhang2021JCRSignalProcessing}. Energy metrics include harvested energy, tag activation probability, UAV propulsion energy, energy efficiency, and network lifetime \cite{Lu2015RFEH,Bi2016WPCN,Yang2021UAVBackscatter}. AI-specific metrics include training overhead, inference complexity, convergence speed, generalization, communication overhead, privacy leakage, and explainability \cite{Zhang2019DeepLearningWireless,Kairouz2021FederatedLearning}.

\section{UAV-Assisted Backscatter Communication for Zero-Energy IoT}
\label{sec:uav_backscatter_zero_energy_iot}
A key issue in this section is that UAV mobility can be both beneficial and costly. Moving closer to passive tags improves incident RF power and backscatter reception, but movement consumes propulsion energy and increases mission time \cite{Zeng2016UAVComm,Mozaffari2019UAVTutorial}. Hovering improves channel stability but may reduce the number of devices served. These effects make UAV-assisted BackCom fundamentally different from fixed-reader BackCom and from conventional UAV data collection with active devices.

UAV-assisted BackCom combines aerial mobility with passive or near-passive data transmission, making it promising for zero-energy IoT. UAVs can operate as carrier emitters, data collectors, relays, mobile access points, or aerial receivers, while ground tags transmit by modulating and reflecting incident RF signals. Compared with fixed gateways, UAVs can improve coverage, reduce propagation distance, enhance line-of-sight probability, and support scattered IoT devices in remote or infrastructure-limited environments \cite{Gu2023UAVSurvey}. However, UAV-assisted BackCom also creates coupled design problems involving air-to-ground propagation, cascaded backscatter channels, energy harvesting, UAV flight energy, resource allocation, and interference management \cite{Yang2019UAVBackscatterLCOMM,Yang2021UAVBackscatter,Wang2022UAVBackscatterRA}. Fig.~\ref{fig:uav_backscatter_zero_energy_iot} illustrates a representative architecture of UAV-assisted backscatter communication for zero-energy IoT.

\subsection{System and Channel Models}
In more realistic models, the UAV channel may include probabilistic LoS/NLoS propagation, shadowing, small-scale fading, antenna orientation, and Doppler effects \cite{khalil2025robust}. The backscatter link may also depend on tag orientation, polarization mismatch, structural-mode scattering, and impedance-switching loss. These factors are often simplified in analytical studies to make optimization tractable. However, simplified models can overestimate UAV-mobility gains if they ignore the sensitivity of weak reflected signals to practical propagation and hardware effects.

A typical UAV-assisted BackCom system consists of a UAV, passive tags, one or more carrier emitters, and a receiver or terrestrial base station. Depending on the deployment, the UAV may transmit an RF carrier, receive backscattered signals, relay data, or perform multiple functions in the same mission. In carrier-emitter-assisted architectures, ground- or aerial-based RF sources illuminate tags, while the UAV collects modulated reflections \cite{Yang2021UAVBackscatter}. In data-gathering architectures, the UAV collects information from terrestrial tags and uploads it to a base station \cite{Yang2019UAVBackscatterLCOMM}. In relay-assisted architectures, the UAV forwards weak backscatter signals to improve throughput when direct links are unreliable \cite{Hua2020ThroughputUAVBackscatter}. 

The channel model includes air-to-ground UAV links, cascaded backscatter links, and direct interference links. The UAV-to-tag or UAV-to-receiver channel is commonly modeled using distance-dependent path loss with LoS/NLoS probability, fading, antenna gain, and altitude effects \cite{khalil2023beyond,Khuwaja2018UAVChannelModeling,Yan2019UAVChannelModeling}. For a UAV at $\mathbf{q}[n]=[x[n],y[n],H[n]]^T$, the distance to tag $k$ is
\begin{equation}
    d_{u,k}[n] = \sqrt{\|\mathbf{q}_{u}^{2D}[n]-\mathbf{w}_{k}\|^2 + H^2[n]},
\label{eq:uav_tag_distance_section_v}
\end{equation}
where $\mathbf{q}_{u}^{2D}[n]=[x[n],y[n]]^T$ and $\mathbf{w}_{k}$ is the horizontal position of tag $k$. The large-scale channel gain is often modeled as
\begin{equation}
    g_{u,k}[n] = \beta_{0} d_{u,k}^{-\alpha}[n].
\label{eq:uav_tag_gain_section_v}
\end{equation}
When a complex channel coefficient is required, it can be expressed as $h_{u,k}[n]=\sqrt{g_{u,k}[n]}\tilde{h}_{u,k}[n]$, where $\tilde{h}_{u,k}[n]$ captures small-scale fading and phase. For BackCom, the effective channel is cascaded because the signal propagates from the RF source to the tag and then to the receiver or UAV \cite{Boyer2014BackscatterRFID,VanHuynh2018AmbientSurvey}. Thus, performance is highly sensitive to UAV altitude, tag location, RF-source placement, reflection efficiency, and direct-link interference.

\subsection{Energy Harvesting and Passive Operation}
Energy causality creates temporal coupling across slots. A tag that harvests energy in one slot may use it later for sensing or backscattering, so scheduling decisions should account for the tag's battery or capacitor state \cite{Lu2015RFEH}. This differs from conventional active IoT scheduling, where device energy is usually assumed to be available. In UAV-assisted BackCom, the UAV may need to revisit poorly energized tags, hover near clusters that require activation, or adapt carrier emission according to tag energy states.

\begin{figure}[!t]
    \centering
    \includegraphics[width=1.0\columnwidth]{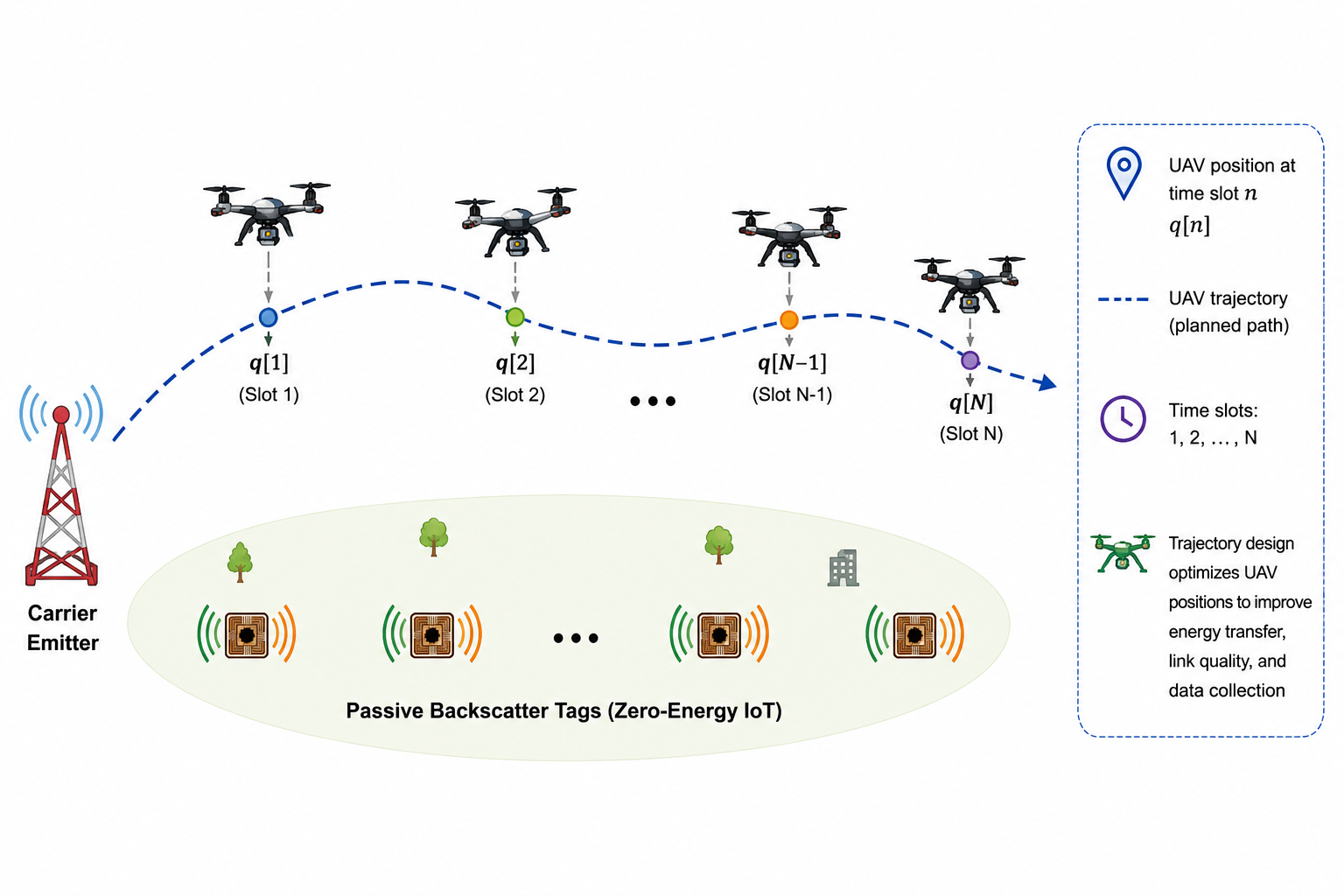}
    \caption{UAV trajectory design for backscatter communication in zero-energy IoT networks.}
    \label{fig:uav_trajectory_optimization_backscatter}
\end{figure}

Zero-energy IoT devices rely on harvested energy for sensing, switching, impedance modulation, control logic, and limited computation. In UAV-assisted BackCom, energy may be harvested from UAV-generated carriers, terrestrial RF sources, ambient signals, or hybrid mechanisms \cite{Bi2016WPCN,VanHuynh2018AmbientSurvey}. A common linear approximation expresses harvested energy at tag $k$ in slot $n$ as
\begin{equation}
    E_{k}^{\mathrm{har}}[n] = \eta_{k} P_{k}^{\mathrm{in}}[n] \tau_{n},
\label{eq:harvested_energy_slot}
\end{equation}
where $\eta_{k}$ is the RF-to-DC conversion efficiency, $P_{k}^{\mathrm{in}}[n]$ is incident RF power, and $\tau_{n}$ is the slot duration. When the tag splits incident power between reflection and harvesting, a simplified model is
\begin{equation}
    E_{k}^{\mathrm{har}}[n] = \eta_{k}\bigl(1-\rho_k[n]\bigr)P_{k}^{\mathrm{in}}[n]\tau_{n},
\label{eq:harvested_energy_reflection_split}
\end{equation}
where $\rho_k[n]$ is the reflected-power ratio. This model is a tractable approximation; practical rectifiers may require nonlinear energy-harvesting models and a minimum activation threshold. Thus, reflection coefficient, harvesting time, circuit power, and activation threshold must be jointly considered \cite{Zhou2013SWIPT,Xu2023AIEmpoweredBackscatter}. Energy-neutral operation requires that the energy consumed for sensing, switching, and modulation remain below the harvested and stored energy over the considered horizon.

\subsection{UAV Trajectory, Placement, and Resource Allocation}
Another issue is the interaction between communication scheduling and trajectory discretization. 
Many studies divide the UAV mission into time slots and assume that the UAV position is fixed within each slot. 
This approximation simplifies optimization and is commonly adopted in recent UAV-assisted BackCom designs that jointly optimize UAV trajectory, transmission power, and backscatter-device scheduling \cite{Xie2024UserTrajectoryBackscatter}. 
However, the selected slot length can affect modeling accuracy. 
Short slots better capture channel variation but increase the number of optimization variables, while long slots reduce complexity but may hide important mobility effects. 
Practical designs should therefore consider both algorithmic tractability and the physical time scale of UAV motion, tag switching, and sensing measurements.

The placement problem is also affected by deployment density. 
In sparse networks, the UAV may hover near each tag or cluster to improve energy transfer and data collection. 
In dense networks, serving all tags individually may be infeasible, and the UAV must balance sum throughput, max--min fairness, latency, and energy neutrality. 
Recent multi-UAV backscatter data-collection studies further show that tag selection, reflection control, UAV trajectory, and UAV energy consumption must be jointly considered when fairness and secure energy efficiency are important design objectives \cite{Zeng2025FairnessSecureEE}. 
Cluster-based scheduling, path segmentation, and priority-aware service can reduce complexity but introduce additional decisions about cluster formation, service order, and update frequency.

\begin{figure}[h!]
    \centering
    \includegraphics[width=1.0\columnwidth]{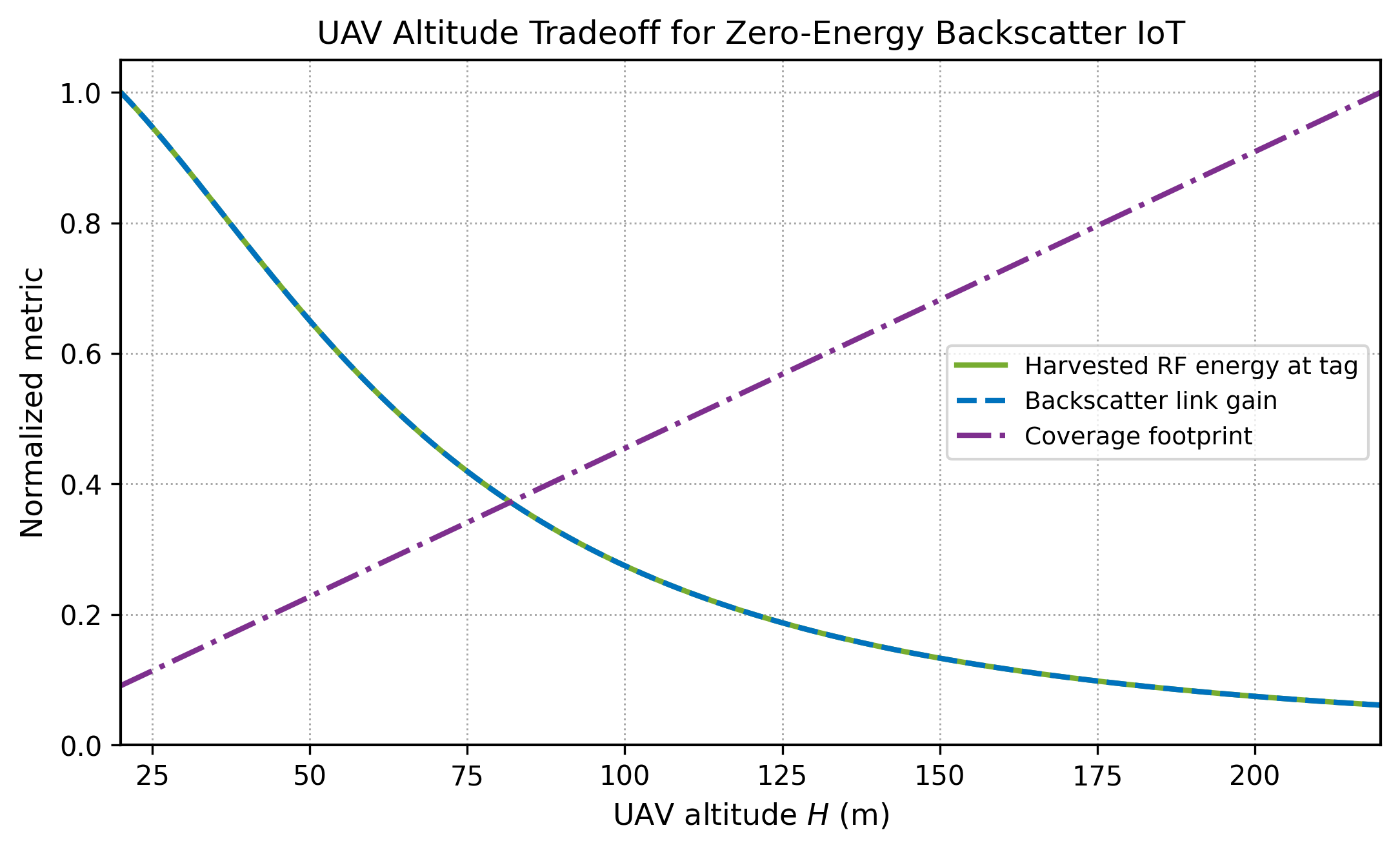}
    \caption{UAV altitude tradeoff for zero-energy backscatter IoT under the normalized geometry-based parameters in Table~\ref{tab:tutorial_simulation_parameters}.}
    \label{fig:uav_based_altitude_tradeoff}
\end{figure}

The main advantage of UAV-assisted BackCom is controllable aerial mobility. By adjusting trajectory and hovering locations, the UAV can improve channel quality, increase harvested energy, reduce outage probability, and enhance data collection efficiency \cite{Han2021UAVAidedBackscatter}. Fig.~\ref{fig:uav_trajectory_optimization_backscatter} illustrates UAV trajectory design over multiple time slots. For a mission divided into $N$ slots, trajectory design optimizes $\{\mathbf{q}[n]\}_{n=1}^{N}$ under maximum-speed, altitude, initial/final-position, and flight-time constraints. Fig.~\ref{fig:uav_based_altitude_tradeoff} illustrates the altitude tradeoff: increasing altitude expands coverage but reduces harvested energy and backscatter link gain.

\begin{table*}[!t]
\centering
\caption{Simulation Parameters for the Tutorial-Style Numerical Illustrations.}
\label{tab:tutorial_simulation_parameters}
\footnotesize
\renewcommand{\arraystretch}{1.25}
\setlength{\tabcolsep}{3pt}
\begin{tabularx}{\textwidth}{|p{0.11\textwidth}|p{0.22\textwidth}|p{0.29\textwidth}|X|}
\hline
\hline
\textbf{Figure} 
& \textbf{Purpose} 
& \textbf{Main Parameters} 
& \textbf{Model Used to Generate the Curves} \\
\hline
\hline
Fig.~\ref{fig:uav_based_altitude_tradeoff}
& UAV altitude tradeoff for zero-energy BackCom
& UAV altitude $H\in[20,220]$ m; reference horizontal distance $r_{0}=60$ m; path-loss exponent $\alpha=2.2$; normalized transmit power and RF-to-DC efficiency; half-beam coverage indicator normalized by the maximum altitude
& The UAV--tag distance is $d(H)=\sqrt{r_{0}^{2}+H^{2}}$. 
The normalized coverage indicator is $C_{\mathrm{norm}}(H)=H/H_{\max}$. 
The normalized harvested-energy and link-gain indicators are computed as
$G_{\mathrm{norm}}(H)=\left(d(H)/d(H_{\min})\right)^{-\alpha}$.
All values are normalized by their maximum values. \\
\hline
Fig.~\ref{fig:uav_based_single_tag_spatial_spectrum}
& Single passive-tag AoA localization
& ULA receiver with $N_{r}=16$ antennas; inter-element spacing $d_{a}=\lambda/2$; true passive-tag AoA $\theta_{0}=-24.9^{\circ}$; scan range $[-90^{\circ},90^{\circ}]$; $K=50$ snapshots; SNR $=20$ dB
& The received snapshots follow \eqref{eq:spatial_snapshot_tutorial} with $L=1$. 
The plotted curve is the conventional/Bartlett spectrum in \eqref{eq:conventional_spectrum_tutorial}, normalized as
$10\log_{10}\left(P(\theta)/\max_{\theta}P(\theta)\right)$. \\
\hline
Fig.~\ref{fig:uav_based_capon_vs_conventional_multitag}
& Angular separability of two close passive tags
& ULA receiver with $N_{r}=10$ antennas; inter-element spacing $d_{a}=\lambda/2$; two passive tags at $\theta_{1}=10^{\circ}$ and $\theta_{2}=15^{\circ}$; equal reflection/scattering power; $K=20$ snapshots; SNR $=20$ dB; diagonal loading $\epsilon=10^{-3}\mathrm{tr}(\widehat{\mathbf{R}})/N_{r}$
& The conventional spectrum is computed using \eqref{eq:conventional_spectrum_tutorial}, while the Capon/MVDR spectrum is computed using \eqref{eq:mvdr_spectrum_tutorial}. 
Both spectra are normalized to their peak values and plotted in dB. \\
\hline
Fig.~\ref{fig:uav_based_beampattern_ground_nodes}
& UAV-mounted array beampattern toward ground nodes
& Transmit ULA with $N_{t}\in\{8,16,32\}$ antennas; inter-element spacing $d_{a}=\lambda/2$; representative ground-node directions $\{-40^{\circ},-10^{\circ},25^{\circ},55^{\circ}\}$; equal beam weights
& The normalized beampattern is computed as
$P(\theta)=\sum_{i=1}^{4}
\left|
\mathbf{a}_{N_{t}}^{H}(\theta)
\mathbf{a}_{N_{t}}(\theta_{i})
\right|^{2}$,
and plotted as
$10\log_{10}\left(P(\theta)/\max_{\theta}P(\theta)\right)$. \\
\hline
\hline
\end{tabularx}
\end{table*}

Resource allocation involves transmit power, tag scheduling, reflection coefficients, time allocation, bandwidth, and sometimes NOMA decoding order. Energy-efficient UAV BackCom has been studied by jointly optimizing UAV trajectory, tag scheduling, and carrier-emitter power under throughput and harvested-energy constraints \cite{Yang2021UAVBackscatter}. Max--min resource allocation improves fairness by jointly optimizing scheduling, UAV power control, trajectory, and reflection coefficients \cite{Wang2022UAVBackscatterRA}. Throughput maximization also shows that trajectory and protocol design strongly affect UAV-aided BackCom performance \cite{Hua2020ThroughputUAVBackscatter}. 

\begin{figure}[!t]
    \centering
    \includegraphics[width=1.0\columnwidth]{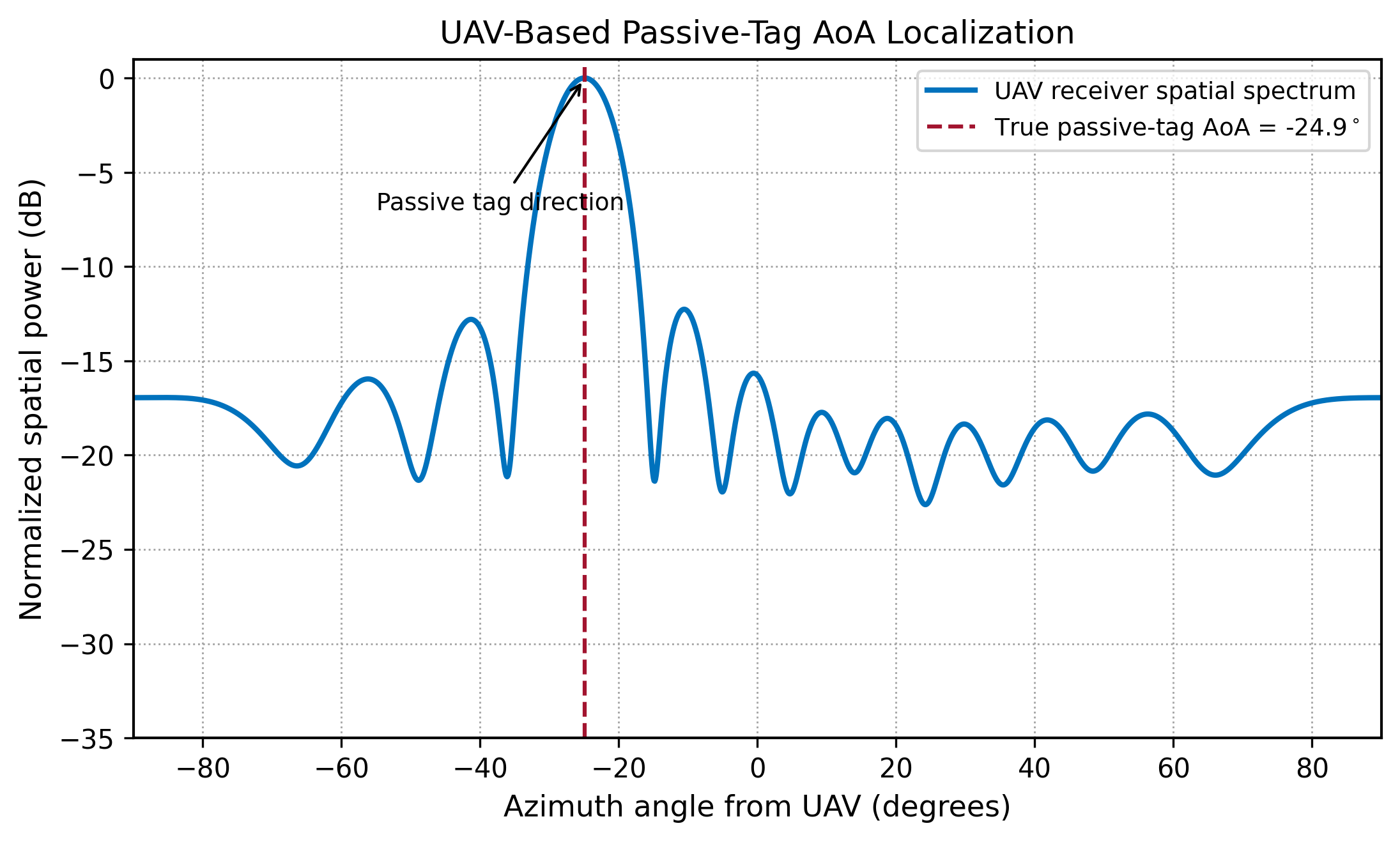}
    \caption{UAV-based passive-tag AoA localization using the ULA spatial-spectrum model and parameters in Table~\ref{tab:tutorial_simulation_parameters}.}
    \label{fig:uav_based_single_tag_spatial_spectrum}
\end{figure}

\noindent\textit{Numerical-illustration setup for Figs.~\ref{fig:uav_based_altitude_tradeoff}, 
\ref{fig:uav_based_single_tag_spatial_spectrum}, 
\ref{fig:uav_based_capon_vs_conventional_multitag}, and 
\ref{fig:uav_based_beampattern_ground_nodes}:}
The numerical plots in Figs.~\ref{fig:uav_based_altitude_tradeoff}, 
\ref{fig:uav_based_single_tag_spatial_spectrum}, 
\ref{fig:uav_based_capon_vs_conventional_multitag}, and 
\ref{fig:uav_based_beampattern_ground_nodes} are tutorial-style illustrative simulations generated from simplified geometry-based and array-processing models. 
They are not intended to reproduce the results of a specific experimental platform or to benchmark a particular algorithm. 
All curves are normalized to highlight the main trends: altitude increases coverage but weakens received power; array-domain processing enables AoA-based passive-tag localization; Capon/MVDR processing improves angular separability; and larger UAV-mounted arrays provide narrower beams. 
Unless otherwise stated, the array response is modeled using a uniform linear array (ULA) with half-wavelength spacing, i.e.,
\begin{equation}
    \mathbf{a}_{N}(\theta)
    =
    \frac{1}{\sqrt{N}}
    \left[
    1,
    e^{-j\pi\sin\theta},
    \ldots,
    e^{-j\pi(N-1)\sin\theta}
    \right]^{T}.
\label{eq:ula_response_tutorial}
\end{equation}
For the spatial-spectrum examples, the received snapshot model is
\begin{equation}
    \mathbf{y}[q]
    =
    \sum_{\ell=1}^{L}
    \beta_{\ell}\mathbf{a}_{N}(\theta_{\ell})s_{\ell}[q]
    +
    \mathbf{n}[q],
    \quad q=1,\ldots,K,
\label{eq:spatial_snapshot_tutorial}
\end{equation}
where $\theta_{\ell}$ is the direction of the $\ell$th passive tag or sensing target, $\beta_{\ell}$ is its complex reflection/scattering coefficient, $s_{\ell}[q]\sim\mathcal{CN}(0,1)$, and $\mathbf{n}[q]\sim\mathcal{CN}(\mathbf{0},\sigma^{2}\mathbf{I})$. 
The sample covariance matrix is computed as
\begin{equation}
    \widehat{\mathbf{R}}
    =
    \frac{1}{K}
    \sum_{q=1}^{K}
    \mathbf{y}[q]\mathbf{y}^{H}[q].
\label{eq:sample_covariance_tutorial}
\end{equation}
The conventional spatial spectrum and Capon/MVDR spectrum are respectively evaluated as
\begin{equation}
    P_{\mathrm{conv}}(\theta)
    =
    \mathbf{a}_{N}^{H}(\theta)
    \widehat{\mathbf{R}}
    \mathbf{a}_{N}(\theta),
\label{eq:conventional_spectrum_tutorial}
\end{equation}
and
\begin{equation}
    P_{\mathrm{MVDR}}(\theta)
    =
    \frac{1}{
    \mathbf{a}_{N}^{H}(\theta)
    \left(
    \widehat{\mathbf{R}}+\epsilon\mathbf{I}
    \right)^{-1}
    \mathbf{a}_{N}(\theta)},
\label{eq:mvdr_spectrum_tutorial}
\end{equation}
where $\epsilon$ is a small diagonal-loading factor used for numerical stability. 
Table~\ref{tab:tutorial_simulation_parameters} summarizes the parameters used to generate the tutorial plots.

Many UAV-assisted BackCom designs lead to mixed-integer, nonlinear, and non-convex optimization problems due to mobility-dependent channels, binary scheduling, coupled energy-rate constraints, reflection coefficients, and UAV propulsion energy. Common tools include block coordinate descent, SCA, Dinkelbach's method, alternating optimization, and decomposition \cite{Yang2021UAVBackscatter,Wang2022UAVBackscatterRA,Zhang2021HDRLUAVBackscatter}. Learning-based methods are attractive for online trajectory and resource control under uncertainty.

\subsection{Interference Management and Reliability}
Reliability is also linked to sensing and localization. A weak backscatter signal can simultaneously cause packet errors, missed tag detections, and inaccurate localization measurements \cite{yang2025joint}. Similarly, direct-link interference can mask the tag signal, reducing both decoding reliability and sensing visibility. Therefore, interference cancellation and robust detection should be evaluated not only by communication BER or outage but also by their impact on localization accuracy and sensing reliability.

Interference management is critical because the desired backscattered signal is usually much weaker than the carrier or direct-link signal. In monostatic and bistatic systems, the receiver may observe strong carrier leakage in addition to weak tag reflections. In ambient BackCom, the receiver must separate tag information from legacy transmissions, co-channel interference, and noise \cite{VanHuynh2018AmbientSurvey,Zhong2024InterferenceUAVBackscatter}. Time-division scheduling reduces tag collisions, while NOMA can increase connectivity by allowing multiple tags to share the same resource block with SIC \cite{AlJubayrin2022NOMABackscatterUAV}. Trajectory control can move the UAV toward locations where the desired backscatter link is strengthened and interference is reduced \cite{Han2021UAVAidedBackscatter}. Reliability should be evaluated using outage probability, BER, packet delivery ratio, minimum user rate, fairness, latency, and energy efficiency.

\section{Backscatter Localization in UAV-Assisted IoT Networks}
\label{sec:backscatter_localization_uav_iot}
Another important difference is that the localization feature may be generated by the same passive modulation process used for data transmission. For example, changes in the tag's reflection state can affect RSS, phase, and CSI measurements, while UAV motion can change the delay, angle, and Doppler structure of the received signal. Therefore, localization algorithms should jointly account for tag state, UAV state, RF-source geometry, and measurement noise.

Backscatter localization in UAV-assisted IoT networks estimates the positions of passive devices using measurements collected by UAVs, terrestrial anchors, RF sources, or cooperative nodes \cite{VanHuynh2018AmbientSurvey}. Unlike active IoT devices, backscatter tags modulate and reflect incident carriers. Therefore, localization relies on a cascaded RF source--tag--receiver path, making the signal weak and sensitive to geometry, tag orientation, multipath, and interference \cite{Boyer2014BackscatterRFID}. UAVs can improve localization by acting as mobile readers, anchors, aerial receivers, or RF illuminators.

\subsection{Signal-Feature-Based Localization}
No single feature is universally optimal. AoA-based localization can be accurate but requires antenna arrays, calibration, and sufficient angular resolution. RSS-based methods are easier to implement but are sensitive to shadowing and model mismatch. CSI-based methods can exploit richer multipath information but require reliable channel estimation, which is difficult for weak backscatter links. Doppler and phase-based methods can benefit from UAV motion, but they may suffer from ambiguity, oscillator instability, and synchronization errors. Practical systems may therefore combine multiple features.

Signal-feature-based localization estimates passive-tag position from RSS, CSI, phase, AoA, ToA, TDoA, Doppler, or RF fingerprints \cite{Patwari2005LocatingNodes,Mao2007Localization,Zafari2019IndoorLocalization}. For tag $k$ observed by UAV $m$ in slot $n$, a generic feature vector is
\begin{equation}
    \mathbf{z}_{k,m}[n] = [P_{r,k,m}[n],\phi_{k,m}[n],\tau_{k,m}[n],\theta_{k,m}[n],f_{D,k,m}[n]]^T.
\end{equation}
In a UAV-assisted backscatter link, RSS in dB can be approximated as
\begin{equation}
\begin{split}
    P^{\mathrm{bs}}_{r,k,m}[n] = & P_s + G_s + G_m + 20\log_{10}|\Gamma_k[n]|  \\
    & -10\alpha_1\log_{10} d_{s,k}
      -10\alpha_2\log_{10} d_{k,m}[n]
      +\xi_{k,m}[n],
\end{split}
\label{eq:backscatter_rss}
\end{equation}
where $P_s$ is the source transmit power in dB scale, $G_s$ and $G_m$ are antenna gains, $d_{s,k}$ is the source-to-tag distance, $d_{k,m}[n]$ is the tag-to-UAV distance, $\alpha_1$ and $\alpha_2$ are path-loss exponents, and $\xi_{k,m}[n]$ captures shadowing and modeling errors. This model illustrates why BackCom localization is harder than active localization: the received power depends on two propagation distances and the tag's reflection efficiency \cite{Boyer2014BackscatterRFID,VanHuynh2018AmbientSurvey}. 

If synchronized receivers $i$ and $j$ observe the same tag response from the same RF source, the common source-to-tag delay cancels and a simplified TDoA measurement is
\begin{equation}
    \Delta \tau_{i,j,k}[n]=
    \frac{\|\mathbf{p}_k-\mathbf{q}_i[n]\|-\|\mathbf{p}_k-\mathbf{q}_j[n]\|}{c}
    +\epsilon_{i,j,k}[n],
\label{eq:tdoa_model}
\end{equation}
where $c$ is the speed of light and $\epsilon_{i,j,k}[n]$ is measurement noise. If the receivers are not synchronized, or if different RF sources are used, additional clock-bias and source-delay terms must be included.

\subsection{Cooperative and Multi-UAV Localization}
The cooperation architecture can be centralized or distributed. In a centralized design, UAVs forward measurements to an edge server that performs data fusion and position estimation. This can improve accuracy but increases backhaul traffic and latency \cite{zheng2023multi}. In a distributed design, UAVs exchange partial estimates or compressed measurements, reducing central processing but requiring consensus or distributed filtering. The best architecture depends on mission time, available backhaul capacity, UAV computational resources, tag density, and privacy constraints.

Cooperative localization combines measurements from multiple UAVs, anchors, RF sources, or neighboring devices. In conventional networks, cooperation improves localization by exploiting spatial diversity and geometry \cite{Patwari2005LocatingNodes,Wymeersch2009CooperativeLocalization}. Multi-UAV localization improves measurement diversity by allowing several UAVs to observe the same tag or target from different viewpoints. Recent UAV-based passive localization studies show that geometry strongly affects RSS- and TDoA-based accuracy, and CRLB-based criteria can guide UAV placement or path planning \cite{Cheng2023DroneSwarmRSS,Khalil2023UAVPassiveTDOA}. For a measurement model $\mathbf{z}_{k}=\mathbf{f}(\mathbf{p}_k)+\mathbf{v}$ with noise covariance $\mathbf{R}$, let
\begin{equation}
    \mathbf{H}_{k,m}[n]=
    \frac{\partial \mathbf{f}_{k,m}[n]}{\partial \mathbf{p}_k}
\label{eq:localization_jacobian}
\end{equation}
denote the measurement Jacobian. The Fisher information matrix (FIM) can be written as
\begin{equation}
    \mathbf{J}(\mathbf{p}_k)=
    \sum_{m,n}
    \mathbf{H}_{k,m}^{T}[n]\mathbf{R}_{k,m}^{-1}[n]\mathbf{H}_{k,m}[n],
\label{eq:fim}
\end{equation}
and the localization error covariance satisfies
\begin{equation}
    \mathrm{Cov}(\hat{\mathbf{p}}_k) \succeq \mathbf{J}^{-1}(\mathbf{p}_k).
\label{eq:crlb}
\end{equation}
Cooperation introduces overhead in coordination, synchronization, collision avoidance, data fusion, and edge/cloud communication. Thus, cooperative localization must jointly optimize UAV geometry, measurement scheduling, overhead, and tag energy availability.

\subsection{AI-Based Localization}
Hybrid model-driven learning is a promising direction. Instead of training a black-box model from scratch, physical relationships such as distance-dependent path loss, array response, or TDoA geometry can be embedded into the learning pipeline \cite{khalil2025robust}. This can reduce data requirements, improve generalization, and provide uncertainty estimates. For UAV-assisted systems, online adaptation is particularly important because the UAV changes its viewpoint and the channel distribution may vary rapidly across space.

AI-based localization is attractive when analytical propagation models are inaccurate or too complex for real-time deployment. A general supervised localization model is
\begin{equation}
    \hat{\mathbf{p}}_k = f_{\boldsymbol{\theta}}\big(\mathbf{Z}_k\big),
\label{eq:ai_mapping}
\end{equation}
where $\mathbf{Z}_k$ contains RSS, CSI, phase, AoA, Doppler, or fingerprint features. The model can be trained by minimizing
\begin{equation}
    \min_{\boldsymbol{\theta}}\sum_{k}\left\|f_{\boldsymbol{\theta}}(\mathbf{Z}_k)-\mathbf{p}_k\right\|_2^2+\lambda \Omega(\boldsymbol{\theta}).
\label{eq:ai_loss}
\end{equation}
Fingerprinting was popularized in WiFi systems such as RADAR \cite{Bahl2000RADAR}, and later works used CSI and deep neural networks to improve representation and accuracy \cite{Wang2017DeepFi,Chen2017ConFi,Ayyalasomayajula2020DLoc}. In UAV-assisted IoT, AI can learn nonlinear relationships among UAV position, RSS/CSI patterns, tag orientation, multipath, and environmental features. However, labeled data are expensive to collect, models may not generalize, and deep models may be too complex for onboard or edge processors.

\subsection{Accuracy, Mobility, and Scalability Tradeoffs}
Scalability also depends on how often localization is required. Static tags may only need occasional position refinement, while mobile passive objects or temporary tags require frequent updates \cite{khalil2025robust}. Frequent localization increases measurement overhead and UAV energy consumption, but infrequent localization may degrade trajectory planning and beam control. Thus, future systems may need adaptive localization policies that trigger measurements only when uncertainty becomes too large or when sensing/communication performance degrades.

Backscatter localization performance depends on accuracy, UAV mobility, energy consumption, latency, and scalability. Higher accuracy usually requires more measurements, better UAV geometry, larger bandwidth, antenna arrays, or repeated observations. A compact joint mobility-localization objective is
\begin{equation}
\begin{split}
    \min_{\{\mathbf{q}[n]\},\{\mathcal{S}[n]\}} \quad
    & \sum_k w_k \mathrm{tr}\!\left(\mathbf{J}^{-1}(\mathbf{p}_k)\right)
      +\lambda_1E_{\mathrm{UAV}}
      +\lambda_2T_{\mathrm{loc}} \\
    \mathrm{s.t.}\quad
    & \|\mathbf{q}[n+1]-\mathbf{q}[n]\|\leq V_{\max}\Delta t,\quad \forall n,\\
    & \sum_{\ell=1}^{n} E_k^{\mathrm{con}}[\ell]
      \leq E_k^{0}+\sum_{\ell=1}^{n}E_k^{\mathrm{har}}[\ell],
      \quad \forall k,n,\\
    & \mathcal{S}[n]\subseteq\{1,\ldots,K\},\quad \forall n.
\end{split}
\label{eq:localization_tradeoff}
\end{equation}
Here, $\mathcal{S}[n]$ is the set of scheduled tags in slot $n$, $E_k^{0}$ is the initial stored energy of tag $k$, and the second constraint enforces cumulative energy causality.

\begin{figure}[!t]
    \centering
    \includegraphics[width=1.0\columnwidth]{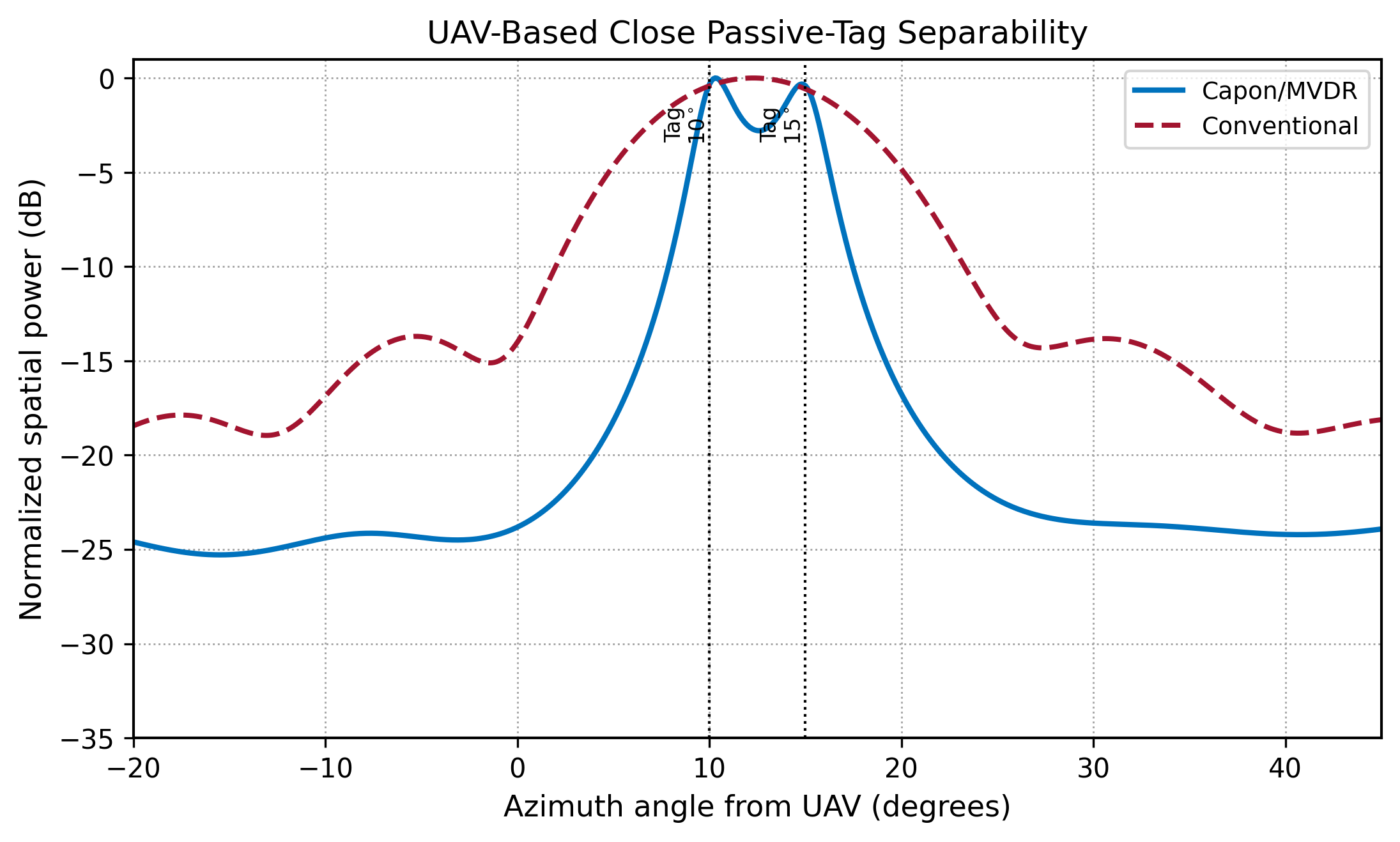}
    \caption{Comparison of conventional and Capon/MVDR spatial spectra for two closely spaced passive backscatter tags using the parameters in Table~\ref{tab:tutorial_simulation_parameters}.}
    \label{fig:uav_based_capon_vs_conventional_multitag}
\end{figure}

In dense IoT deployments, a UAV may need to localize hundreds or thousands of passive tags, making exhaustive measurement collection and centralized processing impractical. Scalable designs may use clustering, compressed sensing, sparse measurement selection, multi-UAV cooperation, edge processing, or learning-based inference \cite{Hassanien2004GeneralizedCapon}. Fig.~\ref{fig:uav_based_capon_vs_conventional_multitag} shows that Capon/MVDR processing can provide sharper angular peaks than conventional processing \cite{Bjornson2024MultipleAntenna}.

\begin{figure*}[!t]
    \centering
    \includegraphics[width=0.95\textwidth]{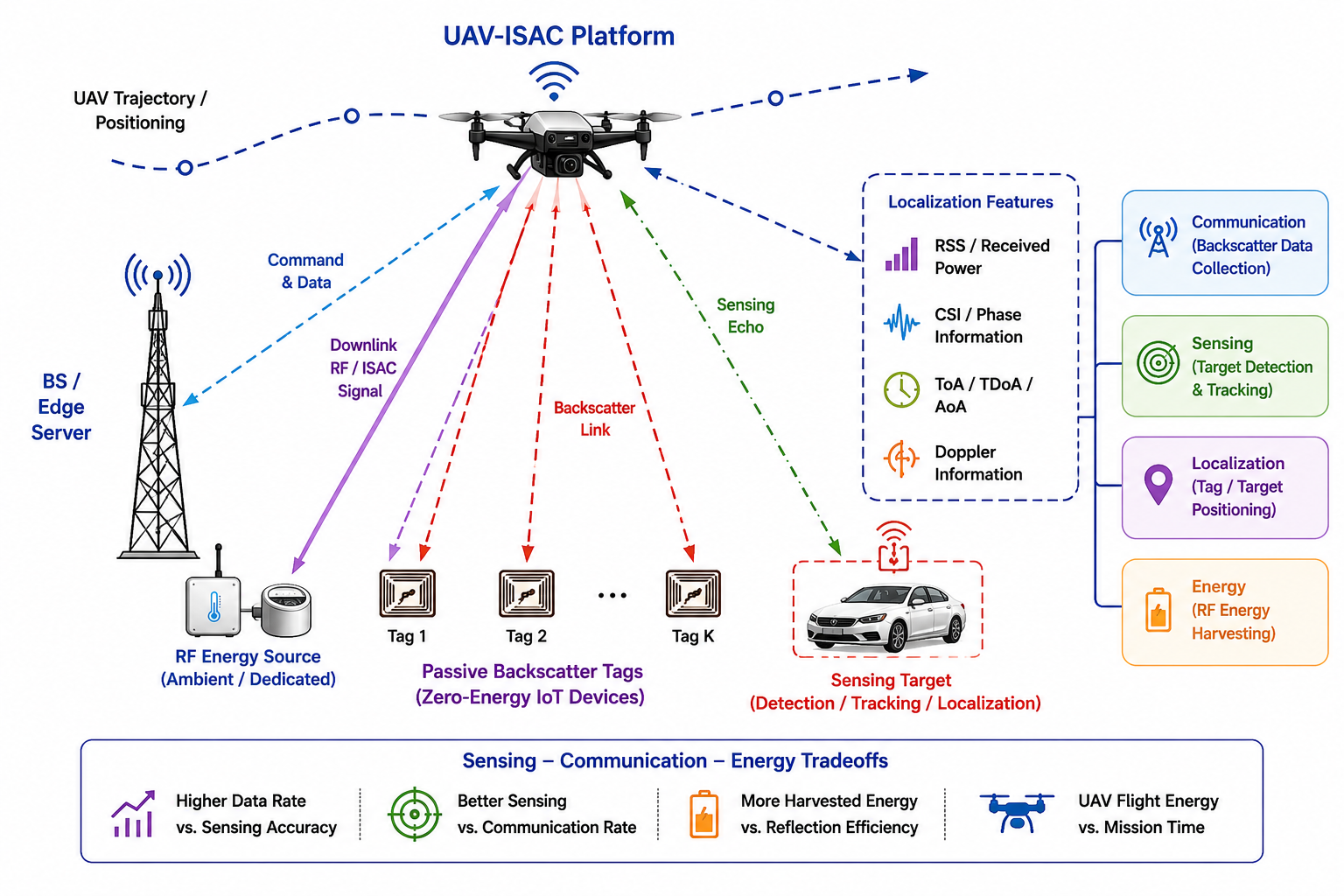}
    \caption{Illustration of an ISAC-enabled UAV-backscatter system for zero-energy IoT.}
    \label{fig:isac_uav_backscatter_nobg}
\end{figure*}

\section{ISAC-Enabled UAV-Backscatter Systems}
\label{sec:isac_uav_backscatter}
The main promise of UAV-backscatter-ISAC is that one platform can provide RF illumination, passive data collection, sensing, and localization. However, combining these functions also increases receiver complexity because the received signal may contain a direct carrier, tag-modulated reflections, target echoes, clutter, and interference. The receiver must therefore separate useful backscatter information from sensing echoes and direct-link components while maintaining low latency and manageable computational cost.

ISAC-enabled UAV-backscatter systems integrate passive BackCom, aerial mobility, and wireless sensing into a unified framework \cite{Liu2022ISACSurvey}. The UAV can transmit or receive dual-functional signals, illuminate passive tags, collect backscattered data, sense targets, and support localization using shared spectrum and hardware. Compared with conventional UAV communication or standalone sensing, this integration improves spectrum utilization and enables environment-aware zero-energy IoT operation \cite{Liu2020JointRadarComm,Zhang2021JCRSignalProcessing}. However, passive tags introduce cascaded channels, weak reflections, control of reflection coefficients, and energy-neutral constraints \cite{Zargari2023BackscatterSensingIntegration,Zhao2024BISAC}. Fig.~\ref{fig:isac_uav_backscatter_nobg} illustrates the considered architecture.

\subsection{Backscatter-ISAC Architectures}
In UAV-enabled settings, the distinction between monostatic and bistatic operation is important. A monostatic UAV can control both illumination and reception, but must contend with self-interference and payload limitations. A bistatic architecture can place the transmitter on the ground and the receiver on the UAV, reducing the UAV's transmit power burden but requiring synchronization and coordination \cite{hu2022trajectory}. Hybrid terrestrial--aerial architectures may offer stronger coverage and sensing diversity, but they also require joint scheduling of terrestrial and aerial resources.

Backscatter-ISAC architectures can be classified according to the roles of the UAV, RF source, passive tags, and receiver. In aerial monostatic operation, the UAV transmits a dual-functional waveform and receives both tag backscatter and target echoes \cite{Meng2024UAVEnabledISAC}. In aerial bistatic operation, the UAV may act as a receiver or sensing platform while a terrestrial node illuminates passive tags \cite{Ahmed2025UAVISACSurvey}. A generic received signal model for a UAV-backscatter-ISAC receiver can be written as
\begin{equation}
\begin{aligned}
    y[n] &= y_d[n]
    +\sum_{k=1}^{K} y_k^{\mathrm{bs}}[n]
    +\sum_{\ell=1}^{L} y_{\ell}^{\mathrm{s}}[n]
    +z[n],\\
    y_d[n] &= h_d[n]s[n],\\
    y_k^{\mathrm{bs}}[n] &=
    \Gamma_k[n] b_k[n] h_{k,r}[n]
    \mathbf{h}_{s,k}^{H}[n]\mathbf{w}[n]s[n],\\
    y_{\ell}^{\mathrm{s}}[n] &=
    \alpha_{\ell}[n]\mathbf{a}_{r}^{H}(\theta_{\ell})
    \mathbf{A}_{\ell}[n]\mathbf{w}[n]s[n].
\end{aligned}
\label{eq:bisac_signal_model}
\end{equation}
Here, $y_d[n]$, $y_k^{\mathrm{bs}}[n]$, and $y_{\ell}^{\mathrm{s}}[n]$ denote the direct-link component, the backscatter-tag component, and the sensing-echo component, respectively. The remaining variables are defined as follows: $s[n]$ is the transmitted waveform, $\mathbf{w}[n]$ is the transmit beamforming vector, $b_k[n]$ is the information symbol of tag $k$, $h_d[n]$ is the direct-link channel, $\mathbf{h}_{s,k}[n]$ is the source-to-tag channel vector, $h_{k,r}[n]$ is the tag-to-receiver channel, $\alpha_{\ell}[n]$ is the sensing-target reflection coefficient, $\mathbf{A}_{\ell}[n]$ captures the sensing-path response, and $z[n]$ is receiver noise. This compact expression is intended as a tutorial model; monostatic, bistatic, and multi-antenna systems may require additional self-interference, clutter, synchronization, and calibration terms.

\subsection{Joint Communication, Sensing, and Localization}
The same UAV position may not be optimal for all objectives. A low altitude can strengthen backscatter reception and energy transfer but reduce coverage and sensing field of view. A high altitude can increase coverage and improve visibility but weaken the cascaded backscatter channel. Similarly, a trajectory that maximizes sensing diversity may increase data collection delay. These tradeoffs suggest adaptive mission planning with sensing-intensive, localization-intensive, and communication-intensive phases.

For communication, passive tags convey data by modulating and reflecting the incident signal. For sensing, the same transmitted or reflected signal can detect targets, estimate parameters, or monitor the environment. For localization, the UAV exploits RSS, CSI, delay, Doppler, angle, or phase features from backscattered signals. A common communication metric is
\begin{equation}
    R_k[n]=B\log_2(1+\gamma_k[n]),
\label{eq:bisac_rate}
\end{equation}
and a beampattern-based sensing metric is
\begin{equation}
    G_s(\theta,n)=\mathbf{a}^{H}(\theta)\mathbf{R}_{x}[n]\mathbf{a}(\theta),\quad \mathbf{R}_{x}[n]=\mathbb{E}\{\mathbf{x}[n]\mathbf{x}^{H}[n]\}.
\label{eq:isac_beampattern}
\end{equation}
In UAV-assisted ISAC, trajectory and scheduling affect sensing frequency, target visibility, and communication coverage; hence, UAV location becomes a sensing--communication variable \cite{Meng2022TrajectoryBeamformingIPSAC,Meng2023ThroughputUAVIPSAC,Liu2024UAVISACIoT}.

\subsection{Beamforming, Waveform, and Reflection Design}
Waveform design adds another layer of complexity. A waveform suitable for communication may not provide the desired delay or Doppler resolution for sensing, while a sensing-oriented waveform may not be optimal for passive tag modulation. For passive IoT, the waveform should also provide sufficient RF energy to activate the tag \cite{hanna2022distributed,khalil2023energy}. Practical waveform design should therefore consider information rate, sensing ambiguity, energy-transfer efficiency, receiver complexity, and compatibility with ambient or standardized RF sources.

\begin{figure}[!t]
    \centering
    \includegraphics[width=1.0\columnwidth]{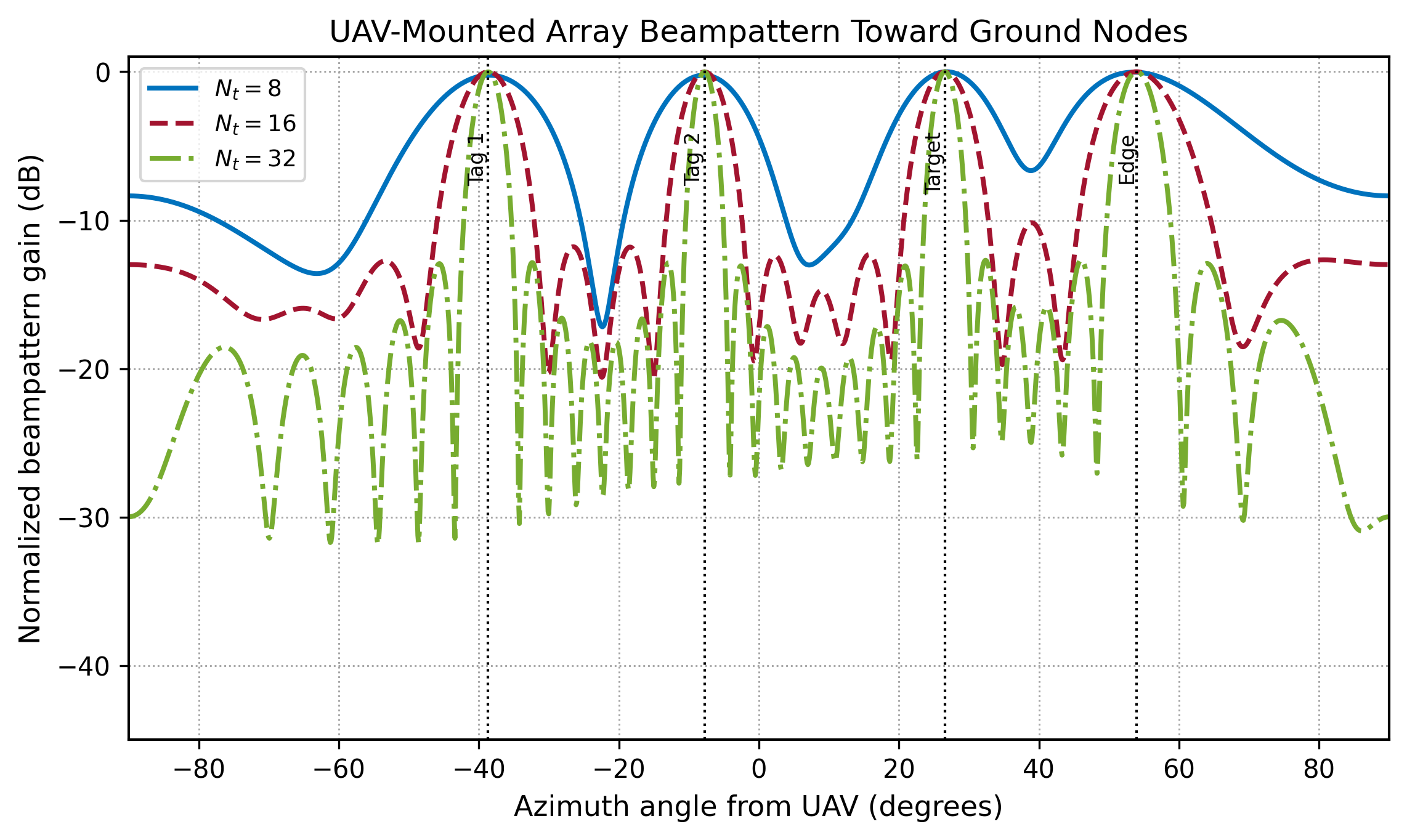}
    \caption{UAV-mounted array beampattern toward representative ground nodes for $N_t=\{8,16,32\}$ using the parameters in Table~\ref{tab:tutorial_simulation_parameters}.}
    \label{fig:uav_based_beampattern_ground_nodes}
\end{figure}

Beamforming and waveform design are central because the transmitted signal must satisfy both communication and sensing requirements. In B-ISAC, passive tag reflection can enhance communication, create sensing signatures, or introduce interference. Recent B-ISAC works study joint beamforming for tag detection, tag estimation, and communication enhancement \cite{Zhao2024BISAC,Zhao2024JointBeamformingBISAC}. Fig.~\ref{fig:uav_based_beampattern_ground_nodes} shows a beampattern example for a UAV-mounted array.

The tag reflection coefficient is modeled as
\begin{equation}
    \Gamma_k[n]=\sqrt{\rho_k[n]}e^{j\phi_k[n]},\quad 0\leq \rho_k[n]\leq 1.
\label{eq:reflection_coefficient}
\end{equation}
A typical joint beamforming and reflection-control problem is
\begin{equation}
\begin{split}
    \max_{\mathbf{w}[n],\{\Gamma_k[n]\}}\quad
    &\sum_{k=1}^{K}R_k[n]\\
    \mathrm{s.t.}\quad
    & G_s(\theta_\ell,n)\geq G_{\ell}^{\min},\quad \forall \ell,\\
    & \|\mathbf{w}[n]\|^2\leq P_{\max},\\
    & 0\leq \rho_k[n]\leq 1,\quad \forall k.
\end{split}
\label{eq:beam_reflection_problem}
\end{equation}
In UAV systems, this problem is coupled with trajectory, altitude, task scheduling, and sensing-period design \cite{Meng2022TrajectoryBeamformingIPSAC,Meng2023ThroughputUAVIPSAC}.

The beampattern in Fig.~\ref{fig:uav_based_beampattern_ground_nodes} is therefore a normalized array-domain illustration rather than a full link-level simulation; it is used to show how increasing the number of UAV-mounted antenna elements improves angular selectivity for RF illumination, backscatter reception, and sensing.

\subsection{Sensing--Communication--Energy Tradeoffs}
These trade-offs also imply that optimization constraints should be selected based on application requirements. In structural monitoring, the system may prioritize reliable periodic tag activation and long-term operation \cite{wu2019fundamental}. In disaster response, the UAV may prioritize rapid sensing and localization, even at the cost of increased energy consumption. In industrial inspection, low latency and high reliability may be more important than maximizing the number of served tags. Therefore, future work should avoid one-size-fits-all objective functions and instead define scenario-specific requirements.

The weights in multi-objective formulations should not be treated as fixed constants in all scenarios. Emergency sensing may prioritize detection and localization accuracy, while agricultural monitoring may prioritize energy sustainability and coverage \cite{li2023adaptive}. Industrial inspection may require low latency and high reliability, whereas environmental monitoring may tolerate longer delays. Adaptive weighting, constraint-based formulations, and Pareto-front analysis can provide more informative design insights than a single weighted-sum objective.

The key challenge in ISAC-enabled UAV-backscatter systems is balancing sensing quality, communication performance, and energy efficiency. Increasing tag reflection improves backscatter communication and sensing visibility but reduces harvested energy. Similarly, hovering near tags improves energy transfer and backscatter reception but increases mission time and propulsion energy \cite{Bi2016WPCN,Yang2021UAVBackscatter}. A compact normalized multi-objective formulation is
\begin{equation}
\begin{aligned}
    \max_{\boldsymbol{\Omega}}\quad
    & \omega_c\!\sum_{n,k}\bar{R}_k[n]
      +\omega_s\!\sum_{n,\ell}\bar{G}_s(\theta_\ell,n) \\
    & {}-\omega_e\bar{E}_{\mathrm{tot}} \\
    \mathrm{s.t.}\quad
    & \sum_{t=1}^{n}E_k^{\mathrm{con}}[t]
      \leq E_k^{0}
      +\sum_{t=1}^{n}E_k^{\mathrm{har}}[t],
      \quad \forall k,n,\\
    & \|\mathbf{q}[n+1]-\mathbf{q}[n]\|
      \leq V_{\max}\Delta t,
      \quad \forall n,\\
    & \|\mathbf{w}[n]\|^2\leq P_{\max},
      \quad \forall n,\\
    & 0\leq \rho_k[n]\leq 1,
      \quad \forall k,n.
\end{aligned}
\label{eq:isac_energy_tradeoff}
\end{equation}
where 
$\boldsymbol{\Omega}=\{\mathbf{q}[n],\mathbf{w}[n],\Gamma_k[n],\mathcal{S}[n]\}_{n,k}$ 
collects the UAV trajectory, beamforming vectors, tag reflection coefficients, and scheduling variables. Here, $\bar{R}_k[n]$, $\bar{G}_s(\theta_\ell,n)$, and $\bar{E}_{\mathrm{tot}}$ denote normalized communication, sensing, and total-energy metrics. This formulation shows that the UAV must jointly decide where to fly, which tags to serve, how to beamform, how much energy to allocate, and how to satisfy sensing requirements.

\begin{figure*}[!t]
    \centering
    \includegraphics[width=0.8\textwidth]{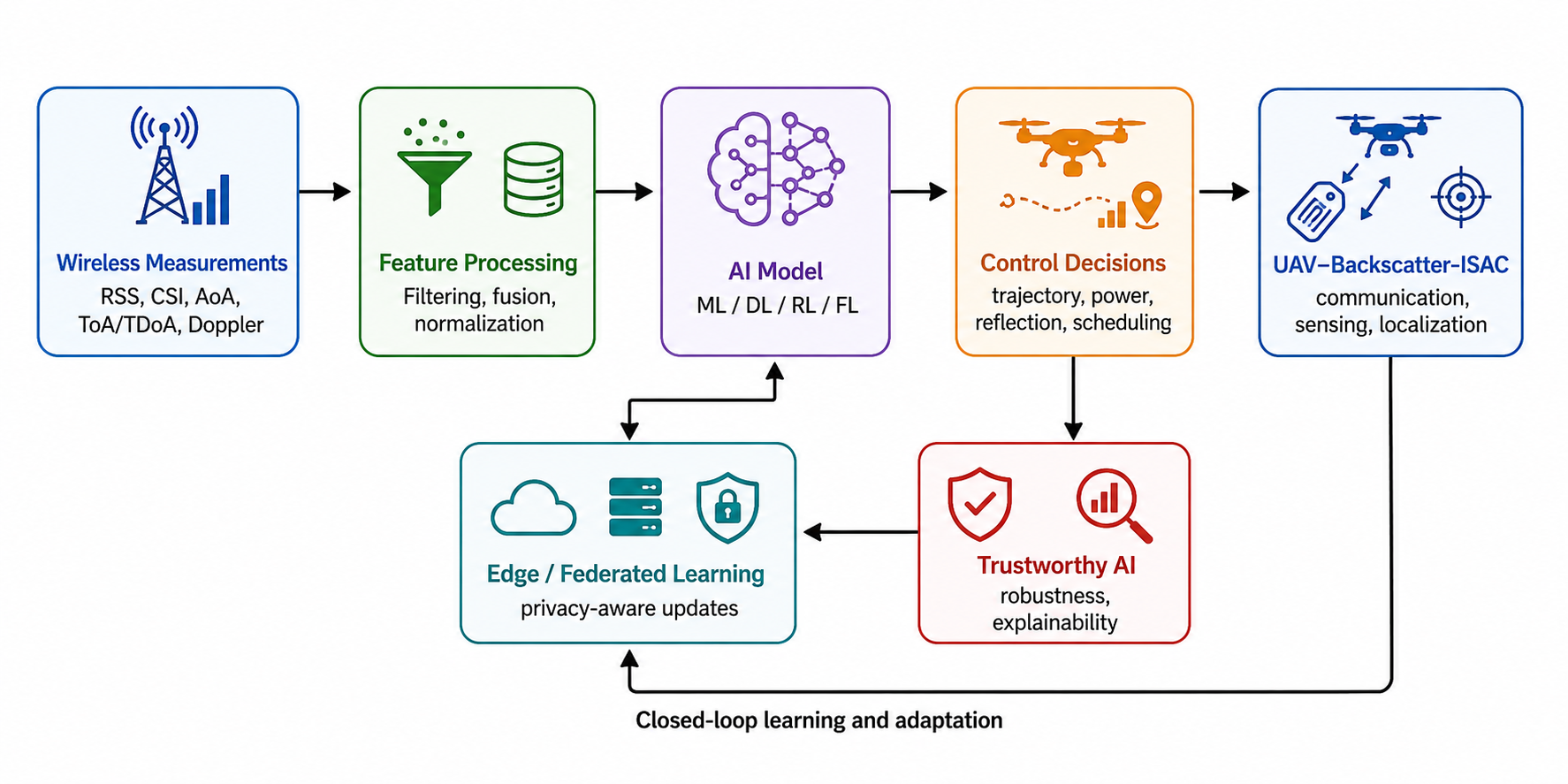}
    \caption{AI-enabled closed-loop design pipeline for UAV-assisted backscatter localization and ISAC.}
    \label{fig:ai_design_pipeline}
\end{figure*}

\section{AI-Empowered Design and Optimization}
\label{sec:ai_design_optimization}
AI is most useful when it complements rather than replaces domain knowledge. Model-based optimization provides feasibility guarantees and physical interpretability, while learning methods can adapt to unknown channels, mobility, interference, and tag distributions. Hybrid methods can therefore combine the reliability of analytical models with the adaptability of data-driven control. This is especially important in UAV-assisted systems, where unsafe exploration may waste energy or violate mission constraints.

AI-empowered design provides a data-driven and adaptive framework for UAV-assisted backscatter localization and ISAC. The considered networks involve coupled variables such as UAV trajectory, channel state, tag reflection coefficient, RF-source power, sensing accuracy, localization error, harvested energy, and interference \cite{OShea2017DeepLearningPhysical}. Classical optimization is effective when models are accurate and stationary, but it can be computationally expensive or inaccurate in dynamic deployments. AI can complement model-based methods by learning from measurements, predicting channel and mobility characteristics, adapting resource-control policies, and supporting real-time decisions \cite{Zhang2019DeepLearningWireless}. Fig.~\ref{fig:ai_design_pipeline} summarizes the AI-enabled design pipeline.

\subsection{AI for Channel Estimation and Signal Detection}
\label{subsec:ai_channel_detection}
Training-data generation is a particular challenge in BackCom because ground-truth cascaded channels and tag states are difficult to measure separately. Simulation data can provide labels, but models trained only on idealized simulation may not transfer to real deployments. Field measurements improve realism, but collecting them for many UAV altitudes, tag orientations, RF sources, and environments is expensive. Semi-supervised learning, transfer learning, and synthetic-to-real adaptation are therefore important.

Channel estimation and signal detection are fundamental in UAV-assisted BackCom because the desired tag signal is weak, indirect, and often masked by interference. Deep learning has been used to jointly learn channel estimation and signal detection in the presence of nonlinear distortion, imperfect CSI, and complex interference \cite{OShea2017DeepLearningPhysical,Ye2018PowerDL}. For AmBC, deep residual learning and transfer learning have been applied to channel estimation and tag detection \cite{Liu2021DeepResidualAmBC,Liu2021DeepTransferAmBC,Zargari2023AmBCJointDNN}. A generic channel estimator is
\begin{equation}
    \hat{\mathbf{h}}=f_{\boldsymbol{\theta}}(\mathbf{Y}),
\label{eq:ai_channel_estimation}
\end{equation}
with training objective
\begin{equation}
    \min_{\boldsymbol{\theta}}\mathbb{E}\left[\|\mathbf{h}-f_{\boldsymbol{\theta}}(\mathbf{Y})\|_2^2\right].
\label{eq:ai_channel_loss}
\end{equation}
For detection, the receiver can learn
\begin{equation}
    \hat{\mathbf{b}}=g_{\boldsymbol{\psi}}(\mathbf{Y},\hat{\mathbf{h}}).
\label{eq:ai_detection}
\end{equation}
In UAV-assisted settings, inputs should include UAV position, altitude, velocity, distance, Doppler, and angle.

\subsection{Learning-Based UAV Trajectory and Resource Optimization}
\label{subsec:learning_uav_resource}

Reward design is another important issue. A reward based only on throughput may cause the UAV to ignore poorly energized tags or sensing tasks. A reward based solely on localization error may lead to excessive hovering and high propulsion energy consumption. Constraint-aware rewards should include penalties for energy-causality violations, collision risk, missed sensing tasks, and latency. For safety-critical missions, learning should be combined with rule-based constraints or model predictive control.

UAV trajectory and resource optimization are sequential decision-making problems. At each slot, the UAV must decide where to move, which tags to serve, how much RF power to transmit, how to allocate resources, and how to balance communication, sensing, localization, and energy objectives. RL and DRL are useful when the UAV must adapt online to uncertain channels, mobile targets, time-varying interference, and incomplete CSI \cite{Sutton2018RL,Chang2023MultiUAVDRL}. The UAV control problem can be modeled as an MDP with objective
\begin{equation}
    \max_{\pi}\mathbb{E}_{\pi}\left[\sum_{n=0}^{N-1}\gamma^n r(s_n,a_n)\right],
\label{eq:rl_objective}
\end{equation}
The reward may combine normalized communication, sensing, localization, and energy terms as
\begin{equation}
    r_n=
    \omega_c\bar{R}_n
    +\omega_s\bar{S}_n
    -\omega_l\bar{e}_n^{\mathrm{loc}}
    -\omega_e\bar{E}_n^{\mathrm{UAV}}
    -\omega_v\mathcal{V}_n,
\label{eq:ai_reward}
\end{equation}
where $\mathcal{V}_n$ penalizes constraint violations such as energy-causality violation, collision risk, missed sensing tasks, or latency violations. Reward terms should be normalized or otherwise scaled to avoid one objective dominating the learning process only because of units \cite{Zhang2023DRLUAVComm,Qin2023DRLUAVISAC,Li2024MultiUAVMEC,Huang2019RISUsingDeepRL}.

\subsection{Edge, Federated, and Lightweight Learning}
\label{subsec:edge_federated_lightweight}
Lightweight learning is also needed for onboard deployment. Model pruning and quantization can reduce inference cost, while split learning can divide computation between UAVs and edge servers. Over-the-air aggregation may reduce federated learning overhead, but it requires synchronization and is sensitive to channel noise. Personalized federated learning can improve performance when UAVs operate across different environments, but it adds complexity to model management.

\begin{figure*}[!t]
    \centering
    \includegraphics[width=0.85\textwidth]{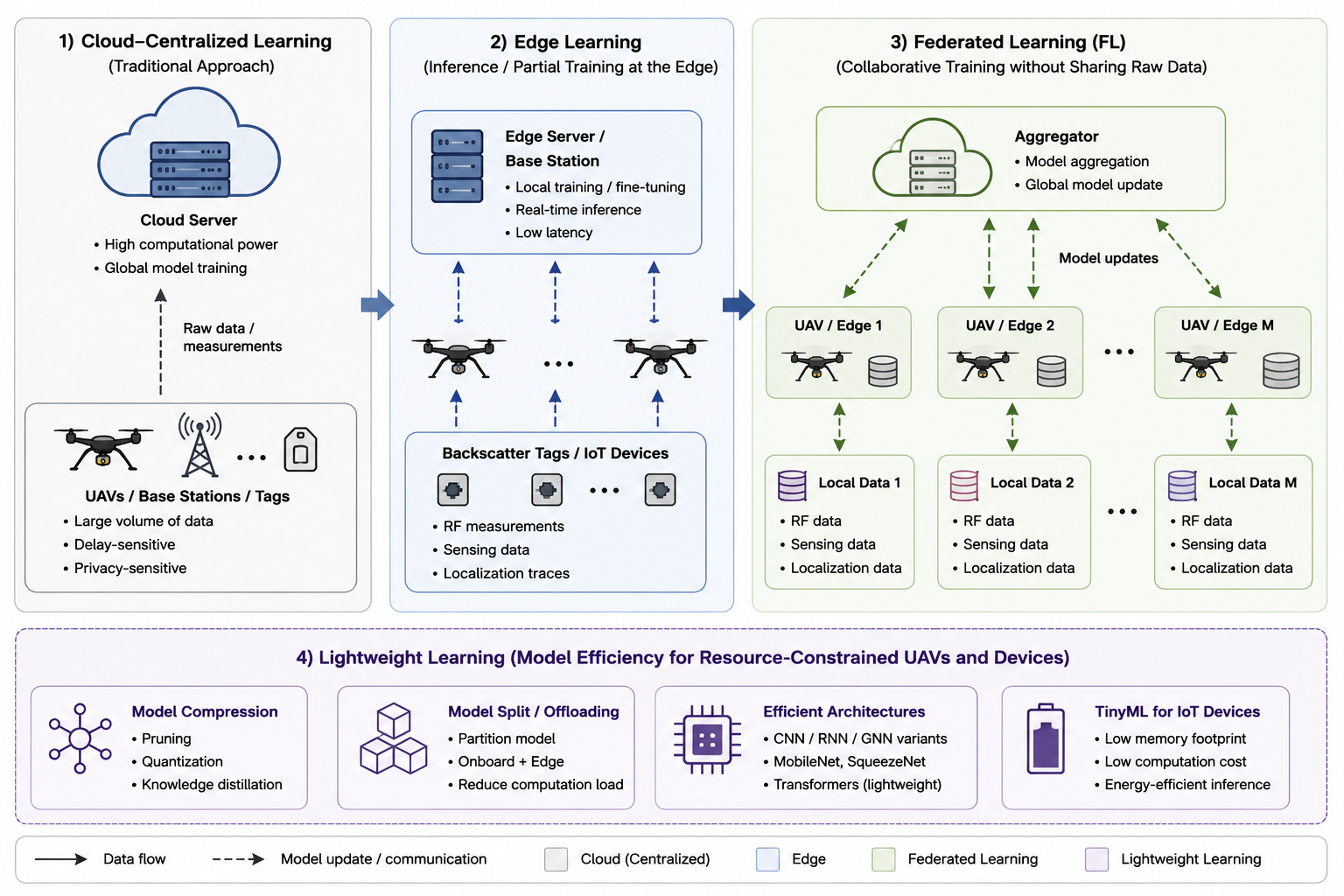}
    \caption{Cloud, edge, federated, and lightweight learning for UAV-assisted backscatter localization and ISAC.}
    \label{fig:edge_federated_lightweight_learning}
\end{figure*}

Cloud-based AI can provide high computation, but it may be unsuitable for UAV-assisted zero-energy IoT because raw RF measurements, sensing data, and localization traces are large, delay-sensitive, and privacy-sensitive. Edge learning moves inference or partial training closer to UAVs, edge servers, or base stations, reducing latency and backhaul load \cite{Mao2017MEC,Xu2020EdgeIntelligence}. TinyML can enable lightweight inference on resource-constrained devices, although zero-energy tags may support only minimal sensing and control logic \cite{Dutta2021TinyML}. Fig.~\ref{fig:edge_federated_lightweight_learning} illustrates cloud, edge, federated, and lightweight learning.

Federated learning is useful when UAVs, edge servers, or gateways collect local data but cannot share raw samples. The standard federated averaging update is
\begin{equation}
    \boldsymbol{\theta}^{(t+1)}=\sum_{m=1}^{M}\frac{D_m}{D}\boldsymbol{\theta}_m^{(t+1)},
\label{eq:fedavg}
\end{equation}
where $D=\sum_mD_m$ \cite{McMahan2017FedAvg,Kairouz2021FederatedLearning}. However, unreliable links, non-IID data, stragglers, limited bandwidth, poisoning, and energy-constrained updates remain barriers \cite{Nguyen2021FL6G,Kairouz2021FederatedLearning}.

\subsection{Trustworthy and Explainable AI}
\label{subsec:trustworthy_xai}
Trustworthiness also includes data governance. Localization traces, sensing maps, and RF fingerprints may reveal sensitive information about infrastructure, users, assets, or operational routines. Even if raw data are not shared, model updates in federated learning may leak information through gradient inversion or membership inference attacks. Therefore, privacy-preserving learning, secure aggregation, differential privacy, and access control should be considered together with physical-layer security and robust sensing.

Trustworthy AI is particularly important when learning controls UAV motion. An incorrect trajectory decision can reduce coverage, violate flight constraints, waste energy, or miss critical sensing events. Similarly, an unreliable localization model can misguide beamforming and scheduling. Future evaluation should include confidence calibration, failure-case analysis, and robustness testing under unseen propagation environments, malicious interference, spoofed measurements, and hardware mismatch.

Trustworthy AI is essential because AI decisions affect UAV movement, sensing coverage, localization, energy delivery, and communication reliability. Explainable AI provides tools to interpret model outputs, identify important features, and improve trust in automated decisions \cite{Arrieta2020XAI}. In the system under consideration, XAI can show whether RSS, CSI, phase, Doppler, or AoA dominates localization; whether trajectory decisions improve energy, reduce uncertainty, or avoid interference; and whether scheduling prioritizes fairness, latency, activation, or sensing accuracy. Trustworthy operation requires robustness to adversarial perturbations, spoofed localization features, poisoned federated updates, RF jamming, and distribution shifts \cite{Brik2023XAI6GORAN}.

\section{Comparative Analysis, Open Challenges, and Future Directions}
\label{sec:comparative_challenges_future}
The comparative analysis in this section is intended to show both maturity and fragmentation. Some areas, such as UAV-assisted BackCom resource allocation, already have several optimization-based studies. Other areas, such as UAV-assisted passive localization, B-ISAC with UAV mobility, and reproducible AI-enabled benchmarks, remain much less developed. This imbalance motivates the quantitative trend figures and coverage matrix presented below.

This section synthesizes the reviewed literature and highlights the main gaps in AI-empowered UAV-assisted backscatter localization and ISAC for zero-energy IoT. Existing studies provide foundations for UAV-assisted BackCom, passive localization, AI-based wireless optimization, and ISAC, but most treat these components separately. UAV-assisted BackCom studies mainly focus on trajectory, scheduling, throughput, and energy efficiency \cite{Hua2020ThroughputUAVBackscatter,Yang2021UAVBackscatter,Wang2022UAVBackscatterRA}, while ISAC studies often focus on waveform, beamforming, sensing accuracy, and communication rate without passive zero-energy tags \cite{Liu2022ISACSurvey,Liu2024UAVISACIoT}. AI-enabled BackCom studies improve channel estimation, detection, and resource allocation, but rarely provide a unified treatment of UAV mobility, localization, ISAC, and energy-neutral operation \cite{Xu2023AIEmpoweredBackscatter,Zhang2025IntelligentISAC}.

\subsection{Comparison of Representative Studies}
\label{subsec:comparison_representative_studies}
The representative studies reviewed in this survey can be grouped into five main streams: UAV-assisted BackCom, backscatter localization, interference-aware UAV BackCom, UAV-ISAC, and backscatter-ISAC. 
UAV-assisted BackCom studies show that UAV mobility improves passive-device access by reducing propagation distance and improving link geometry \cite{Hua2020ThroughputUAVBackscatter,Han2021UAVAidedBackscatter}. 
Energy-efficient UAV-BackCom designs further show that trajectory planning, tag scheduling, carrier-emitter power control, and reflection-coefficient design are strongly coupled \cite{Yang2021UAVBackscatter,Wang2022UAVBackscatterRA}. 
Backscatter localization studies show that passive tags can be localized using RSS, phase, CSI, antenna arrays, or mobility-assisted measurements \cite{Zhang2020RobotBackscatterLocalization,Pettorru2024TrustworthyLocalization}. 
UAV-ISAC studies address trajectory design, sensing-task scheduling, power allocation, and communication constraints \cite{Meng2023ThroughputUAVIPSAC,Liu2024UAVISACIoT}, but most of them assume active users rather than passive zero-energy tags. 
Backscatter-ISAC works introduce passive reflection into ISAC \cite{Zargari2023BackscatterSensingIntegration,Zhao2024BISAC}, but UAV mobility, passive localization, and scalable AI are not yet fully integrated.

The comparison also shows that existing studies use different channel models, UAV altitude constraints, tag energy models, RF-source placements, and performance metrics. 
This makes direct numerical comparison difficult. 
Therefore, Table~\ref{tab:representative_studies_comparison} focuses on the scope, key techniques, reported metrics, and remaining gaps rather than ranking algorithms under incompatible assumptions.

\begin{table*}[!t]
\centering
\caption{Comparison of Representative Studies Related to UAV-Assisted Backscatter Localization and ISAC.}
\label{tab:representative_studies_comparison}
\footnotesize
\renewcommand{\arraystretch}{1.25}
\setlength{\tabcolsep}{3pt}
\begin{tabularx}{\textwidth}{|p{0.12\textwidth} |p{0.20\textwidth} |p{0.21\textwidth} |p{0.15\textwidth} |X|}
\hline\hline
\textbf{Study} & \textbf{Main Focus} & \textbf{Key Techniques} & \textbf{Main Metrics} & \textbf{Limitations / Gaps} \\
\hline\hline
Hua \textit{et al.} \cite{Hua2020ThroughputUAVBackscatter} 
& UAV-aided BackCom throughput maximization 
& UAV relay assistance, TB/TBR protocols, trajectory design 
& Throughput, link feasibility 
& Does not jointly address localization, sensing, AI, or zero-energy ISAC operation. \\
\hline
Yang \textit{et al.} \cite{Yang2021UAVBackscatter} 
& Energy-efficient UAV-assisted BackCom 
& Joint trajectory, scheduling, carrier-emitter power, and resource optimization 
& Energy efficiency, throughput, harvested energy 
& Communication-centric; localization and sensing are not central. \\
\hline
Han \textit{et al.} \cite{Han2021UAVAidedBackscatter} 
& UAV-aided BackCom performance and trajectory optimization 
& Analytical modeling and UAV path design 
& Outage, throughput 
& Limited integration with ISAC, localization, and learning-based optimization. \\
\hline
Wang \textit{et al.} \cite{Wang2022UAVBackscatterRA} 
& Resource allocation for UAV-assisted BackCom 
& Power control, scheduling, reflection design, trajectory optimization 
& Max--min rate, fairness, reliability 
& Does not include sensing, localization, or AI-enabled closed-loop control. \\
\hline
Zhong \textit{et al.} \cite{Zhong2024InterferenceUAVBackscatter} 
& Interference cancellation in UAV ambient BackCom 
& Signal extraction and interference mitigation 
& Detection reliability, interference suppression 
& Broader UAV trajectory, ISAC, and localization coupling remains open. \\
\hline
Zhang \textit{et al.} \cite{Zhang2020RobotBackscatterLocalization} 
& Mobility-assisted backscatter localization 
& Robot-assisted measurement and localization 
& Localization accuracy 
& Does not consider UAV-ISAC or energy-neutral network-wide optimization. \\
\hline
Liu \textit{et al.} \cite{Liu2024UAVISACIoT} 
& UAV-assisted ISAC for IoT 
& 3D trajectory, sensing-task scheduling, and power allocation 
& Radar estimation rate, communication rate 
& Does not include passive BackCom or zero-energy tags. \\
\hline
Zhao \textit{et al.} \cite{Zhao2024BISAC} 
& Backscatter-ISAC 
& Joint beamforming for tag detection, estimation, and communication 
& SINR, rate, detection, estimation error 
& Does not include UAV mobility or multi-UAV localization. \\
\hline
Xu \textit{et al.} \cite{Xu2023AIEmpoweredBackscatter} 
& AI-empowered BackCom survey 
& AI for detection, channel estimation, interference mitigation, resource allocation 
& Detection, estimation, communication performance 
& UAV-assisted localization and ISAC are not unified. \\
\hline
\textbf{This survey} 
& AI-empowered UAV-assisted backscatter localization and ISAC 
& Unified taxonomy, comparison, UAV mobility, passive BackCom, localization, ISAC, AI 
& Communication, sensing, localization, energy, robustness, reproducibility 
& Provides an integrated perspective and identifies cross-layer gaps. \\
\hline\hline
\end{tabularx}
\end{table*}

\subsection{Quantitative Trends and Comparative Insights}
\label{subsec:quantitative_trends}
\begin{figure*}[!t]
    \centering
    \includegraphics[width=0.65\textwidth]{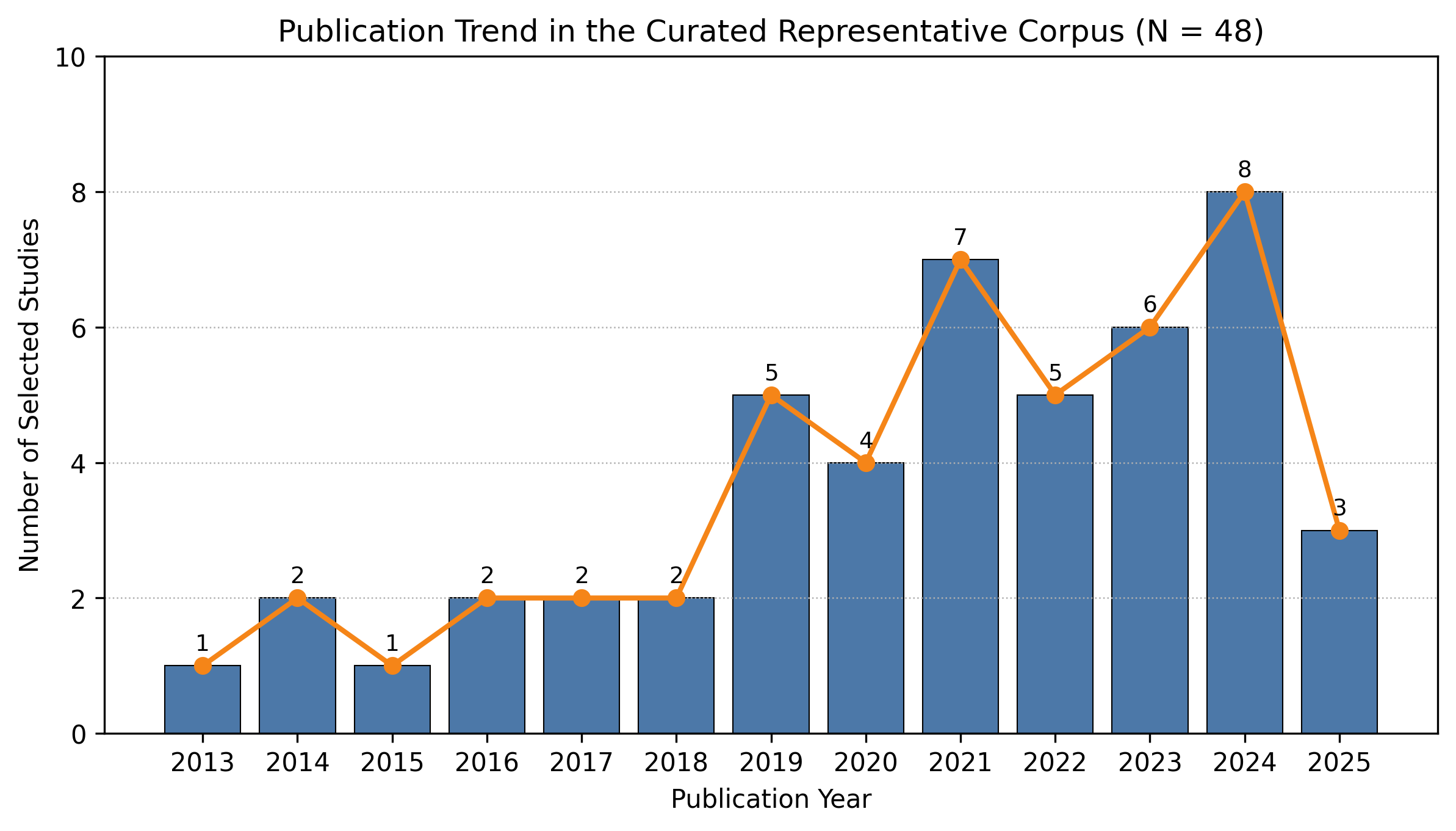}
    \caption{Publication trend of the curated representative corpus used in this survey.}
    \label{fig:publication_trend_exact_corpus}
\end{figure*}

\begin{figure*}[!t]
    \centering
    \includegraphics[width=0.65\textwidth]{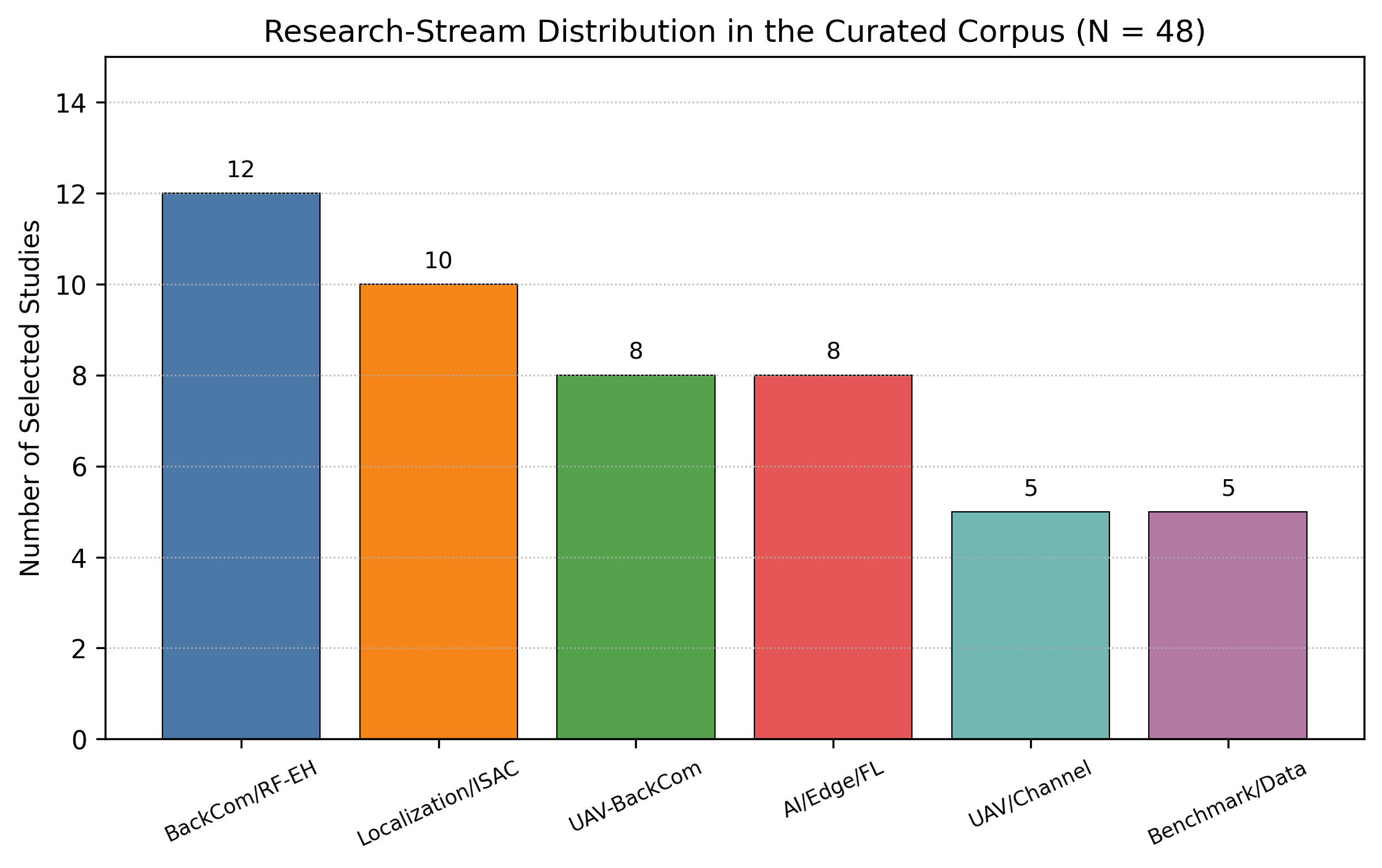}
    \caption{Research-stream distribution of the curated representative corpus.}
    \label{fig:research_stream_distribution_exact_corpus}
\end{figure*}

\begin{figure*}[!t]
    \centering
    \includegraphics[width=0.65\textwidth]{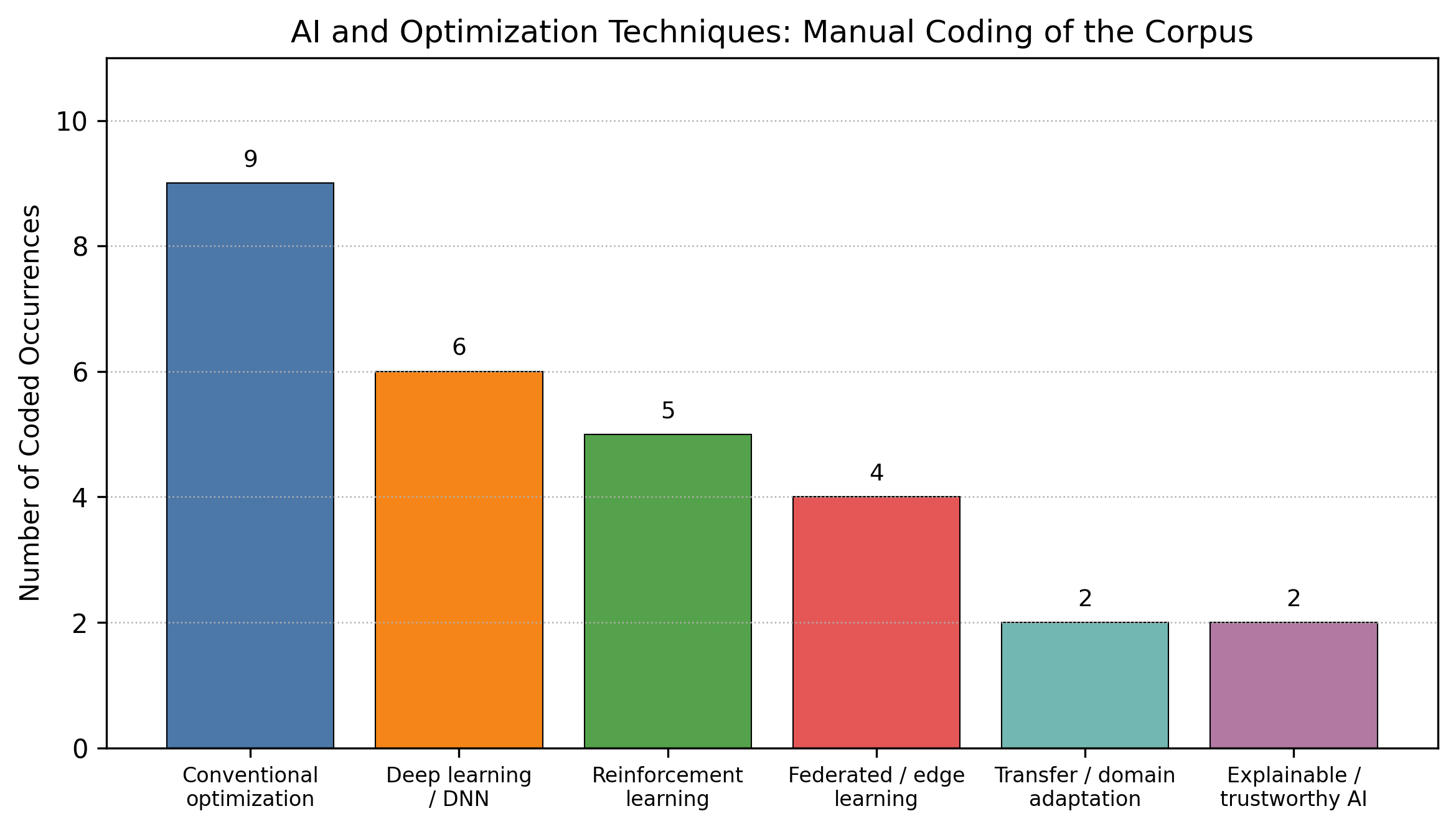}
    \caption{Distribution of AI and optimization techniques after manual coding of the curated corpus.}
    \label{fig:ai_technique_distribution_exact_corpus}
\end{figure*}

\begin{figure*}[!t]
    \centering
    \includegraphics[width=0.65\textwidth]{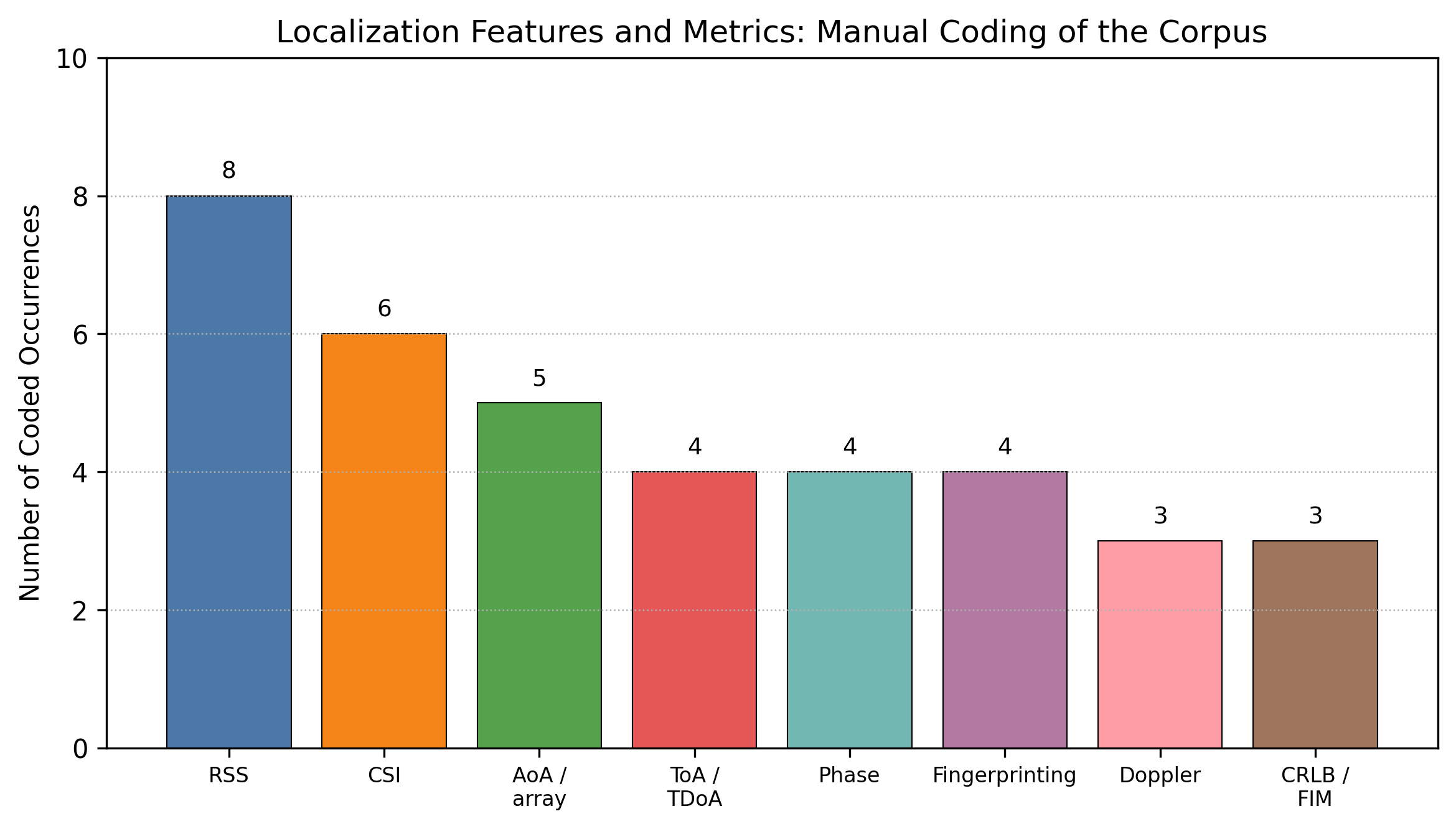}
    \caption{Distribution of localization features and metrics after manual coding of the curated corpus.}
    \label{fig:localization_feature_distribution_exact_corpus}
\end{figure*}

\begin{figure*}[!t]
    \centering
    \includegraphics[width=0.75\textwidth]{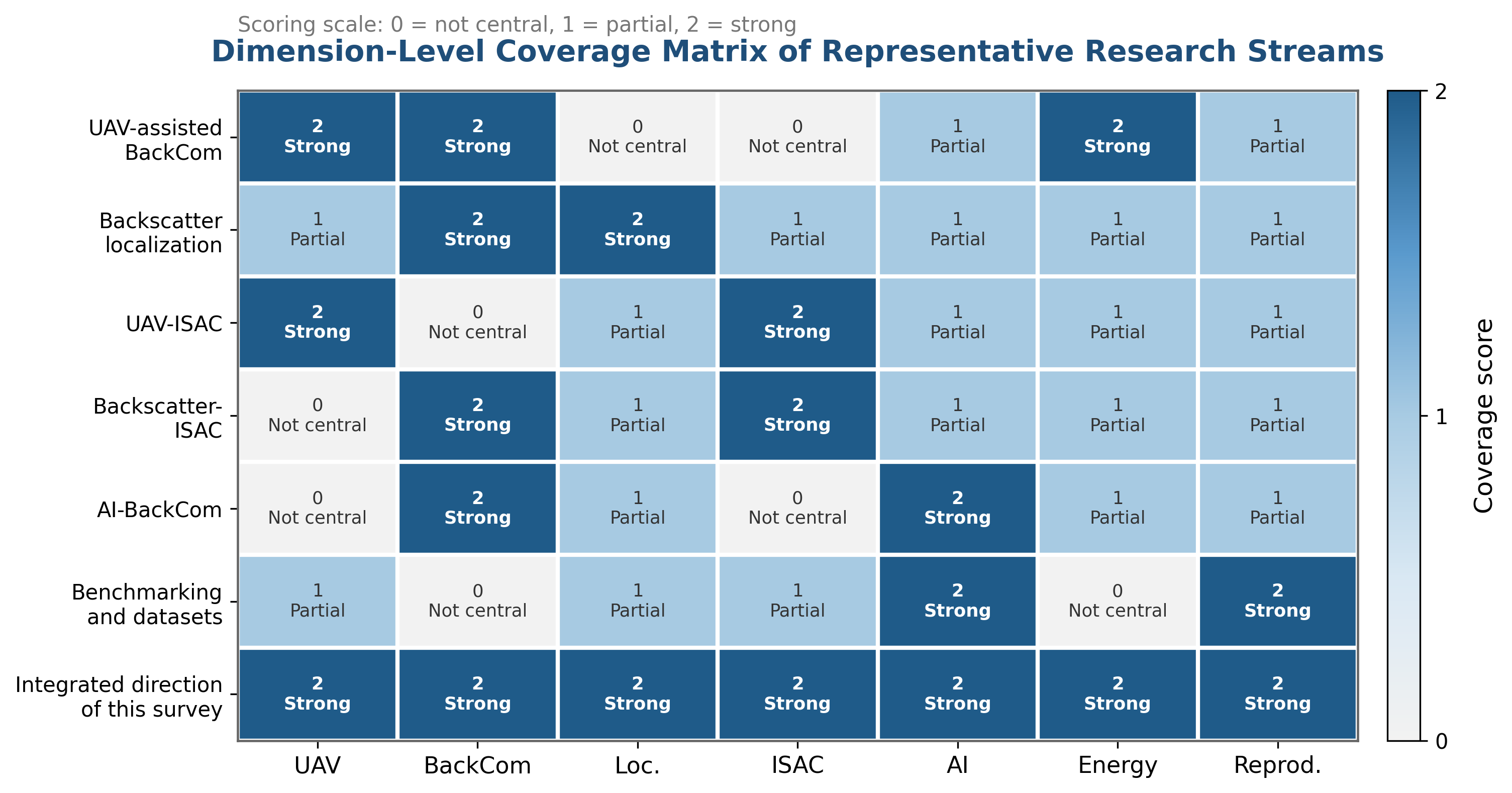}
    \caption{Dimension-level coverage heatmap of representative research streams.}
    \label{fig:dimension_coverage_matrix_heatmap}
\end{figure*}

The quantitative synthesis in this subsection is based on the coded corpus provided in Supplementary Tables~S1--S4, which makes the study selection and coding process behind Figs.~\ref{fig:publication_trend_exact_corpus}--\ref{fig:dimension_coverage_matrix_heatmap} explicit and reproducible. The quantitative synthesis is intended to strengthen the paper's survey-style contribution by complementing the qualitative review with a structured mapping of the selected literature. Instead of only describing individual studies, it provides a compact view of how the selected works are distributed across publication years, research streams, AI/optimization methods, localization features, and dimension-level coverage. This type of structured synthesis is consistent with systematic review and systematic mapping practices, where a curated corpus is analyzed to reveal trends, coverage gaps, and research concentrations \cite{Page2021PRISMA,Rethlefsen2021PRISMAS,Petersen2015Mapping}. 
It helps identify which aspects of the field are relatively mature and which remain fragmented. 
For example, the presence of many communication- and energy-oriented BackCom studies does not imply that backscatter localization or B-ISAC with UAV mobility is equally mature \cite{Yang2021UAVBackscatter,Wang2022UAVBackscatterRA,Zhang2020RobotBackscatterLocalization,Zhao2024BISAC}. 
Similarly, the growth of AI-based methods does not automatically imply reproducible, scalable, or hardware-ready solutions \cite{Xu2023AIEmpoweredBackscatter,Henderson2018DRLMatters,Pineau2021Reproducibility}.

The quantitative results should be interpreted as a snapshot from a structured survey rather than a complete bibliometric census. 
Their purpose is to show how the selected representative studies are distributed across the main technical streams and where coverage gaps remain. 
The analysis is based on a curated corpus of 48 representative studies selected and cited in this survey. 
The corpus includes foundational works, surveys, standards-oriented reports, datasets, and representative technical papers on RF energy harvesting, BackCom, UAV-assisted BackCom, UAV channels, localization, ISAC, AI-enabled optimization, edge/federated learning, benchmarking, and reproducibility \cite{Lu2015RFEH,Bi2016WPCN,VanHuynh2018AmbientSurvey,Zeng2016UAVComm,Liu2022ISACSurvey,Xu2023AIEmpoweredBackscatter}. 
The reported values are exact counts within this selected corpus rather than database-wide bibliometric counts. 
For reproducibility, the coded corpus should be provided in the manuscript appendix or as supplementary material with the fields: BibTeX key, year, primary research stream, AI/optimization codes, localization-feature codes, reported metrics, and dimension-level coverage scores.

Fig.~\ref{fig:publication_trend_exact_corpus} shows the publication trend of the curated representative corpus. 
The figure indicates that early works primarily established the foundations of RF energy harvesting, ambient BackCom, UAV communication, and localization, while relevant studies increased after 2019 with the emergence of UAV-assisted BackCom, ISAC, AI-enabled optimization, and reproducible wireless datasets \cite{Yang2021UAVBackscatter,Wang2022UAVBackscatterRA,Liu2024UAVISACIoT,Alkhateeb2019DeepMIMO,Hoydis2022Sionna}. 
The recent growth around 2023--2024 reflects increasing attention to UAV-assisted BackCom, B-ISAC, intelligent wireless optimization, trustworthy localization, and interference-aware UAV-backscatter systems.


Fig.~\ref{fig:research_stream_distribution_exact_corpus} summarizes the distribution of the selected studies across major research streams. 
BackCom/RF energy harvesting represents a large portion of the corpus because it provides the foundation for passive and zero-energy operation \cite{Lu2015RFEH,Bi2016WPCN,VanHuynh2018AmbientSurvey}. 
Localization/ISAC forms another major group, showing the increasing importance of sensing-aware and position-aware wireless networks \cite{Liu2022ISACSurvey,Pettorru2024TrustworthyLocalization,Zhao2024BISAC}. 
UAV-assisted BackCom and AI/edge/federated learning also contribute strongly, while UAV/channel modeling and benchmarking/data studies provide supporting foundations for realistic and reproducible evaluation.

Table~\ref{tab:supp_curated_corpus_part1} and Table~\ref{tab:supp_curated_corpus_part2} complement Table~\ref{tab:representative_studies_comparison} by adding the publication year and emphasizing reported quantitative metrics. 
While Table~\ref{tab:representative_studies_comparison} is mainly qualitative, Table~\ref{tab:supp_curated_corpus_part1} and Table~\ref{tab:supp_curated_corpus_part2} highlight what each representative study actually measures and what remains missing from a joint UAV-backscatter-localization-ISAC perspective. 
The comparison shows that existing works usually report strong results for a specific objective, such as throughput, energy efficiency, outage probability, localization accuracy, radar estimation rate, or sensing/communication SINR. 
However, only a limited number of studies jointly evaluate communication, sensing, localization, energy, AI complexity, and reproducibility within a single framework.

\begin{table*}[h!]
\centering
\caption{Curated 48-Study Corpus Used for Quantitative Coding and Coverage Analysis (Part I of II).}
\label{tab:supp_curated_corpus_part1}
\footnotesize
\renewcommand{\arraystretch}{1.08}
\setlength{\tabcolsep}{2.2pt}
\begin{tabularx}{\textwidth}{|p{0.04\textwidth}|p{0.17\textwidth}|p{0.055\textwidth}|p{0.18\textwidth}|X|}
\hline\hline
\textbf{ID} & \textbf{Study} & \textbf{Year} & \textbf{Primary Stream} & \textbf{Reason for Inclusion / Coding Tags} \\
\hline\hline
1 & Lu \textit{et al.} \cite{Lu2015RFEH} & 2015 & RF energy harvesting / WPCN & Foundational RF energy harvesting survey; coded for energy harvesting, WPCN, energy-neutral operation, and RF power transfer. \\
\hline
2 & Bi \textit{et al.} \cite{Bi2016WPCN} & 2016 & RF energy harvesting / WPCN & Wireless-powered communication network foundation; coded for energy causality, harvest-then-transmit operation, and energy-aware communication. \\
\hline
3 & Zhou \textit{et al.} \cite{Zhou2013SWIPT} & 2013 & SWIPT / RF energy transfer & Foundational SWIPT work; coded for rate-energy tradeoff, receiver constraints, and simultaneous information-energy transfer. \\
\hline
4 & Liu \textit{et al.} \cite{Liu2013AmbientBackscatter} & 2013 & Ambient BackCom & Foundational ambient backscatter system; coded for passive communication, ambient RF source reuse, and zero-energy IoT relevance. \\
\hline
5 & Kellogg \textit{et al.} \cite{Kellogg2014WiFiBackscatter} & 2014 & WiFi BackCom & Practical WiFi backscatter demonstration; coded for ambient/wireless backscatter feasibility and low-power communication. \\
\hline
6 & Boyer and Roy \cite{Boyer2014BackscatterRFID} & 2014 & RFID / BackCom fundamentals & Foundational RFID/backscatter modeling; coded for passive tags, cascaded channel behavior, modulation, coding, and MIMO-related BackCom. \\
\hline
7 & Van Huynh \textit{et al.} \cite{VanHuynh2018AmbientSurvey} & 2018 & Ambient BackCom survey & Survey on ambient BackCom architectures and challenges; coded for BackCom taxonomy, RF sources, and interference issues. \\
\hline
8 & Zhang \textit{et al.} \cite{Zhang2019NextGenBackscatter} & 2019 & Next-generation BackCom survey & Survey on BackCom principles, applications, and future directions; coded for BackCom modes, architecture, and IoT relevance. \\
\hline
9 & Xu \textit{et al.} \cite{Xu2023AIEmpoweredBackscatter} & 2023 & AI-empowered BackCom & AI-BackCom survey; coded for AI methods, signal detection, channel estimation, interference mitigation, and resource allocation. \\
\hline
10 & Zeng \textit{et al.} \cite{Zeng2016UAVComm} & 2016 & UAV communications & Foundational UAV communication study; coded for UAV placement, air-to-ground links, and trajectory-related communication design. \\
\hline
11 & Mozaffari \textit{et al.} \cite{Mozaffari2019UAVTutorial} & 2019 & UAV wireless networks & UAV tutorial/survey; coded for UAV roles, trajectory optimization, energy limits, and aerial network architecture. \\
\hline
12 & Yang \textit{et al.} \cite{Yang2019UAVBackscatterLCOMM} & 2019 & UAV-assisted BackCom & Early UAV-assisted BackCom work; coded for UAV data collection, trajectory design, and passive-device access. \\
\hline
13 & Hua \textit{et al.} \cite{Hua2020ThroughputUAVBackscatter} & 2020 & UAV-aided BackCom & Throughput maximization in UAV-aided BackCom; coded for protocol design, relay assistance, trajectory, and throughput metrics. \\
\hline
14 & Han \textit{et al.} \cite{Han2021UAVAidedBackscatter} & 2021 & UAV-aided BackCom & UAV-aided BackCom performance and trajectory optimization; coded for outage, throughput, and mobility-assisted link improvement. \\
\hline
15 & Yang \textit{et al.} \cite{Yang2021UAVBackscatter} & 2021 & Energy-efficient UAV BackCom & Energy-efficient UAV-assisted BackCom; coded for trajectory, scheduling, carrier-emitter power, harvested energy, and energy efficiency. \\
\hline
16 & Wang \textit{et al.} \cite{Wang2022UAVBackscatterRA} & 2022 & UAV BackCom resource allocation & Resource allocation for UAV-assisted BackCom; coded for power control, scheduling, reflection design, trajectory, fairness, and max--min rate. \\
\hline
17 & Zhong \textit{et al.} \cite{Zhong2024InterferenceUAVBackscatter} & 2024 & Interference-aware UAV BackCom & UAV ambient BackCom with interference cancellation; coded for direct-link interference, detection reliability, and signal extraction. \\
\hline
18 & Patwari \textit{et al.} \cite{Patwari2005LocatingNodes} & 2005 & Wireless localization & Foundational localization survey; coded for RSS, geometry-aware positioning, localization metrics, and sensor-network localization. \\
\hline
19 & Mao \textit{et al.} \cite{Mao2007Localization} & 2007 & Wireless localization & Foundational wireless localization survey; coded for localization features, measurement models, and positioning methods. \\
\hline
20 & Wymeersch \textit{et al.} \cite{Wymeersch2009CooperativeLocalization} & 2009 & Cooperative localization & Foundational cooperative localization work; coded for cooperation, multi-anchor geometry, and network localization principles. \\
\hline
21 & Zafari \textit{et al.} \cite{Zafari2019IndoorLocalization} & 2019 & Indoor / IoT localization & IoT localization survey; coded for RSS, CSI, AoA, ToA/TDoA, fingerprinting, and localization evaluation metrics. \\
\hline
22 & Pettorru \textit{et al.} \cite{Pettorru2024TrustworthyLocalization} & 2024 & Trustworthy IoT localization & Recent trustworthy localization survey; coded for robustness, privacy, trustworthiness, and secure localization concerns. \\
\hline\hline
\end{tabularx}
\end{table*}

\begin{table*}[h!]
\centering
\caption{Curated 48-Study Corpus Used for Quantitative Coding and Coverage Analysis (Part II of II).}
\label{tab:supp_curated_corpus_part2}
\footnotesize
\renewcommand{\arraystretch}{1.08}
\setlength{\tabcolsep}{2.2pt}
\begin{tabularx}{\textwidth}{|p{0.04\textwidth}|p{0.17\textwidth}|p{0.055\textwidth}|p{0.18\textwidth}|X|}
\hline\hline
\textbf{ID} & \textbf{Study} & \textbf{Year} & \textbf{Primary Stream} & \textbf{Reason for Inclusion / Coding Tags} \\
\hline\hline
23 & Zhang \textit{et al.} \cite{Zhang2020RobotBackscatterLocalization} & 2020 & Backscatter localization & Mobility-assisted backscatter localization; coded for passive-tag localization, mobile-anchor measurements, and localization accuracy. \\
\hline
24 & Cheng \textit{et al.} \cite{Cheng2023DroneSwarmRSS} & 2023 & UAV/drone passive localization & Drone-swarm/RSS localization study; coded for UAV geometry, RSS-based passive localization, and CRLB/accuracy-related analysis. \\
\hline
25 & Khalil \textit{et al.} \cite{Khalil2023UAVPassiveTDOA} & 2023 & UAV passive TDoA localization & UAV-based passive localization work; coded for TDoA, multi-UAV geometry, and localization accuracy under UAV mobility. \\
\hline
26 & Liu \textit{et al.} \cite{Liu2020JointRadarComm} & 2020 & Joint radar-communication / ISAC & Foundational joint radar-communication survey; coded for ISAC foundation, beamforming, waveform, and sensing-communication tradeoff. \\
\hline
27 & Zhang \textit{et al.} \cite{Zhang2021JCRSignalProcessing} & 2021 & JRC / ISAC signal processing & ISAC signal-processing foundation; coded for waveform design, beampattern, sensing gain, and radar-communication processing. \\
\hline
28 & Liu \textit{et al.} \cite{Liu2022ISACSurvey} & 2022 & ISAC survey & Comprehensive ISAC survey; coded for ISAC taxonomy, metrics, beamforming, waveform design, and communication-sensing tradeoffs. \\
\hline
29 & Zargari \textit{et al.} \cite{Zargari2023BackscatterSensingIntegration} & 2023 & Backscatter sensing integration & BackCom and sensing integration; coded for sensing-assisted BackCom, passive sensing, and B-ISAC relevance. \\
\hline
30 & Zhao \textit{et al.} \cite{Zhao2024BISAC} & 2024 & Backscatter-ISAC & B-ISAC framework; coded for tag detection, tag estimation, joint beamforming, SINR, rate, and estimation error. \\
\hline
31 & Liu \textit{et al.} \cite{Liu2024UAVISACIoT} & 2024 & UAV-assisted ISAC for IoT & UAV-ISAC for IoT; coded for 3D trajectory, sensing-task scheduling, power allocation, radar estimation rate, and communication rate. \\
\hline
32 & Meng \textit{et al.} \cite{Meng2022TrajectoryBeamformingIPSAC} & 2022 & UAV-enabled periodic ISAC & UAV trajectory and beamforming for periodic ISAC; coded for UAV mobility, sensing periodicity, and communication-sensing tradeoff. \\
\hline
33 & Meng \textit{et al.} \cite{Meng2023ThroughputUAVIPSAC} & 2023 & UAV-enabled periodic ISAC & UAV-enabled periodic ISAC throughput study; coded for trajectory, sensing scheduling, throughput, and sensing-communication allocation. \\
\hline
34 & O'Shea and Hoydis \cite{OShea2017DeepLearningPhysical} & 2017 & Deep learning for physical layer & Foundational physical-layer deep learning work; coded for end-to-end learning, signal detection, and channel-related AI. \\
\hline
35 & Zhang \textit{et al.} \cite{Zhang2019DeepLearningWireless} & 2019 & Deep learning for wireless networks & Wireless deep learning survey; coded for AI-enabled resource management, channel estimation, signal processing, and wireless optimization. \\
\hline
36 & Sutton and Barto \cite{Sutton2018RL} & 2018 & Reinforcement learning & Foundational RL reference; coded for sequential decision-making, UAV control, trajectory optimization, and policy learning. \\
\hline
37 & Mao \textit{et al.} \cite{Mao2017MEC} & 2017 & Mobile edge computing & MEC survey/foundation; coded for edge computing, low-latency processing, and edge-supported UAV/IoT intelligence. \\
\hline
38 & McMahan \textit{et al.} \cite{McMahan2017FedAvg} & 2017 & Federated learning & Foundational federated averaging work; coded for FL, privacy-preserving distributed learning, and model-update aggregation. \\
\hline
39 & Kairouz \textit{et al.} \cite{Kairouz2021FederatedLearning} & 2021 & Federated learning survey & Comprehensive FL survey; coded for federated optimization, privacy, communication overhead, non-IID data, and edge learning. \\
\hline
40 & Xu \textit{et al.} \cite{Xu2020EdgeIntelligence} & 2020 & Edge intelligence & Edge intelligence survey; coded for edge inference, split/cloud-edge learning, latency, and resource-constrained intelligence. \\
\hline
41 & Zhang \textit{et al.} \cite{Zhang2025IntelligentISAC} & 2025 & Intelligent ISAC & AI-enabled ISAC study/survey; coded for intelligent ISAC, learning-assisted sensing/communication, and model generalization. \\
\hline
42 & 3GPP TR~38.901 \cite{ThreeGPP38901} & 2020 & Channel models/standardization & Standardized channel model; coded for terrestrial propagation, simulation reproducibility, and benchmark channel assumptions. \\
\hline
43 & 3GPP TR~36.777 \cite{ThreeGPP36777} & 2017 & Aerial / UAV channel models & Aerial vehicle channel-modeling study; coded for UAV channel assumptions, altitude effects, and air-to-ground modeling. \\
\hline
44 & Alkhateeb \cite{Alkhateeb2019DeepMIMO} & 2019 & Wireless dataset/benchmarking & DeepMIMO dataset; coded for ray-tracing channels, ML benchmarks, mmWave/MIMO learning, and reproducibility. \\
\hline
45 & Klautau \textit{et al.} \cite{Klautau2021Raymobtime} & 2021 & Wireless dataset/mobility & Raymobtime dataset; coded for mobility-aware ray tracing, time-varying channels, and learning benchmarks. \\
\hline
46 & Levie \textit{et al.} \cite{Levie2021RadioUNet} & 2021 & Radio-map learning dataset & RadioUNet/RadioMapSeer-related work; coded for radio-map estimation, localization-related learning, and reproducible datasets. \\
\hline
47 & Hoydis \textit{et al.} \cite{Hoydis2022Sionna} & 2022 & Differentiable wireless simulation & Sionna simulator; coded for link-level reproducibility, differentiable PHY simulation, and open wireless-learning evaluation. \\
\hline
48 & Pineau \textit{et al.} \cite{Pineau2021Reproducibility} & 2021 & ML reproducibility & ML reproducibility checklist/work; coded for reproducibility, reporting standards, hyperparameters, datasets, and experimental protocols. \\
\hline\hline
\end{tabularx}
\end{table*}

Fig.~\ref{fig:ai_technique_distribution_exact_corpus} presents the manually coded distribution of AI and optimization techniques in the curated corpus. 
Conventional optimization remains prevalent because many UAV-assisted BackCom and ISAC studies formulate trajectory, reflection, beamforming, power control, and scheduling as non-convex optimization problems \cite{Yang2021UAVBackscatter,Wang2022UAVBackscatterRA,Liu2024UAVISACIoT}. 
Deep learning and reinforcement learning are increasingly used for channel estimation, signal detection, UAV control, and adaptive resource allocation, while federated/edge learning, transfer learning, and trustworthy AI remain less frequent but important for deployment-oriented systems \cite{Xu2023AIEmpoweredBackscatter,Kairouz2021FederatedLearning,Nguyen2021FL6G,Zhang2025IntelligentISAC}.

Fig.~\ref{fig:localization_feature_distribution_exact_corpus} shows the manually coded distribution of localization features and metrics. 
RSS remains common because it is simple and low-cost, while CSI, AoA, phase, ToA/TDoA, fingerprinting, Doppler, and CRLB/FIM-based metrics appear in more advanced localization and sensing studies \cite{Patwari2005LocatingNodes,Wymeersch2009CooperativeLocalization,Zafari2019IndoorLocalization,Win2018NetworkLocalization}. 
This confirms that UAV-assisted backscatter localization is moving from simple power-based methods toward richer geometry-aware, array-assisted, and learning-assisted localization frameworks.

Table~\ref{tab:quantitative_dimension_matrix} reports the dimension-level coverage of representative research streams using a three-level scale: $0$ = not central, $1$ = partially covered, and $2$ = strongly covered. 
Fig.~\ref{fig:dimension_coverage_matrix_heatmap} visualizes the same coverage pattern as a heatmap. 
Together, Table~\ref{tab:quantitative_dimension_matrix} and Fig.~\ref{fig:dimension_coverage_matrix_heatmap} show that no existing research stream fully covers UAV mobility, passive BackCom, localization, ISAC, AI, zero-energy operation, and reproducibility at the same time. 
This motivates the integrated perspective adopted in this survey.

\begin{table*}[!t]
\centering
\caption{Dimension-Level Quantitative Coverage Matrix of Representative Research Streams.}
\label{tab:quantitative_dimension_matrix}
\small
\renewcommand{\arraystretch}{1.25}
\setlength{\tabcolsep}{4pt}
\begin{tabularx}{\textwidth}{|p{0.19\textwidth}|c|c|c|c|c|c|c|X|}
\hline\hline
\textbf{Research Stream} & \textbf{UAV} & \textbf{BackCom} & \textbf{Loc.} & \textbf{ISAC} & \textbf{AI} & \textbf{Energy} & \textbf{Reprod.} & \textbf{Main Observation} \\
\hline\hline
UAV-assisted BackCom & 2 & 2 & 0 & 0 & 1 & 2 & 1 & Strong in UAV mobility and passive communication, but weak in sensing and localization. \\
\hline
Backscatter localization & 1 & 2 & 2 & 1 & 1 & 1 & 1 & Strong in passive positioning, but limited in UAV-ISAC and full energy-neutral optimization. \\
\hline
UAV-ISAC & 2 & 0 & 1 & 2 & 1 & 1 & 1 & Strong in trajectory-aware sensing and communication, but generally assumes active devices. \\
\hline
Backscatter-ISAC & 0 & 2 & 1 & 2 & 1 & 1 & 1 & Strong in passive sensing and communication, but UAV mobility is usually absent. \\
\hline
AI-BackCom & 0 & 2 & 1 & 0 & 2 & 1 & 1 & Strong in learning-assisted BackCom, but not focused on UAV localization and ISAC. \\
\hline
Benchmarking and datasets & 1 & 0 & 1 & 1 & 2 & 0 & 2 & Useful for reproducibility, but current datasets rarely model passive UAV-backscatter-ISAC. \\
\hline
\textbf{Integrated direction of this survey} & 2 & 2 & 2 & 2 & 2 & 2 & 2 & Calls for unified modeling, benchmarking, and evaluation across all dimensions. \\
\hline\hline
\end{tabularx}
\end{table*}

The quantitative synthesis provides three main insights. 
First, the number of selected studies has increased rapidly in recent years, especially in UAV-assisted BackCom, ISAC, AI-enabled optimization, and trustworthy localization. 
Second, localization and sensing metrics are reported inconsistently, making direct comparison difficult. 
Third, AI-based methods are increasingly visible, but reproducible datasets, open simulation platforms, and hardware validation remain limited \cite{Alkhateeb2019DeepMIMO,Klautau2021Raymobtime,Hoydis2022Sionna,Pineau2021Reproducibility}.

\subsection{Benchmarking, Datasets, and Reproducibility}
\label{subsec:benchmarking_reproducibility}

Benchmarking should reflect the multi-objective nature of UAV-assisted backscatter localization and ISAC. 
A benchmark that evaluates only throughput may favor aggressive reflection and short UAV-to-tag distances while ignoring sensing accuracy, harvested energy, and UAV propulsion cost. 
A benchmark that evaluates only localization accuracy may favor repeated measurements and dense UAV trajectories while ignoring latency and tag energy. 
Therefore, future benchmarks should include scenarios in which communication, localization, sensing, and energy requirements compete, so that algorithms can be compared under realistic trade-offs.

A useful benchmark should include both baseline algorithms and baseline scenarios. 
Baseline algorithms may include fixed UAV trajectories, nearest-neighbor service, equal reflection coefficients, communication-only designs, sensing-only designs, model-based optimization, and learning-based control. 
Baseline scenarios should vary tag density, RF-source placement, UAV altitude limits, sensing-target distribution, interference level, and energy-harvesting threshold. 
Reporting results over such scenarios would make it easier to assess robustness.

Benchmarking remains weak in UAV-assisted backscatter localization and ISAC. 
Most studies use customized simulations, channel assumptions, UAV mobility constraints, and energy-harvesting models, making cross-paper comparison difficult. 
A reproducible benchmark should define common scenarios, RF-source locations, tag distributions, UAV altitude ranges, channel models, reflection coefficients, sensing targets, energy-harvesting models, interference assumptions, and metrics. 
Standardized models such as 3GPP TR~38.901 and aerial-vehicle studies such as 3GPP TR~36.777 provide useful starting points \cite{ThreeGPP38901,ThreeGPP36777}, but they do not fully capture cascaded RF-source--tag--UAV channels, tag orientation, reflection states, and passive energy constraints.

Datasets are another bottleneck. 
DeepMIMO provides ray-tracing-based channels for mmWave and massive MIMO learning \cite{Alkhateeb2019DeepMIMO}. 
Raymobtime provides mobility-aware ray-tracing datasets \cite{Klautau2021Raymobtime}. 
RadioMapSeer and RadioUNet support radio-map estimation and localization-related tasks \cite{Levie2021RadioUNet,Yapar2023RadioMapDataset}. 
Sionna supports reproducible link-level simulation and differentiable physical-layer research \cite{Hoydis2022Sionna}. 
Future datasets should include UAV trajectories, RF-source signals, passive tag responses, sensing targets, localization labels, energy-harvesting traces, and interference conditions. 
A normalized benchmarking score can be written as
\begin{equation}
    \mathcal{B}=
    \omega_c\bar{R}
    +\omega_s\bar{S}
    +\omega_l(1-\bar{e}_{\mathrm{loc}})
    +\omega_e\bar{\eta}_{\mathrm{E}}
    -\omega_{\tau}\bar{\tau}
    -\omega_{\kappa}\bar{C}.
\label{eq:benchmark_score}
\end{equation}
Here, all metrics are normalized before aggregation: $\bar{R}$ is communication performance, $\bar{S}$ is sensing performance, $\bar{e}_{\mathrm{loc}}$ is localization error, $\bar{\eta}_{\mathrm{E}}$ is energy efficiency, $\bar{\tau}$ is latency, and $\bar{C}$ is computational or communication overhead. 
The score in \eqref{eq:benchmark_score} is not intended to replace detailed metric-by-metric reporting; rather, it provides a compact way to summarize multi-objective performance under common assumptions.

\subsection{Open Challenges}
\label{subsec:open_challenges}
The following challenges are interdependent rather than isolated. 
For example, realistic channel modeling affects AI generalization, benchmarking quality, and hardware validation. Security and privacy affect the trustworthiness of federated learning, localization, and sensing. Energy-neutral operation affects trajectory planning, tag scheduling, and reflection design. Therefore, future progress will require integrated evaluation rather than isolated improvements in one metric.

\begin{figure*}[!t]
    \centering
    \includegraphics[width=0.8\textwidth]{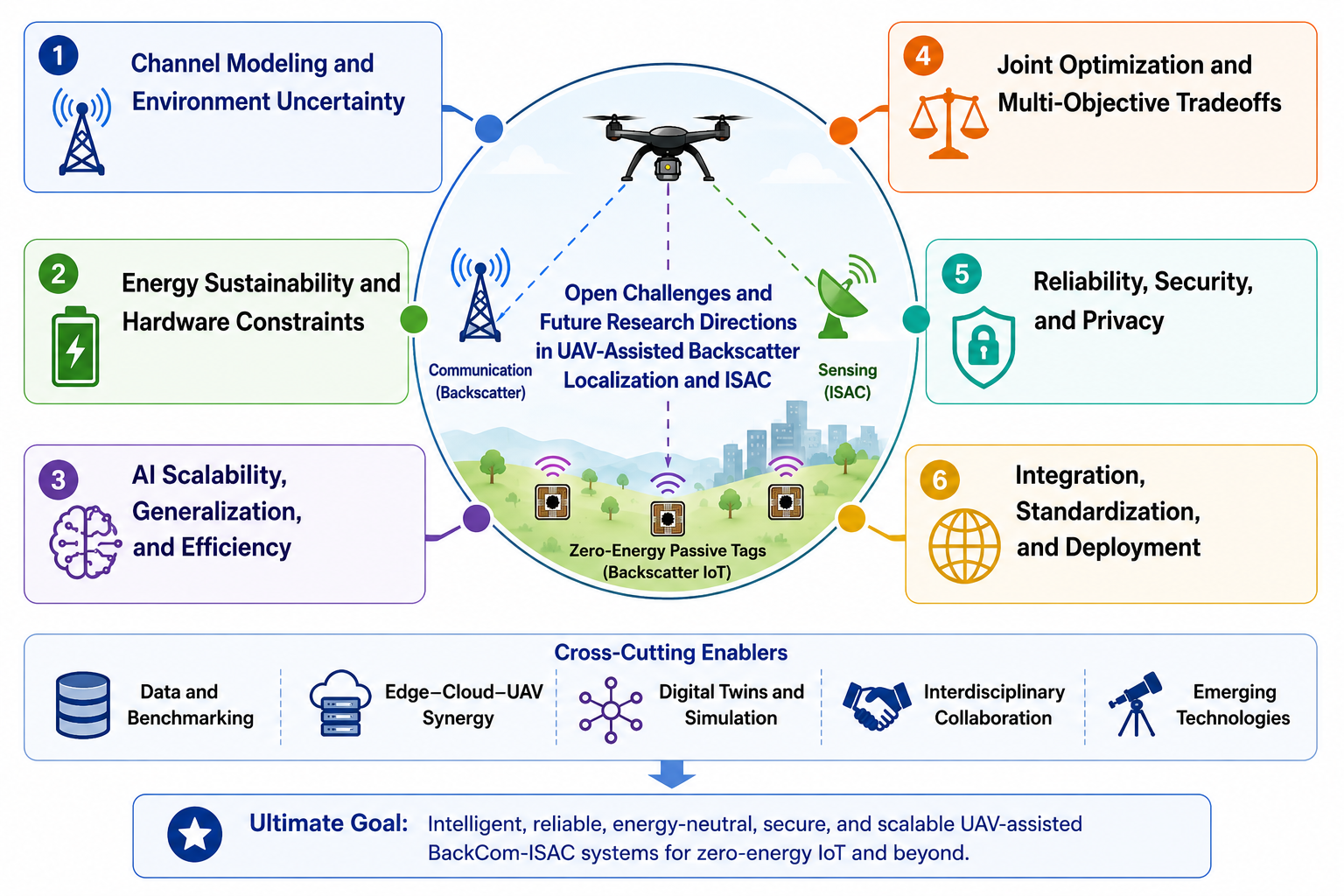}
    \caption{Open challenges and future research directions for AI-empowered UAV-assisted backscatter localization and ISAC.}
    \label{fig:open_challenges_future_directions}
\end{figure*}

Despite rapid progress, AI-empowered UAV-assisted backscatter localization and ISAC still face several technical and practical challenges. These arise from UAV mobility, passive backscatter links, localization uncertainty, sensing requirements, energy-neutral operation, and AI-driven optimization. Fig.~\ref{fig:open_challenges_future_directions} summarizes the major open challenges, future directions, and cross-cutting enablers.

\subsubsection{Realistic Channel Modeling for UAV-Backscatter-ISAC}
Model validation should include hardware effects such as antenna radiation pattern, UAV body shadowing, RF front-end nonlinearity, impedance-switching loss, oscillator drift, and synchronization errors. These effects may be negligible in active radio simulations but can dominate in weak passive backscatter links. Future channel models should therefore be calibrated using measurements that include UAV motion, passive tags, multiple RF sources, and representative environments.

Realistic channel modeling remains fundamental. 
UAV-assisted BackCom depends on aerial channels, terrestrial channels, cascaded backscatter channels, tag reflection properties, and direct-link interference. Aerial propagation is affected by altitude, LoS probability, non-stationarity, airframe shadowing, antenna orientation, and environment geometry \cite{Khuwaja2018UAVChannelModeling,Yan2019UAVChannelModeling}. 
Accurate and experimentally validated models are needed for sub-6 GHz, mmWave, and THz backscatter in the presence of UAV mobility, multipath, interference, and passive-tag constraints.

\subsubsection{Joint Optimization of Communication, Sensing, Localization, and Energy}
A promising approach is to combine model-based optimization with learning. Model-based methods can enforce physics and constraints, while learning can adapt to unknown channels, mobility, and interference. Such hybrid methods can reduce sample complexity compared with purely model-free reinforcement learning. 
However, they must be evaluated not only for final performance but also for convergence speed, computational cost, and feasibility on UAV or edge hardware.

Another challenge is joint optimization across communication, sensing, localization, and energy. In zero-energy UAV-backscatter-ISAC, the design must jointly consider tag activation, harvested energy, reflection control, localization uncertainty, sensing accuracy, rate, latency, and UAV flight energy. Increasing reflection improves reception and sensing visibility but may reduce harvested energy \cite{Liu2024UAVISACIoT}. 
Future frameworks must support multi-objective, cross-layer, and real-time optimization.

\subsubsection{Scalable and Generalizable AI}
Generalization should be tested across environments, not only across random seeds. For example, a model trained in an open rural area may fail in urban or industrial environments due to different multipath, blockage, and interference. Domain adaptation and meta-learning can help, but they should be paired with uncertainty estimation so that the system can identify unreliable predictions and switch to conservative policies.

AI can improve signal detection, channel prediction, trajectory control, localization, sensing inference, and resource allocation \cite{Xu2023AIEmpoweredBackscatter}. However, training data may be scarce, expensive to label, environment-specific, or affected by hardware non-idealities. Scalable AI must be lightweight, transferable, uncertainty-aware, robust, and capable of online adaptation with limited feedback \cite{Zhang2025IntelligentISAC}.

\subsubsection{Security, Privacy, and Trustworthiness}
Security is also tied to energy constraints. 
Passive tags may not support conventional cryptographic handshakes or frequent authentication messages. Lightweight physical-layer authentication, challenge-response backscatter, RF fingerprinting, and anomaly detection may therefore be more suitable. At the same time, UAVs and edge nodes should protect localization data and model updates because compromised AI models can affect communication and sensing decisions.

Backscatter links are vulnerable to eavesdropping, spoofing, replay attacks, jamming, and malicious RF illumination due to the limited cryptographic and computational capabilities of passive tags. 
Localization information is sensitive because it can reveal positions, movements, behavior, and operational patterns \cite{Pettorru2024TrustworthyLocalization}. Secure and trustworthy systems require lightweight authentication, physical-layer security, robust localization, privacy-preserving learning, adversarially robust AI, and explainable decisions.

\subsubsection{Experimental Validation and Hardware Prototyping}
Such platforms do not need to begin with full-scale deployments. 
Small indoor testbeds with programmable RF sources, mobile robots or small UAVs, passive tags, and software-defined radios can provide valuable insight into cascaded channels, tag response, and sensing/localization behavior. Outdoor field trials can then validate scalability, mobility, and environmental robustness. Open-source hardware descriptions and measurement datasets would greatly improve reproducibility.

Experimental validation remains limited. Most UAV-assisted BackCom and B-ISAC work relies on idealized channel models, simplified mobility models, or numerical simulations. Practical experiments must account for UAV vibration, orientation changes, antenna patterns, synchronization errors, regulations, hardware impairments, and real RF interference. Future platforms should jointly test UAV mobility, passive BackCom, localization, ISAC, energy harvesting, and AI-based control.

\subsection{Future Research Directions}
\label{subsec:future_research_directions}
The future directions below are intended to support a practical research roadmap. The first direction focuses on shared benchmarks; the second on joint physical-layer and mobility design; the third on sustainable operation; the fourth on trustworthy AI; and the fifth on integration with 6G technologies. Together, they point toward a system-level research agenda rather than isolated algorithmic improvements.

Future studies should move beyond isolated optimization problems and develop scalable, reproducible, and experimentally validated solutions that jointly consider passive operation, UAV mobility, localization, sensing, communication, and AI.

\subsubsection{Open Benchmark Platforms and Reproducible Evaluation}
Benchmark platforms should also provide standardized evaluation scripts. In many AI-enabled wireless studies, differences in preprocessing, train/test splitting, reward normalization, and stopping criteria make results hard to reproduce. A shared benchmark with fixed training and testing scenarios and standardized reporting templates would enable fair comparisons among optimization-based, learning-based, and hybrid approaches.

Future research should develop realistic benchmark platforms that combine ray tracing, UAV mobility, tag reflection models, energy harvesting, sensing targets, localization labels, and interference \cite{Levie2021RadioUNet}. Integrating DeepMIMO-like ray-tracing, RadioMapSeer-like radio-map learning, and Sionna-like differentiable simulation can support reproducible research \cite{Alkhateeb2019DeepMIMO,Hoydis2022Sionna}.

\subsubsection{Joint UAV Trajectory, Reflection, Beamforming, and Sensing Design}
Future multi-UAV designs should also consider heterogeneous UAV capabilities. Some UAVs may carry stronger transmitters, while others may carry better sensing hardware or edge processors. 
Assigning roles dynamically according to energy state, channel quality, sensing priority, and tag distribution can improve mission efficiency. This creates a new class of task-allocation problems for heterogeneous UAV-backscatter-ISAC networks.

Future UAV-backscatter-ISAC systems should jointly control UAV position, transmit beam, reflection coefficient, energy harvesting, localization uncertainty, sensing waveform, and scheduling. 
Hybrid methods combining convex optimization, alternating optimization, model predictive control, and safe reinforcement learning may be more reliable than purely data-driven methods \cite{Sutton2018RL,Qin2023DRLUAVISAC}. For multi-UAV deployments, distributed optimization and multi-agent learning can support cooperative sensing, localization, and data collection \cite{Chang2023MultiUAVDRL,Zhang2021HDRLUAVBackscatter}.

\subsubsection{Energy-Neutral and Sustainable Network Operation}
Sustainability should also include network maintenance and environmental impact. Battery replacement is difficult in large-scale passive deployments, and frequent UAV charging or replacement can offset the benefits of zero-energy tags. Future studies should therefore report not only instantaneous energy efficiency but also mission-level energy consumption, tag activation probability, network lifetime, and maintenance cost.

Future systems should guarantee tag energy causality, long-term lifetime, and sustainable RF energy use rather than only maximizing throughput or sensing accuracy. RF energy harvesting and wireless-powered communication provide foundations for modeling harvested energy, circuit power, and rate-energy tradeoffs \cite{Lu2015RFEH,Bi2016WPCN,Clerckx2019WIPTFundamentals}. 
Net-zero ISAC in backscatter systems further confirms that communication, sensing, and energy sustainability should be evaluated jointly \cite{Zhang2024NetZeroISACBackscatter}.

\subsubsection{Lightweight, Explainable, and Trustworthy AI}
Explainability should be connected to engineering decisions. 
For example, an explainable UAV-control model should indicate whether a trajectory action was selected to improve harvested energy, reduce localization uncertainty, avoid interference, or enhance sensing gain. Such explanations can help debug learning policies, support regulatory review, and increase user trust in autonomous UAV-assisted sensing systems.

Edge intelligence can reduce latency and backhaul overhead by moving inference closer to UAVs, edge servers, and base stations \cite{Mao2017MEC,Xu2020EdgeIntelligence}. Federated learning can preserve privacy by allowing local training without sharing raw RF, sensing, or localization data \cite{McMahan2017FedAvg,Kairouz2021FederatedLearning}. 
TinyML, pruning, quantization, split learning, and knowledge distillation can reduce inference cost in resource-constrained systems \cite{Dutta2021TinyML}. Explainable AI can clarify whether decisions are driven by channel quality, localization uncertainty, harvested energy, sensing needs, or interference \cite{Arrieta2020XAI}.

\subsubsection{Integration with 6G, RIS, MEC, and Digital Twins}
The integration with 6G technologies should be considered gradually. The IMT-2030 vision identifies ubiquitous connectivity, integrated sensing and communication, AI-native operation, and sustainability as key directions for future wireless systems \cite{ITUR2023IMT2030}. These objectives are closely aligned with UAV-assisted backscatter localization and ISAC, in which passive operation, sensing awareness, edge intelligence, and flexible coverage must be jointly designed. Ambient IoT and related standardization efforts can further provide practical assumptions on passive-device classes, connectivity topologies, energy levels, and service requirements \cite{Qu2024AmbientIoT3GPP,ThreeGPP2025AmbientIoT}.

RIS can be introduced as a programmable propagation-control layer to improve RF illumination, mitigate coverage holes, and strengthen cascaded source--tag--receiver links in passive IoT environments \cite{Basar2019RIS,Wu2019RISWirelessNetworks,DiRenzo2020SmartRadioRIS}. In UAV-assisted backscatter systems, RISs may help redirect incident RF energy toward passive tags, improve backscatter-link reliability, and create more favorable sensing and localization geometries. However, RIS-assisted designs must account for phase-control overhead, channel-estimation difficulty, deployment cost, and the additional coupling among UAV trajectory, RIS configuration, tag reflection, and sensing requirements.

MEC can support real-time signal processing, localization, learning inference, and UAV coordination by moving computation closer to UAVs, gateways, and edge servers \cite{Mao2017MEC}. This is important because passive tags generally cannot execute complex learning or signal-processing algorithms. Therefore, computationally demanding tasks such as cascaded channel estimation, data fusion, beam control, localization inference, and reinforcement learning policy updates are more realistically executed on UAVs, access points, or nearby edge servers rather than on zero-energy tags.

Digital twins are especially promising for this integration because they can emulate UAV trajectories, RF illumination, passive tag responses, sensing targets, and AI-control decisions before field deployment \cite{Khan2022DigitalTwinWireless}. A digital twin can also support safer reinforcement learning by allowing policies to be trained and stress-tested under rare or risky events such as sudden blockage, UAV energy shortage, tag failure, spoofed measurements, or strong interference. The main challenge is to keep the digital twin sufficiently realistic while maintaining computational efficiency and ensuring that simulated policies transfer reliably to physical deployments.

Non-terrestrial network integration can extend UAV-assisted zero-energy IoT to remote, maritime, rural, and disaster-response scenarios where terrestrial infrastructure is unavailable or damaged \cite{Rinaldi2020NTNBeyond,Azari2022EvolutionNTN,Giordani2021NTN6G}. In such scenarios, UAVs may operate as temporary collectors, relays, illuminators, sensing platforms, or edge-intelligence nodes. Combining 6G-oriented requirements, RIS-assisted propagation control, MEC-based intelligence, digital-twin validation, Ambient IoT assumptions, and non-terrestrial connectivity can transform the current collection of separate research prototypes into a more coherent zero-energy IoT architecture.

\section{Conclusion}
\label{sec:conclusion}
This paper presented a comprehensive survey of AI-empowered UAV-assisted backscatter localization and ISAC for zero-energy IoT networks. The survey reviewed the main enabling technologies, including RF energy harvesting, BackCom, UAV-assisted networking, localization, ISAC, and AI-based wireless optimization. A structured review methodology was introduced, followed by a unified taxonomy that classifies existing studies according to network architecture, UAV role, backscatter mode, RF source, localization and sensing function, AI technique, and performance metric. The paper further examined UAV-assisted BackCom, passive-device localization, ISAC-enabled UAV-backscatter architectures, beamforming and reflection design, sensing--communication--energy tradeoffs, and AI-empowered optimization methods. The comparative analysis showed that existing works provide important progress but often treat UAV mobility, BackCom, localization, ISAC, and AI separately. Quantitative trends, coverage analysis, and tutorial-style numerical illustrations further highlighted key tradeoffs among coverage, harvested energy, backscatter link gain, localization accuracy, sensing quality, and array directivity. Major open challenges remain in realistic cascaded channel modeling, joint communication--sensing--localization--energy optimization, energy-neutral operation, scalable and trustworthy AI, benchmarking, security, privacy, and hardware validation. Future research should move toward integrated cross-layer frameworks that combine UAV trajectory control, reflection design, beamforming, edge/federated learning, RIS, MEC, digital twins, and 6G standardization to enable sustainable, intelligent, and sensing-aware zero-energy IoT infrastructure.

\section*{Acknowledgment}
The author would like to thank the Office of the Associate Provost for Research at the United Arab Emirates University (UAEU), UAE, for their support.

\bibliographystyle{IEEEtran}
\bibliography{References}


\end{document}